\documentclass[utf8]{FrontiersinHarvard} % for articles in journals using the Harvard Referencing Style (Author-Date), for Frontiers Reference Styles by Journal: https://zendesk.frontiersin.org/hc/en-us/articles/360017860337-Frontiers-Reference-Styles-by-Journal
\usepackage[]{url,lineno,microtype}%,subcaption}
\usepackage[backref=page]{hyperref}
\usepackage[onehalfspacing]{setspace}
\newcommand{\ca}{\mbox{Ca\,{\sc ii}~K\,}}
\DeclareUnicodeCharacter{2009}{\,} 
\DeclareUnicodeCharacter{03B1}{ }
\usepackage{xcolor}
\definecolor{greentheo}{rgb}{0.15,0.50,0.30}

\def\keyFont{\fontsize{8}{11}\helveticabold }
\def\firstAuthorLast{Chatzistergos {et~al.}} 
\def\Authors{Theodosios~Chatzistergos\,$^{1,*}$, Natalie~A.~Krivova\,$^{1}$, 
and Ilaria~Ermolli\,$^{2}$}
\def\Address{$^{1}$Max Planck Institute for Solar System Research, Justus-von-Liebig-Weg 3,	37077 G\"{o}ttingen, Germany \\
$^{2}$INAF Osservatorio Astronomico di Roma, Monte Porzio Catone, Italy }
\def\corrAuthor{Theodosios Chatzistergos}

\def\corrEmail{chatzistergos@mps.mpg.de}

\begin{document}
%\selectlanguage{English}
\onecolumn      %uncomment for submission
\sloppy
\firstpage{1}

\title[Full-disc \ca observations - a window to past solar magnetism]{Full-disc \ca observations - a window to past solar magnetism}
\author[\firstAuthorLast ]{\Authors} %This field will be automatically populated
\address{} %This field will be automatically populated
\correspondance{} %This field will be automatically populated

\extraAuth{}% If there are more than 1 corresponding author, comment this line and uncomment the next one.
%\extraAuth{corresponding Author2 \\ Laboratory X2, Institute X2, Department X2, Organization X2, Street X2, City X2 , State XX2 (only USA, Canada and Australia), Zip Code2, X2 Country X2, email2@uni2.edu}

\maketitle
\begin{abstract}
\section{Full-disc observations of the Sun in the \ca line provide one of the longest collections of solar data.
First such observations were made in 1892 and since then various sites around the world have carried out regular observations, with Kodaikanal, Meudon, Mt Wilson, and Coimbra being some of the most prominent ones.
By now, \ca observations from over 40 different sites allow an almost complete daily coverage of the last century.
\ca images provide direct information on plage and network regions on the Sun and, through their connection to solar surface magnetic field, offer an excellent opportunity to study solar magnetism over more than a century. This makes them  also extremely important, among others, for solar irradiance reconstructions and studies of the solar influence on Earth's climate.
However, these data also suffer from numerous issues, which for a long time have hampered their analysis.
Without properly addressing these issues, \ca data cannot be used to their full potential.
Here, we first provide an overview of the currently known \ca data archives and sources of the inhomogeneities in the data, before discussing existing processing techniques, followed by a recap of the main results derived with such data so far.}
\tiny
 \keyFont{ \section{Keywords:} solar observations, solar chromosphere, solar activity, plage, irradiance reconstruction} %All article types: you may provide up to 8 keywords; at least 5 are mandatory.
\end{abstract}

\section{Introduction}
The Sun is the closest star to Earth  and its all-dominant energy source \citep{kren_where_2017}. 
The last couple of decades have seen immense advancement in solar monitoring.
Observations made with satellites, such as the Solar and Heliospheric Observatory \citep[SoHO;][]{domingo_soho_1995}, Hinode \citep{kosugi_hinode_2007}, or Solar Dynamics Observatory \citep[SDO;][]{pesnell_solar_2012}, allowed us to study the Sun close-up and  provided time series of many solar parameters with high quality and high cadence, such as with the Geostationary Operational Environmental Satellites \citep[GOES;][]{donnelly_solar_1977}, Variability of solar IRradiance and Gravity Oscillations \citep[VIRGO;][]{frohlich_virgo_1995}, PRoject for On-Board Autonomy \citep[PROBA;][]{teston_proba_1997,hochedez_lyra_2006}, Solar Radiation and Climate Experiment \citep[SORCE;][]{rottman_sorce_2005}, or Solar TErrestrial RElations Observatory \citep[STEREO;][]{kaiser_analysis_2005} among others.
Observations from space have the advantage of being unaffected by Earth's atmosphere, allowing measurement of quantities sensible to the atmospheric absorption, one example being the Total Solar Irradiance (TSI, the total, spectrally-integrated solar radiative energy flux measured at the top of Earth's atmosphere). 
Current and future missions, such as the Solar Orbiter \citep{muller_solar_2020}, Parker Solar Probe \citep{fox_solar_2016}, Aditya-L1 \citep{seetha_aditya-l1_2017,tripathi_solar_2017}, Solar-C \citep{watanabe_solar-c_2014,shimizu_solar-c_2020}, and balloon-borne Sunrise \citep{solanki_sunrise_2010,solanki_second_2017,feller_sunrise_2020} will keep pushing our understanding of the physics of the Sun \citep{kusano_pstep_2021}. 

As marvellous and beneficial as they are, satellite observations of the Sun exist for only a few decades, which is a rather short time interval to assess potential long term changes in various characteristics of the Sun.
Among others, this is particularly critical for assessing the long-term variability in TSI and its effect on Earth’s climate \citep{gray_solar_2010,solanki_solar_2013-1,krivova_solar_2018,intergovernmental_panel_on_climate_change_climate_2021}.

Luckily, a wealth of ground-based solar observations exist that can be used to study solar behaviour over longer periods in the past.
The invention of the telescope in 1600's marked the start of systematic observations of the Sun \citep{vaquero_sun_2009}, which however became regular only in the 1800's.
These early solar observations were limited to the white-light (or visible) part of the solar spectrum, which samples the photosphere bringing out mostly dark regions, called sunspots. 
They revealed the variable nature of solar activity, represented by increases and subsequent decreases in the number of sunspots, the famous 11-year sunspot cycle \citep{wolf_mittheilungen_1850-1}.  

The use of a prism to disperse the solar spectrum, e.g. by Herschel \citep{herschel_investigation_1800} or Secchi \citep{ermolli_legacy_2021,chinnici_angelo_2021}, in the 1800's allowed sampling different heights in the solar atmosphere.
However, systematic photographic observations of different heights of the solar atmosphere only started after the invention of the spectroheliograph by \cite{hale_note_1890,hale_kenwood_1891,hale_solar_1893}.  
The first line that was systematically observed with a spectroheliograph was the singly ionized Calcium line (\ca), at 3933.67 \AA.
The earliest recorded observation goes back to 1892 and since then a plethora of such observations from many places around the globe has been collected \citep{chatzistergos_analysis_2020}. 

Observations in the \ca line sample the lower chromosphere and provide direct information on plage and network regions, which are the chromospheric counterpart of faculae and network in the photosphere. 
These regions are manifestations of the solar surface magnetic field and one of the keys to studies of its evolution. They are also indispensable to reconstructions of past irradiance variations \citep[see, e.g.,][]{solanki_solar_2013-1}.
There are various other chromospheric data \citep{ermolli_solar_2015} that can provide information on plage regions, such as the series of the 10.7 cm radio flux \citep[available since 1947;][]{tapping_next_2013}, Lyman $\alpha$ emission \citep[available since 1969;][]{woods_improved_2000}, Mg II index \citep[available since 1978;][]{heath_mg_1986,snow_comparison_2014}, He~{\sc i} (10830\AA) equivalent width (since 1977); Ca~{\sc ii}~(8542\AA) central depth (since 1978), H$\alpha$ central depth (since 1984), CN (3883~\AA) bandhead index \citep[since 1979;][]{livingston_sun-as--star_2007}, and \ca 0.5 or 1~\AA~ disc-integrated emission index \citep[since 1974;][]{white_variability_1998}. 
However, all these measurements cover significantly shorter periods of time and are disc-integrated quantities, carrying less information than the full-disc \ca observations.
This renders \ca observations a unique dataset for studying past solar magnetism and activity, solar irradiance reconstructions and studies of the solar influence on Earth's climate.

Despite the invaluable information encrypted in \ca data, they are known for suffering from numerous issues which are challenging to be accounted for and eventually lead to their disuse. 
Over the last couple of decades there has been a renewed interest in these data. 
In view of this, considerable efforts have been put into the digitisation of historical archives, which allows a more systematic exploitation of these data. 

Here we provide an overview of recent efforts aimed at a systematic exploitation of the potential of full-disc \ca data.
We first review the currently known \ca archives and describe instruments used for such observations (Sect. \ref{sec:data}).
We then discuss the main techniques that have been developed for processing \ca images (Sect. \ref{sec:processing}).
This is followed by an overview of the main results derived with \ca data so far, specifically focusing on studies regarding Carrington maps (Sect. \ref{sec:carringtonmaps}), \ca plage areas (Sect. \ref{sec:plageareas}), network regions (Sect. \ref{sec:network}), the relationship between magnetic field strength and \ca brightness (Sect. \ref{sec:bvscak}), as well as reconstructions of irradiance variations (Sect. \ref{sec:irradiance}).
Finally, we summarise the current status of studies employing \ca data in Sect. \ref{sec:summary}.

\newcounter{tableid}
\begin{table*}
	\caption{List of spectroheliograph Ca~II~K datasets. Columns are: name of the observatory, type of detector (if both plate and CCD were used then the date in parenthesis refers to the year the transition occurred), estimated period of observations, whether the data have been digitised, total number of digital images (including multiple images on a single day when available), spectral width of the spectrograph, average pixel scale of the images, and the bibliography entry.}             
	\label{tab:observatoriesshg}      
	\centering                                      
	\small
	\begin{tabular}{l*{7}{c}}          
		\hline\hline                        
		Observatory & Detector &Period	&Digitised	  &Images&SW   			          &Pixel scale			  	  & Ref.\\
		&   		   		 &			 	&	      &      &[$\text{\AA}$]		  &[$"/$pixel] 			  	  &	\\
		\hline
Abastumani      &Plate           &1954--    &No       &-    &-         &-       &\addtocounter{tableid}{1}\thetableid\\  
Arcetri 		&Plate	 		 &1931--1974&Yes      &4871 &0.3 	   &2.5	    &\addtocounter{tableid}{1}\thetableid\\  
Cambridge 		&Plate           &1913--1941&No       &-    &-   	   &-       &\addtocounter{tableid}{1} \thetableid, \addtocounter{tableid}{1} \thetableid \\ 
Catania			&Plate           &1908--1977&Partially&1008 &- 	 	   &1.1--5  &\addtocounter{tableid}{1}\thetableid\\
Coimbra	   		&Plate/CCD (2007)&1925--&Yes      &19758&0.16 	   &2.2     &\addtocounter{tableid}{1}\thetableid\\
Crimea		    &Plate           &1955--1979&No       &-    &-  	   &-       &\addtocounter{tableid}{1} \thetableid  \\ 
Ebro 			&Plate           &1905--1937&No       &-    &-  	   &- 	    &\addtocounter{tableid}{1} \thetableid \\  	
Hamburg 	    &Plate           &1943--1958&No       &-    &-	 	   &-       &\addtocounter{tableid}{1} \thetableid \\  
Kenwood			&Plate	 	     &1892--1895&Partially&5    &-	 	   &3.1	    &\addtocounter{tableid}{1}\thetableid\\
Kharkiv			&Plate/CCD (1994)&1951--2021&Partially&564  &3.0 	   &3.3	    &\addtocounter{tableid}{1}\thetableid\\  
Kislovodsk		&Plate/CCD (2002)&1960--&Yes      &9738 &-  	   &1.3--2.3&\addtocounter{tableid}{1}\thetableid\\
Kodaikanal		&Plate	  		 &1904--2007&Yes      &45047&0.5 	   &0.9	  	&\addtocounter{tableid}{1}\thetableid\\ 
Kyoto			&Plate   	     &1928--1969&Yes      &3119 &0.74 	   &2.0		&\addtocounter{tableid}{1}\thetableid\\ 
Madrid 			&Plate           &1912--1917&No       &-	&-  	   &-       &\addtocounter{tableid}{1} \thetableid \\ 
Manila			&Plate	 		 &1968--1978&Partially&162	&0.5 	   &1.2		&\addtocounter{tableid}{1}\thetableid\\ 	
McMath-Hulbert	&Plate	 		 &1948--1979&Yes      &4932 &0.1 	   &3.1     &\addtocounter{tableid}{1}\thetableid\\
Meudon	   		&Plate/CCD (2002)&1893--&Yes      &20117&0.15, 0.09$^{(a)}$&1.1--2.2&\addtocounter{tableid}{1}\thetableid\\ 
Mitaka  		&Plate	 	     &1917--1974&Yes      &4193 &0.5 	   &0.7--0.9&\addtocounter{tableid}{1}\thetableid\\
Mount Wilson 	&Plate	 	     &1915--1985&Yes      &39545&0.2 	   &2.9	  	&\addtocounter{tableid}{1}\thetableid\\
Sacramento Peak &Plate 	 	     &1960--2002&Yes      &7750 &0.5 	   &1.2		&\addtocounter{tableid}{1}\thetableid\\
Schauinsland	&Plate			 &1944--1964&Partially&18   &- 	       &1.7, 2.6&\addtocounter{tableid}{-12} \thetableid\\
South Kensington&Plate           &1902--1912&No       &-    &- 	       &-       &\addtocounter{tableid}{13} \thetableid \\  
Wendelstein  	&Plate			 &1943--1977&Partially&422  &- 	       &1.7, 2.6&\addtocounter{tableid}{-13} \thetableid\\
Yerkes			&Plate	 		 &1899--1907&Partially&7	&- 	       &2.4		&\addtocounter{tableid}{13}\addtocounter{tableid}{1}\thetableid \\ 
		\hline
	\end{tabular}
	{\textbf{Notes: }$^{(a)}$ The two values correspond to the periods before and after 15 June 2017.\\
	\textbf{References: }
\addtocounter{tableid}{-\thetableid}
		(\addtocounter{tableid}{1}\thetableid) \citet{khetsuriani_abastumani_1967}; 
		(\addtocounter{tableid}{1}\thetableid) \citet{ermolli_digitized_2009}; 
		\addtocounter{tableid}{1} (\thetableid) \citet{hubrecht_sun_1912};
		\addtocounter{tableid}{1} (\thetableid) \citet{moss_report_1942};
		(\addtocounter{tableid}{1}\thetableid) \citet{zuccarello_solar_2011};
		(\addtocounter{tableid}{1}\thetableid) \citet{garcia_synoptic_2011};
		\addtocounter{tableid}{1} (\thetableid) \url{http://craocrimea.ru/ru/};
		\addtocounter{tableid}{1} (\thetableid) \citet{curto_historical_2016};
		(\addtocounter{tableid}{1}\thetableid) \citet{wohl_old_2005};
		(\addtocounter{tableid}{1}\thetableid) \citet{hale_solar_1893};
		(\addtocounter{tableid}{1}\thetableid) \citet{belkina_ccd_1996};
		(\addtocounter{tableid}{1}\thetableid) \citet{tlatov_synoptic_2015};
		(\addtocounter{tableid}{1}\thetableid) \citet{priyal_long_2014};
		(\addtocounter{tableid}{1}\thetableid) \citet{kitai_digital_2013};
		\addtocounter{tableid}{1} (\thetableid) \citet{vaquero_spectroheliographic_2007};
		(\addtocounter{tableid}{1}\thetableid) \citet{miller_new_1965};
		(\addtocounter{tableid}{1}\thetableid) \citet{mohler_mcmath-hulbert_1968};  
		(\addtocounter{tableid}{1}\thetableid) \citet{malherbe_new_2019}; 
		(\addtocounter{tableid}{1}\thetableid) \citet{hanaoka_long-term_2013};
		(\addtocounter{tableid}{1}\thetableid) \citet{lefebvre_solar_2005};
		(\addtocounter{tableid}{1}\thetableid) \citet{tlatov_new_2009};
		\addtocounter{tableid}{1} (\thetableid) \citet{lockyer_spectroheliograms_1909};
		(\addtocounter{tableid}{1}\thetableid) \citet{hale_rumford_1903}.
	}
\end{table*}

\newcounter{tableidfilter}
\begin{table*}
	\caption{List of filtergram Ca~II~K datasets. Columns are: name of the observatory, type of detector (if both plate and CCD were used then the date in parenthesis refers to the year the transition occurred), estimated period of observations, whether the data have been digitised, total number of digital images (including multiple images on a single day when available, except for Baikal, Calern, Kanzelhöhe, and Teide ChroTel. The values are approximate for the currently running observatories), spectral width of the interference filter, average pixel scale of the images, and the bibliography entry.}             
	\label{tab:observatoriesfilter}      
	\centering                                      
	\small
	\begin{tabular}{l*{7}{c}}          
		\hline\hline                        
		Observatory & Detector &Period&Digitised		  &Images&SW   			          &Pixel scale			  	  & Ref.\\
		&   		   		 &			& 		      &      &[$\text{\AA}$]		  &[$"/$pixel] 			  	  &	\\
		\hline
Anacapri	      &Plate           &1968--1973&No 	    &-    &-       &&\addtocounter{tableidfilter}{1} \thetableidfilter, \addtocounter{tableidfilter}{1} \thetableidfilter, \addtocounter{tableidfilter}{1} \thetableidfilter \\   
Baikal		      &Plate/CCD (2003)&1995--&No       &846  &1.2	   &2.7 				 	  &\addtocounter{tableidfilter}{1}\thetableidfilter\\
Big Bear  	      &Plate/CCD (1996)&1971--2006&Partially&5027 &3.2, 1.5&4.2, 2.4 &\addtocounter{tableidfilter}{1}\thetableidfilter, \addtocounter{tableidfilter}{1}\thetableidfilter\\
Brussels   		  &CCD	           &2012--    &-        &14699&2.7	   &1.0     				  &\addtocounter{tableidfilter}{1}\thetableidfilter\\   
Calern			  &CCD	           &2011--    &-        &1560 &7	   &1.0						  &\addtocounter{tableidfilter}{1}\thetableidfilter\\
Huairu            &CCD             &1991--2003&-        &3105 &2       &4.0                      &\addtocounter{tableidfilter}{1} \thetableidfilter \\ 
Kandilli          &Plate           &1968--1994&No       & -   &-      &-                         &\addtocounter{tableidfilter}{1} \thetableidfilter \\ 
Kanzelh\"ohe	  &CCD	           &2010--    &-        &8550 &3.0	   &1.0 					  &\addtocounter{tableidfilter}{1}\thetableidfilter\\
Kodaikanal        &CCD	           &1997--2006&-        &9411 &2.5     &2.2&\addtocounter{tableidfilter}{1}\thetableidfilter\\
Kodaikanal Twin	  &CCD 	           &2008--2013&-        &3059 &1.2	   &1.2	  				  	  &\addtocounter{tableidfilter}{1}\thetableidfilter\\ 
Kodaikanal WARM	  &CCD 		       &2017--    &-        &585  &1.0	   &2.4	  				  	  &\addtocounter{tableidfilter}{1}\thetableidfilter\\ 
Locarno           &Plate           &1958--1980&         &     &        &                          &\addtocounter{tableidfilter}{1}\thetableidfilter\\ 
Mauna Loa PSPT    &CCD		       &1998--2015&-        &31933&2.7	   &1.0	  					  &\addtocounter{tableidfilter}{1}\thetableidfilter\\
Mees	   		  &CCD		       &1988--1998&-        &1519 &1.2	   &5.5						  &\addtocounter{tableidfilter}{1}\thetableidfilter\\
Meudon	   		  &CCD  	       &2007--2014&-        &1519 &1.4	   &0.9						  &\addtocounter{tableidfilter}{1}\thetableidfilter\\ 
Mitaka  		  &CCD  	       &2015--    &-        &897  &4.5	   &1.0						  &\addtocounter{tableidfilter}{1}\thetableidfilter\\
Pic du Midi		  &CCD		       &2007--    &-        &3794 &2.5	   &1.2 	  				  &\addtocounter{tableidfilter}{1}\thetableidfilter\\
PICARD/SODISM	  &CCD		       &2010--2014&-        &47046 &7	   &1						  &\addtocounter{tableidfilter}{1}\thetableidfilter\\ 
PICARD/SOL	      &CCD		       &2010--2014&-        &14584 &7	   &1						  &\addtocounter{tableidfilter}{1}\thetableidfilter\\ 
Rome Monte Mario  &Plate	       &1964--1979&Partially&5826 &0.3	   &5.0						  &\addtocounter{tableidfilter}{1}\thetableidfilter\\
Rome PSPT         &CCD		       &1996--    &-        &3449 &2.5	   &2.0	  	  &\addtocounter{tableidfilter}{1}\thetableidfilter\\
Rome PSPT		  &CCD		       &2008--2019&-        &1298 &1.0 	   &2.0	  	  &\thetableidfilter\\ 
San Fernando CFDT1&CCD		       &1988--    &-        &4986 &9	   &5.1						  &\addtocounter{tableidfilter}{1}\thetableidfilter\\
San Fernando CFDT2&CCD		       &1992--    &-        &4065 &9	   &2.6 						  &\thetableidfilter\\ 
Schauinsland	  &Plate	       &1968--1984&No       & -	  &       &-       &\addtocounter{tableidfilter}{1}\thetableidfilter\\
South Pole        &CCD             &1981--1994&-        &-    &6--10       &2.1 &\addtocounter{tableidfilter}{1}\thetableidfilter, \addtocounter{tableidfilter}{1}\thetableidfilter\\
Teide ChroTel	  &CCD		       &2009--    &-        &1843 &0.3	   &1.0 					 	  &\addtocounter{tableidfilter}{1}\thetableidfilter \\  
Upice			  &CCD		       &1998--    &-        &3234 &1.6	   &4.0, 2.4  &\addtocounter{tableidfilter}{1}\thetableidfilter \\  
Valašské Meziříčí &CCD  	       &2011--    &-        &318  &2.4	   &1.8				  		  &\addtocounter{tableidfilter}{1}\thetableidfilter \\ 
		\hline
	\end{tabular}
	{\textbf{References: }
\addtocounter{tableidfilter}{-\thetableidfilter}
	    \addtocounter{tableidfilter}{1}(\thetableidfilter) \citet{kiepenheuer_fraunhofer_1969};
		\addtocounter{tableidfilter}{1} (\thetableidfilter) \citet{kiepenheuer_fraunhofer-institut_1974};
		\addtocounter{tableidfilter}{1} (\thetableidfilter) \citet{antonucci_chromospheric_1977};
		(\addtocounter{tableidfilter}{1}\thetableidfilter) \citet{golovko_data_2002}; 
		(\addtocounter{tableidfilter}{1}\thetableidfilter) \citet{naqvi_big_2010}; 
		\addtocounter{tableidfilter}{1} (\thetableidfilter) \citet{zirin_studies_1974};
		(\addtocounter{tableidfilter}{1}\thetableidfilter) \url{http://www.sidc.be/uset/}; 
		(\addtocounter{tableidfilter}{1}\thetableidfilter) \citet{meftah_solar_2018};
		(\addtocounter{tableidfilter}{1}\thetableidfilter) \citet{deng_reports_1997};
		\addtocounter{tableidfilter}{1} (\thetableidfilter) \citet{dizer_kandilli_1968};
		(\addtocounter{tableidfilter}{1}\thetableidfilter) \citet{potzi_kanzelhohe_2021};
		(\addtocounter{tableidfilter}{1}\thetableidfilter) \citet{singh_application_2022};
		(\addtocounter{tableidfilter}{1}\thetableidfilter) \citet{singh_twin_2012};
		(\addtocounter{tableidfilter}{1}\thetableidfilter) \citet{pruthvi_two-channel_2015};
		(\addtocounter{tableidfilter}{1}\thetableidfilter) \citet{waldmeier_swiss_1968};
		(\addtocounter{tableidfilter}{1}\thetableidfilter) \citet{rast_latitudinal_2008};  
		(\addtocounter{tableidfilter}{1}\thetableidfilter) \url{http://kopiko.ifa.hawaii.edu/KLine/index.shtml};
		(\addtocounter{tableidfilter}{1}\thetableidfilter) \url{http://bass2000.obspm.fr/data_guide.php};
		(\addtocounter{tableidfilter}{1}\thetableidfilter) \citet{hanaoka_past_2016};
		(\addtocounter{tableidfilter}{1}\thetableidfilter) \citet{koechlin_solar_2019};   
		(\addtocounter{tableidfilter}{1}\thetableidfilter) \citet{meftah_picard_2014};   
		(\addtocounter{tableidfilter}{1}\thetableidfilter) \citet{meftah_picard_2012};   
		(\addtocounter{tableidfilter}{1}\thetableidfilter) \citet{chatzistergos_historical_2019};
		(\addtocounter{tableidfilter}{1}\thetableidfilter) \citet{ermolli_romepspt_2022};
		(\addtocounter{tableidfilter}{1}\thetableidfilter) \citet{chapman_solar_1997};
		(\addtocounter{tableidfilter}{1}\thetableidfilter) \citet{wohl_old_2005};
		(\addtocounter{tableidfilter}{1}\thetableidfilter) \citet{jefferies_helioseismology_1988};
		(\addtocounter{tableidfilter}{1}\thetableidfilter) \citet{hagenaar_distribution_1997};
		(\addtocounter{tableidfilter}{1}\thetableidfilter) \citet{bethge_chromospheric_2011};
		(\addtocounter{tableidfilter}{1}\thetableidfilter) \citet{klimes_simultaneous_1999}; 	
		(\addtocounter{tableidfilter}{1}\thetableidfilter) \cite{lenza_system_2014}; 
	}
\end{table*}

\section{Ca II K data}
\label{sec:data}

\begin{figure*}[t!]
\begin{center}
\includegraphics[width=0.95\linewidth]{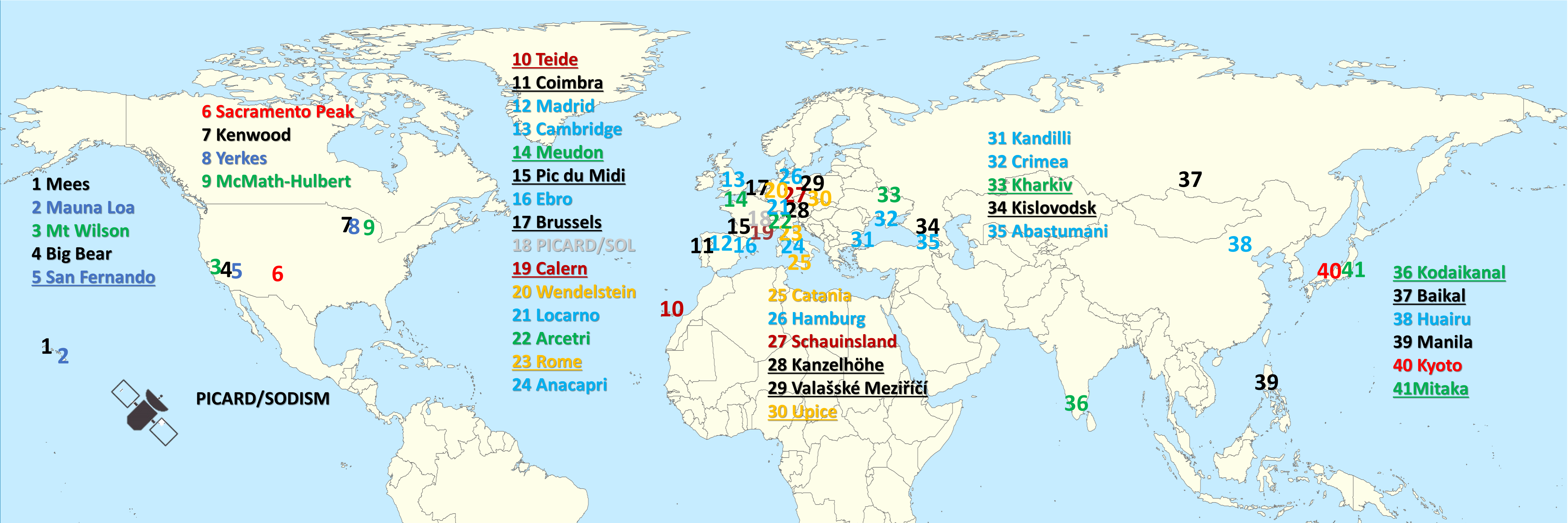}
\end{center}
\caption{Map showing approximate locations of the \ca observatories. Observatories marked in ciel either do not yet have data in digital form or there have been no published results from the digital images. All other archives are shown in various random colours. The locations are marked by numbers, while the names of the sites are listed with the corresponding numbers and colours to help associating the name with the observatory. 
The underlined names refer to sites performing \ca observations at present.
}\label{fig:worldmaparchives}
\end{figure*}

\ca line was the first line to be systematically monitored, with full-disc observations starting in 1892 and continuing up to the present day.
Since then, observations have been performed in \ca by over 40 observatories, covering different periods in time (see Tables \ref{tab:observatoriesshg} and \ref{tab:observatoriesfilter}), with (to the best of our knowledge) 16 sites performing observations in the \ca line at present. 
Figure \ref{fig:worldmaparchives} maps the approximate locations of the ground-based \ca observatories known to us, complemented by the SOlar Diameter Imager and Surface Mapper (SODISM) instrument onboard the PICARD \citep[][]{meftah_picard_2012} satellite.
All observatories are situated in the Northern hemisphere, with the exception of some observations performed at the South Pole (not shown on the map). The majority of sites are in Europe, with 11 currently active observatories (see underlined observatories in Fig. \ref{fig:worldmaparchives}). 
Three other active stations are located in Asia (Kodaikanal, Mitaka, Baikal), one in north America (San Fernando), and one on Canary islands (Teide).
All together, \ca archives provide an excellent temporal coverage of the entire 20th century, with at least one \ca observation per day for 
88\% of all days from 1892 onwards and 98\% from 1907 onwards among the data analysed by \cite{chatzistergos_analysis_2020}.

\begin{figure*}[t!]
\begin{center}
\includegraphics[width=0.95\linewidth]{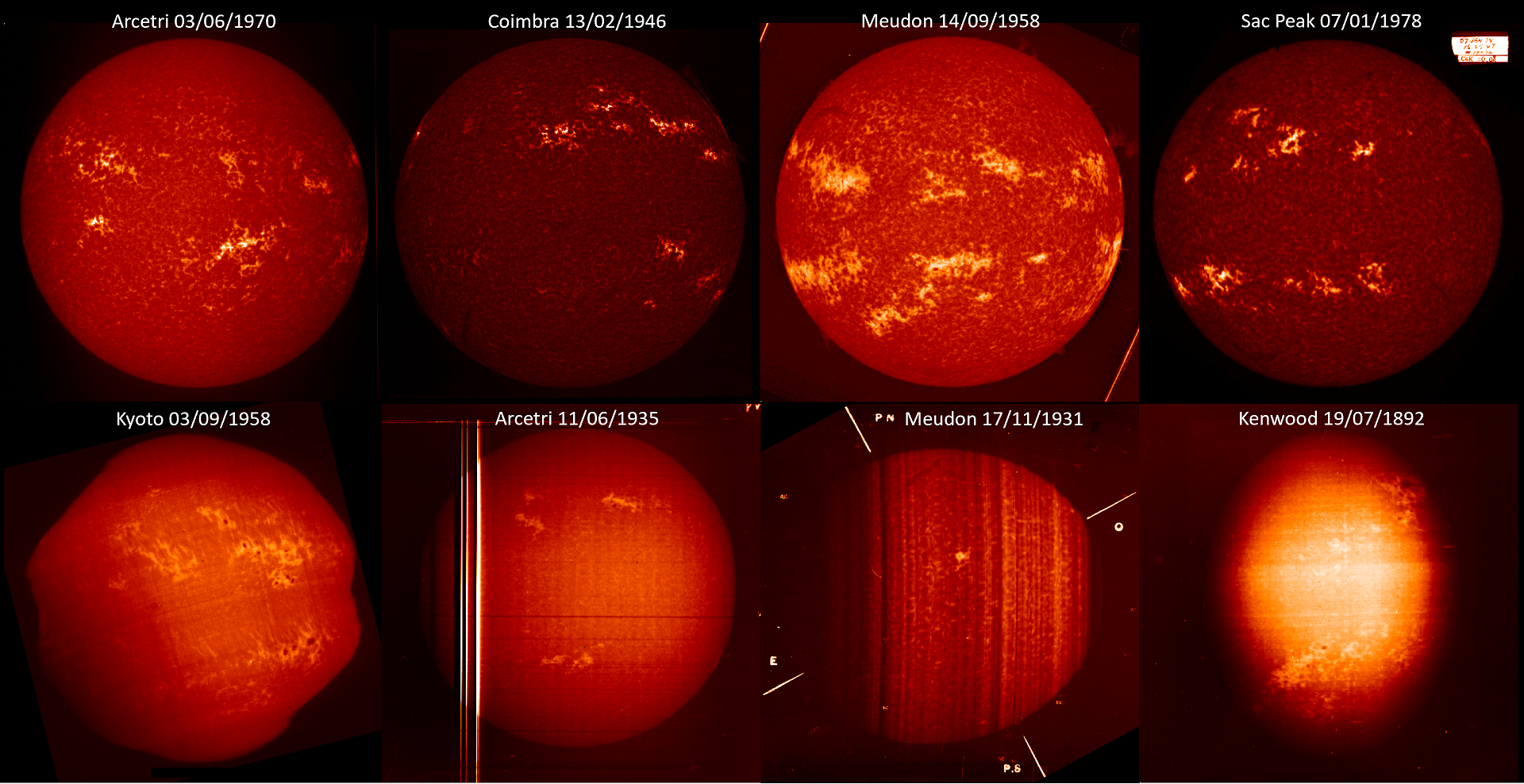}
\includegraphics[width=0.95\linewidth]{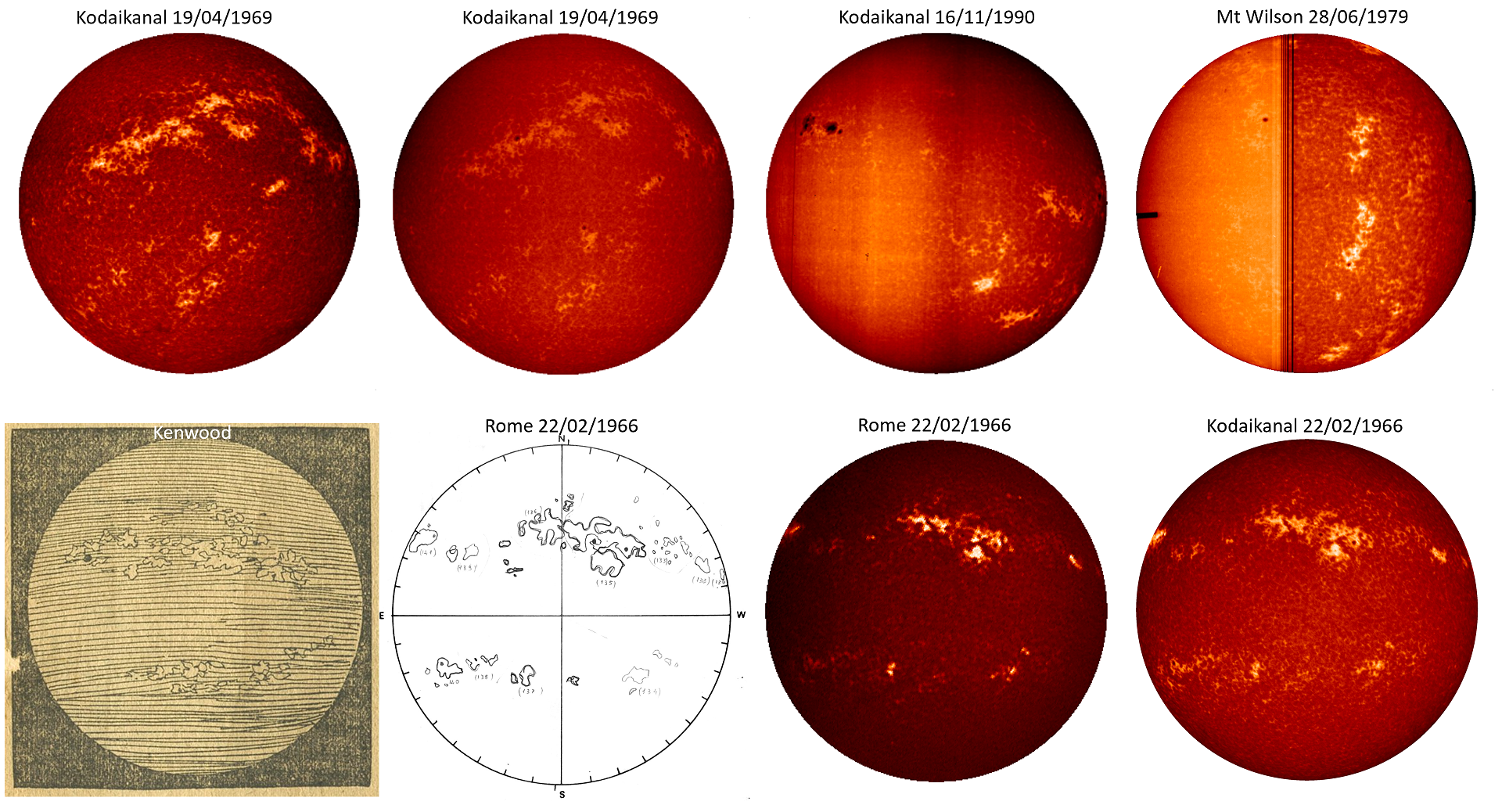}
\end{center}
\caption{Exemplary raw full-disc images of the Sun in the \ca line. The first row shows images of  good quality from four historical photographic archives, indicated at the top of each image together with the date of observation. The next three rows aim at highlighting some characteristics of such data. In particular, the second row shows image distortions due to spectroheliograph problems, while the third row showcases visual differences due to variations in central wavelength or bandpass used for observations. The last row shows \ca drawings from Kenwood and Rome observatories as compared to a filtergram and spectroheliogram from Rome and Kodaikanal taken on the same day as the Rome drawing.}\label{fig:wavelengthartefacts}
\end{figure*}

Tables \ref{tab:observatoriesshg} and \ref{tab:observatoriesfilter} list the main characteristics of all \ca archives known to us.
There are unfortunately only a few long-running archives, with the series from Meudon and Kodaikanal being the only ones extending for more than a century, although, at the time of writing this, Coimbra falls marginally short of a century.
Most archives perform only a few observations per day (typically between one and five), but some more recent ones (such as Rome Monte Mario, Kanzelhöhe, and Brussels) also have high-cadence observations.
For most applications of \ca data one observation per day is typically sufficient, however having multiple observations per day from various sources is extremely important.
On the one hand, there are applications for which the exact time of observation might be needed, for instance for comparing \ca images to magnetograms to select data pairs as close in time as possible \citep{chatzistergos_recovering_2019}, or to recover TSI variations within a day \citep{chatzistergos_reconstructing_2021-1}.
On the other hand, complementary observations from different sites over the same days allow assessment of the quality of the available datasets and identification of potential inconsistencies within the archives. Furthermore, multiple overlapping days from many archives allow a better cross-calibration of parameters extracted from these data, such as plage and network areas, discussed in Sect. \ref{sec:plageareas} and \ref{sec:network}.
Examples of historical \ca observations of good quality are given in the first row of Fig. \ref{fig:wavelengthartefacts}.
Tables \ref{tab:observatoriesshg} and \ref{tab:observatoriesfilter} reveal how different the observational characteristics of the various archives are.
Some sites used spectroheliographs, while others filters. 
More recent set-ups used a charge-coupled device \citep[CCD;][]{janesick_scientific_2001} camera to directly produce digital files, while the historical data were stored on photographic plates some of which are now available also in digital form.
To understand the underlying differences between the archives, we first introduce the instruments used to obtain \ca observations.

\subsection{Instrumentation used for observations}
\subsubsection{Spectroheliograph \& optical filters}
\label{sec:spectroheliograph}

\begin{figure*}[t!]
\begin{center}
\includegraphics[width=0.95\linewidth]{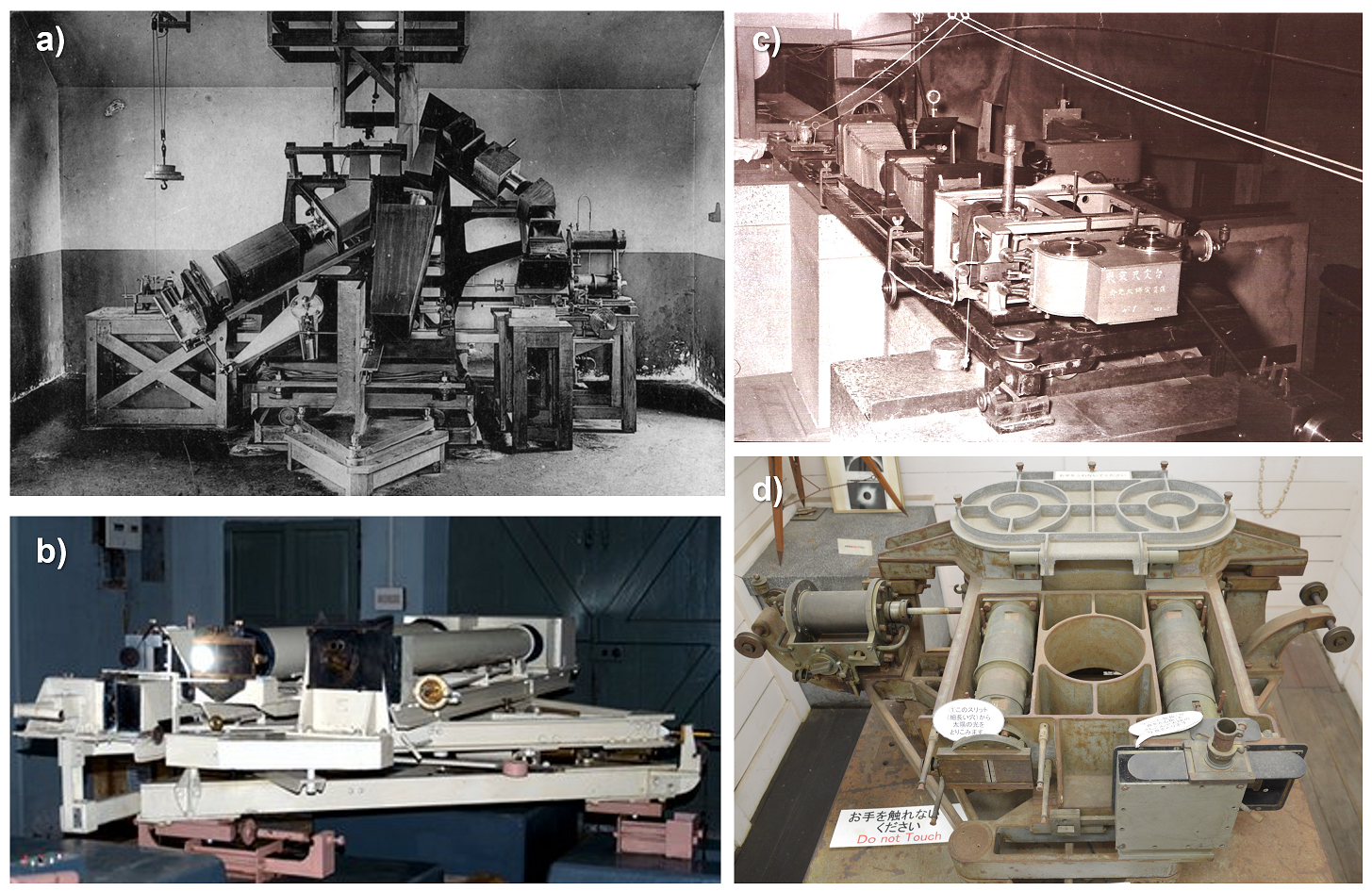}
\end{center}
\caption{Spectroheliographs used at the observatories of Meudon (panel a; \url{https://www.observatoiredeparis.psl.eu/the-meudon-spectroheliograph.html?lang=en}, Kodaikanal (panel b; \url{https://kso.iiap.res.in/new/instruments}, Mitaka (panel c; \url{https://solarwww.mtk.nao.ac.jp/en/topics/topics_0001.html}, and Kyoto (panel d; photo taken by T. Chatzistergos).} \label{fig:spectroheliographs}
\end{figure*}

We first discuss instruments used to isolate a selected limited range of wavelengths around the \ca line, which are the spectroheliograph and optical filters.
Spectroheliographs are in general big and bulky instruments.
A spectroheliograph employs a prism or a diffraction grating to disperse the incident solar light. 
A slit, with a controllable width, is then placed such as to allow only the radiation within the desired wavelength band to pass.
In this way, spectroheliographs allow isolating a spectral window around the \ca line.
\ca was the first line observed with a spectroheliograph, with the H$\alpha$ line following shortly after \citep{chatterjee_long-term_2017}.
It is important to note that a spectroheliograph does not allow observing the entire solar disc instantaneously, rather only a narrow strip imposed by another narrow slit at the entrance of the telescope. 
Various mechanical components are used to smoothly move the entrance slit and/or the photographic plate to scan and image the entire solar disc. 
As a result, a full scan of the solar disc requires a few minutes. 
Figure \ref{fig:spectroheliographs} shows photographs from some of the most prominent spectroheliographs, including Meudon, Kodaikanal, Kyoto, and Mitaka.

Unfortunately, neither the motion of the mechanical parts of the spectroheliograph nor the width and position of the slits are consistently precise. 
The uneven motion of the instrument leads to geometrical distortions of the recorded image, leading to certain strips being sometimes more stretched than others or even overlapping.
Thus brightness of the images might vary over different strips. 
Four such examples are shown in the second row of Fig.~\ref{fig:wavelengthartefacts}. 
In certain archives, such as Kyoto (see the first image of the second row in Fig.~\ref{fig:wavelengthartefacts}), these strips are curved \citep{chatzistergos_historical_2020}, most likely due to the arrangement of the optical parts of the telescopes. 
We note that such brightness variations across strips can also be introduced by changes of the atmospheric transmission during scanning of the entire disc.
Problems with maintaining the position and width of the spectral slit constant affect the altitude of the solar atmosphere that is sampled. 
The third row of Figure \ref{fig:wavelengthartefacts} shows three images from Kodaikanal observatory and one from Mt Wilson, demonstrating the effects of different setups of the spectral slit on the observations. 
The first two of them, from Kodaikanal, were taken on the same day but clearly sample different heights in the atmosphere (as can be seen by the absence of sunspots in the first image and their clear appearance in the second one). 
The third (Kodaikanal) and the fourth (Mt Wilson) images show the cases, where either the width or the centering of the slit changed over the course of scanning the solar disc. 
This produced images that look more photospheric (with large sunspots regions) on the left, while more chromospheric on the right sides of the images. 
Instrumental changes also affect the bandwidth of observations and thus the consistency of the long-running archives.
For instance, the grating system in Mt Wilson series was changed on the 21 of August 1923 \citep{chatzistergos_analysis_2019}, while Meudon changed from a 1-prism to a 3-prism system in 1908 \citep{dazambuja_annales_1930}, both leading to a better dispersed solar spectrum and thus narrower bandwidth observations.
\cite{chatzistergos_analysis_2019,chatzistergos_delving_2019} also discussed potentially degrading quality of Kodaikanal data with time.

Optical filters allow instantaneous observations of the whole solar disc. 
These are wavelength-selective filters placed at the optical path of the telescope, which means they are employed with a much simpler instrumental configuration than spectroheliographs.
Due to that, filtergrams do not exhibit severe image distortions like the ones mentioned for the spectroheliograms, e.g. brightness variations across linear segments.
However, the way the filter is placed \citep[for example if there is a potential tilt;][]{lofdahl_tilted_2011} can still affect the bandpass, while there can also be deterioration with time, thus changing the filter transmission profile.

\begin{figure*}[t!]
\begin{center}
\includegraphics[width=0.95\linewidth]{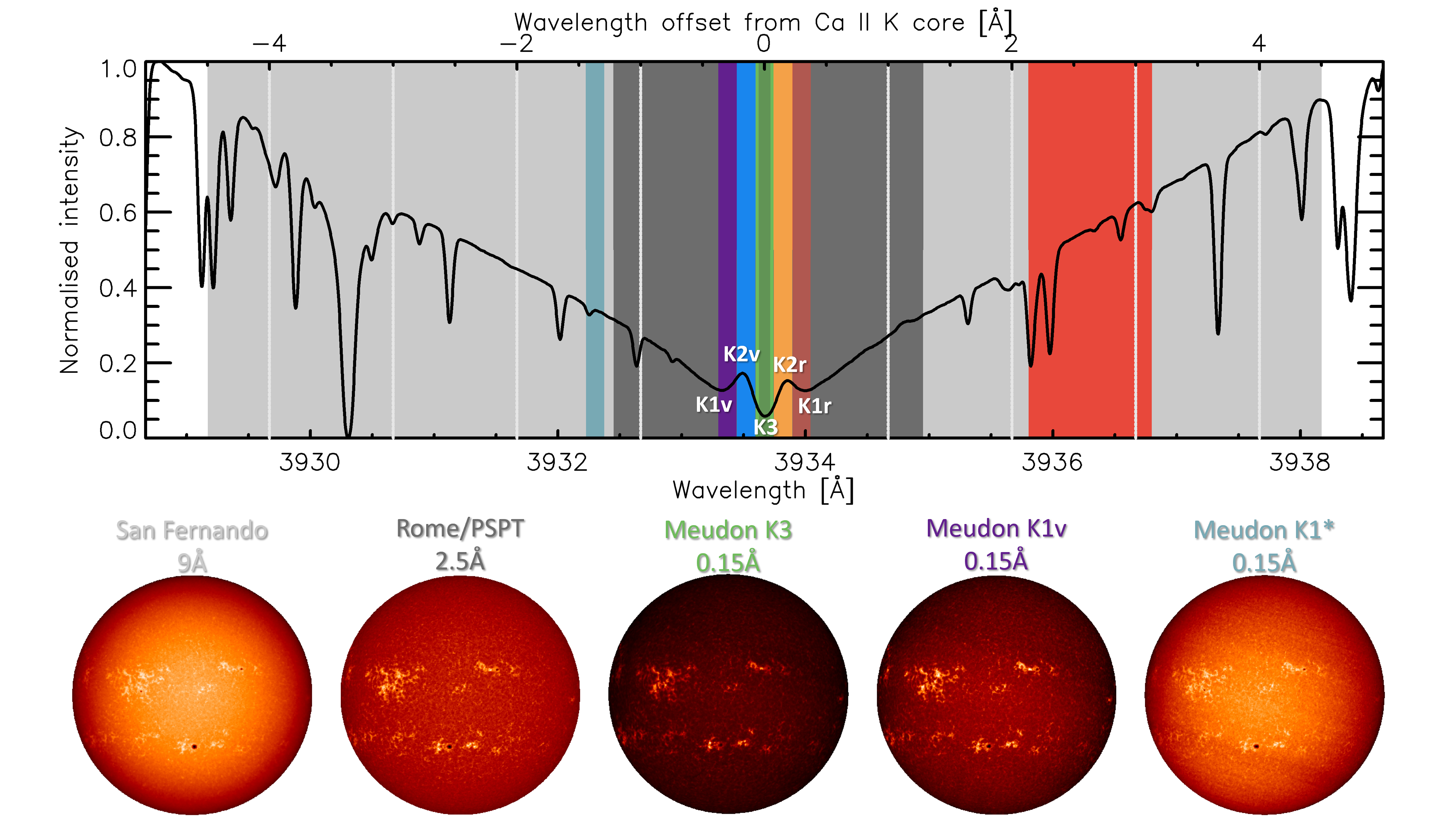}
\end{center}
\caption{The normalised high-resolution disc-integrated solar spectrum over a 10\AA~ interval centred at the core of the \ca line from the atlas of the Hamburg Observatory. Also shown are the
nominal pass-bands of selected \ca archives, including the two extreme cases from San Fernando (light grey) and Meudon K3 (light green up to 2017, darker green since 2017), an average case for CCD-based archives from the Rome/PSPT (dark grey), along with six off-band archives: MLSO/PSPT (red), Meudon K2r (orange), K1r (brown), K2v (blue), K1v (purple), and Meudon K1* (light blue, this is referred to as K1 in the Meudon database even though it is not at the K1 minimum, but centred further away at the wing of the line at 3932.3\AA). The vertical light grey lines mark wavelength intervals of 1 \AA~ from the core of the \ca line. The bottom part of the figure presents five exemplary observations taken on the same day (2013/09/23) at San Fernando, Rome/PSPT, Meudon K3, Meudon K1v, and Meudon K1*. }\label{fig:passbandarchives}
\end{figure*}

One very important aspect to keep in mind is that various observatories use non-identical observational settings, that is either filters with different transmission profiles or spectroheliographs with different slit widths, and thus sample different heights in the solar atmosphere \citep[See Fig. 3 in][]{ermolli_radiative_2010}.
Spectroheliographs were used since 1892, and thus most long-running archives were produced with spectroheliographs.
Filters started being used for \ca observations in the late 1950's \citep[e.g.][]{ohman_solar_1956}, and are employed at most of the currently running observatories. 
Currently, only four sites use a spectroheliograph, namely those at Coimbra, Kharkiv, Kislovodsk, and Meudon.
In general, filtergrams, have broader bandwidths than spectroheliographs.
\cite{chatzistergos_analysis_2020} showed that the average nominal bandwidth of 43 available datasets remained about 0.3\AA~between 1904 and 1987 when mainly spectroheliographs were used, while it increased to about 2.5\AA~on average for the later periods when filtergrams became more common.
Figure \ref{fig:passbandarchives} shows the average quiet Sun \ca line profile from the high-resolution disk-integrated atlas from the Hamburg Observatory\footnote{\url{ftp://ftp.hs.uni-hamburg.de/pub/outgoing/FTS-Atlas/}} \citep{neckel_spectral_1999,doerr_how_2016} highlighting various features of this line: the core denoted as K3, the emission peaks K2v and K2r, as well as the secondary minima K1v and K1r, where ``v'' and ''r'' stand for the violet and red parts of the wing, respectively.
Coloured vertical bars mark wavelength bands of some prominent \ca series, for comparison.
In particular, we show the passbands of spectroheliograph observations at Meudon (green), which since 2017 employs the narrowest known to us nominal bandwidth of 0.09\AA~ (prior to 2017, it was 0.15\AA) and thus includes only part of the K3 minimum. 
However, Meudon also took observations with offsets to the central wavelength (shown in light blue, purple, blue, orange, and brown). 
An average over modern CCD-based data is represented by the Rome/PSPT (precision solar photometric telescope; shown in dark grey) which has a considerably broader bandwidth than Meudon, including all K1, K2, and K3 features of the \ca line, but also extending more to the wing of the line, thus having more photospheric contribution. 
We also show the bandwidth of the San Fernando archive (light grey), which is, to our knowledge, the broadest one and extends even further into the wing than Rome/PSPT. 
We note that San Fernando observatory uses two telescopes called Cartesian full disk telescope, CFDT, 1 and 2.
To highlight the effects of the passband on the observations, in the lower half of Figure \ref{fig:passbandarchives} we show examples of  observations taken on the same day at different sites having different characteristics.
We emphasise that the previous discussion refers to the nominal bandwidth of the various observatories, while the actual ones might differ.

\subsubsection{Image capturing devices}
\label{sec:imagecapturing}

The more recent observations employ a CCD \citep[][]{janesick_scientific_2001} or a complementary metal-oxide semiconductor \citep[CMOS;][]{fossum_active_1993} sensor to store the images directly in digital formats (to simplify the discussion, in the following we will refer to CCD and CMOS collectively as CCD).
These are essentially linear detectors, meaning that the recorded image value is directly proportional to the incident radiation.
However, there can still be saturated or dark regions below the noise level.
CCD-based data are subject to the standard dark, bias, and flat-field calibration of CCD detectors \citep{meurs_flattening_1987}. 
This is an important step to reduce intensity biases and artefacts due to issues with the detector or dust in the telescope components.
To our knowledge, the earliest employment of such devices for \ca observations might have been at the South Pole in 1981 and then at the Mees and San Fernando observatories in 1988. 
All sites that started observations after 1995 employ a CCD  sensor.

\begin{figure*}[t!]
\begin{center}
\includegraphics[width=0.95\linewidth]{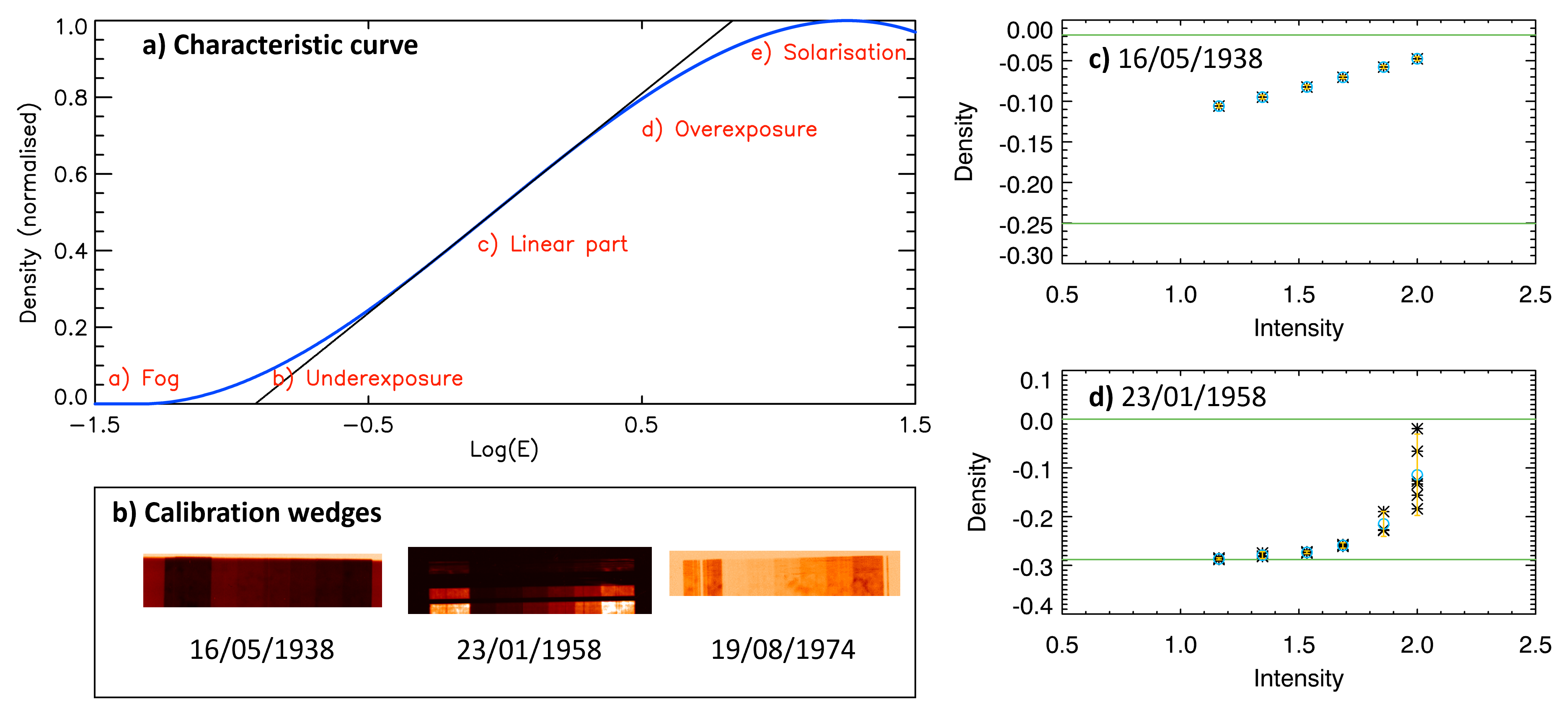}
\end{center}
\caption{a) Schematic of a characteristic curve of photographic observations (blue) and a linear fit (black) to the central part of the curve. b) Examples of calibration wedges from the Arcetri observatory. c-d) Characteristic curves determined from the first two Arcetri calibration wedges shown in panel b). Black asterisks show individual measurements from the wedges, blue circles the mean density value for each intensity value, while yellow marks the symmetric 1$\sigma$ interval. The green horizontal lines denote the minimum and maximum values found within the solar disc for each day. }\label{fig:characteristiccurve}
\end{figure*}

Earlier data were stored on photographic plates, which comprise a photosensitive emulsion coated on a glass plate or celluloid film (we will collectively refer to them as plates in the following). 
The photographic process involves the exposure of the emulsion to create a latent image, which is revealed after the application of developing agents \citep{mees_theory_1942,james_fundamentals_1968,dainty_image_1974}. 
Thus, exposed regions turn darker depending on the incident radiation.
In contrast to CCD sensors, photographic plates are not linear detectors. 
This means the recorded darkening on the plates is not directly proportional to the incident radiation.
The response of the plates, typically referred to as the characteristic curve or Hurter–Driffield curve \citep{hurter_photochemical_1890}, is defined as $d=f(\log{E})$, where $E$ is the exposure (defined as intensity multiplied by time) and $d$ is the density (defined as $d=\log_{10}{(1/T)}$, where $T$ is the transparency of the emulsion, a term describing how darkened the emulsion got).
The characteristic curve has in general a sigmoid shape (Fig. \ref{fig:characteristiccurve}) and is linear only in its central part (part c in Fig. \ref{fig:characteristiccurve}a), which is the region of proper exposure. 
Regions with low intensity of incident radiation usually end up on the underexposure level or even below the fog level (thus escaping being registered), while under excessive exposure,  darkening turns weaker than what is expected from the linear relation.
Knowledge of the characteristic curve is crucial for photometric calibration of photographic observations.
Unfortunately, this curve depends on multiple additional factors \citep[emulsion composition, developing agents, temperature and humidity levels at the time of the observation, storage conditions of the plates etc;][]{chatzistergos_analysis_2017}, with the consequence that each observation has its own unique characteristic curve, which needs to be determined to allow the photometric calibration.

The bulk of the historical \ca data are stored on photographic plates. 
However, there are also some plage drawings in \ca, such as the ones from the Rome observatory (see Fig. \ref{fig:wavelengthartefacts} 4th row).
Rome observatory has a collection of 1564 drawings covering 14 July 1965 to 14 July 1981, and thus ending two years after the last photographic observation from the same site.
Such drawings are of much more limited use than the photographic data, since they do not include information on the brightness of plage regions and carry additional uncertainty due to the manual aspect of the recorded observation (including some subjectivity in the definition and selection of the plage boundaries and the accuracy of drawing).
It is unclear how many observatories maintained drawings in the \ca line. 
We are not aware of any other \ca plage area drawings other than the ones from Rome and Kenwood observatories.
However, considering that many photographic data have been displaced due to shift of scientific interests at various observatories, it would not be surprising if similar fate found datasets of drawings as well.
Furthermore, digitising such collections is time consuming, while priorities would obviously be given to photographic \ca data due to their greater potential yield.
Notwithstanding these limitations, any such drawing collection would also be very important for recovering plage areas in the past, potentially even before the photographic data or to fill temporal gaps in the plage area series from photographic data.

\subsection{Digitisation}
\label{sect:digitisation}
An important step to allow drawings and observations stored on photographic plates to be used for quantitative analyses is to convert them into digital format.
Most early studies with \ca data in digital form were restricted to the authors digitising themselves a small subset of the data they needed for their specific purposes \citep[e.g.][]{antonucci_chromospheric_1977,munzer_pole-equator-difference_1989,mein_spectroheliograms_1990,kariyappa_variability_1994,nesme-ribes_fractal_1996,caccin_variation_1998}. 
It is unclear how many of those data are still available. 
Unfortunately, most of these data do not seem to have survived or, at least, are not publicly available.

A more systematic work on \ca data started only after the first digitisations of large samples or even entire data collections.
This process started in the mid 1990's with the data from Mt Wilson.
In particular, \cite{foukal_behavior_1996} digitised the Mt Wilson data covering 1915--1984 with an 8-bit CCD camera, which were then re-digitised  by \cite{lefebvre_solar_2005} with a 12-bit camera a few years later.
The next long series of \ca data to be digitised were the ones from Kodaikanal and Arcetri.
\cite{makarov_22-years_2004} digitised Kodaikanal data covering 1907--1999 with an 8-bit CCD camera. 
Also in this case, it was recognised that this digitisation was not perfect and also missing a big part of the data, which led to a new digitisation of the Kodaikanal data (now covering 1904--2007) by \cite{priyal_long_2014}  with a 16-bit CCD camera. \cite{chatzistergos_delving_2019} compared the 8 and 16 bit digitisations of Kodaikanal data. They showed that the quality of data in terms of spatial resolution was better in the 16~bit version, which allowed a more accurate extraction of plage areas from these images. 
The series from Arcetri was digitised in 2004 with a 1200 dpi and 16 bit scanner \citep{giorgi_calibration_2005,centrone_image_2005,marchei_digitization_2006,ermolli_digitized_2009}.

The following years saw increased interest in digitisation of \ca archives, with \cite{wohl_old_2005} having digitised a small sample of the Wendelstein and Schauinsland \ca data, \cite{tlatov_new_2009} the Sacramento Peak data with an 8-bit CCD camera, \cite{garcia_synoptic_2011} the Coimbra data, Solar Geophysical data (SGD, hereafter), the McMath-Hulbert data, \cite{kitai_digital_2013} the Kwasan observatory (in Kyoto) data, and \cite{tlatov_synoptic_2015} the Kislovodsk data.
Meudon has the longest collection among all \ca archives, however initially only the data since 1980 were digitised.
More recently, the digitisation of the data prior to 1980 was completed with a commercial scanner (EPSON perfection 4990 photo), while the data since 1980 are currently being redigitised with the same scanner \citep{malherbe_monitoring_2022}.
The Mitaka data have been digitised three times \citep[see also][]{chatzistergos_analysis_2019}. 
The first time the entire data collection was digitised with 8-bit depth \citep{hanaoka_long-term_2013}. 
The next two digitisations were done with 16-bit depth, but included only subsets of the data. 
In particular, the second digitisation included only the data before 02 March 1960, while the third one includes the data after 02 March 1960, but with a different set-up, and the first 10 observations of each year for the periods before 1961.
\cite{chatzistergos_historical_2019} presented the digitised Rome \ca observations which were stored in 35 mm celluloid films. 
These were patrol high-cadence observations with approximately 100 images recorded per day, and only a small sub-sample of the data could be digitised so far, typically two observations for each day when data exist.
% with reflecta RPS 10M commercial film scanner.

The observations from Catania, Kenwood, Kharkiv, Manila, and Yerkes are largely not digitised yet. 
Only four Kenwood images were found as lantern slides and digitised by the Division of History of Science and Technology at Yale University’s Peabody Museum of Natural History.
Similarly, a small subset of Yerkes observations were found and digitised by the University of Chicago Photographic Archive, Special Collections Research Center, University of Chicago Library.
Only a small sample of Kharkiv \ca data  \citep{chatzistergos_analysis_2020}, and
92 observations (between 1970 and 1971) from Catania \citep{chatzistergos_historical_2019} have been digitised so far.

\cite{chatzistergos_analysis_2020} digitised additional samples of Catania, Manila, Wendelstein, Rome, and Kodaikanal observations with the reflecta RPS 10M commercial film scanner.
These observations were part of the Photographic journal of the Sun, which was published between 1967--1978 by the observatory of Rome. 
The Photographic journal of the Sun included one \ca observation for each day in the form of 35 mm celluloid films. 
Rome observatory was the main provider of the data, but days without observations from Rome were filled with data from Catania, Wendelstein, Kodaikanal, or Manila observatories.
These digitised data from Kodaikanal, Rome, Wendelstein, and Catania complement the already available datasets (but only in cases when previously digitised archives did not include the data from the Photographic journal of the Sun). 
For Manila, this is to our knowledge the only available digital record of \ca data. 
We emphasise, however, that a recovery and digitisation of the originals would be beneficial.

Unfortunately, there are still numerous archives that have either not or only partially been digitised.
For example, the data from Kenwood, Kandilli, and Ebro have not yet been digitised.
Furthermore, since finding information about old \ca observations is not straightforward, there is a high chance that the list of archives in Tables \ref{tab:observatoriesshg} and \ref{tab:observatoriesfilter} is not complete.
For instance, \cite{bates_compact_1972} had presented a spectroheliograph for operation on a rocket or balloon and showed example observations in the \ca line. 
However, besides one observation included in their paper, it is unclear how many observations were performed or whether they have survived to this day.
Further, it cannot be excluded that many of the historical \ca observations have been lost. This seems to most likely have been the fate of the data from Madrid and Hamburg sites.
Also, the fate of Kharkiv data is up in the air at the time of writing this review. 
This highlights the importance of recovering such historical photographic archives and digitising them for preservation reasons \citep[see also][]{pevtsov_historical_2019}.

\subsection{Amateur observations}
All datasets listed so far come from professional observatories.
However, in more recent years optical filters and telescopes became also accessible to the public, which led to many observations in the \ca line by amateur astronomers.
There are even dedicated online archives where amateur astronomers host their solar observations, such as the international online solar database (IOSD; \url{http://solardatabase.free.fr/tableau_cak.php}).
The IOSD website hosts a large number of solar observations from all over the world with observations going back to 2005.

To our knowledge, such amateur observations have not been used for scientific purposes yet, for various reasons, including the following.
Different observers have different setups and therefore should be treated as different series (see, e.g., the discussion on the bandwidth effect in Sect.~\ref{sec:spectroheliograph}). 
Observations from a single observer might be very sporadic or too few to allow a proper inter-comparison with data from other sources. 
The intended purpose of such observations is not, in principle, to maintain an archive of scientific data, rather to create a collection of visually appealing images. 
This means there might be image editing, varying between images, aiming at improving the attractiveness of the image at the potential cost of information.
It is unclear if the standard CCD calibration has always been applied, or if it is applied consistently.
The recording or observing equipment might be constantly changing thus making it almost impossible to produce a coherent series of data for scientific use, while the location of observation might be varying too.
Lastly, the images are typically stored in reduced size compressed file formats such as JPG, which are optimised for internet storage of the files, but induce some loss of information.

\begin{figure*}[t!]
\begin{center}
\includegraphics[width=0.95\linewidth]{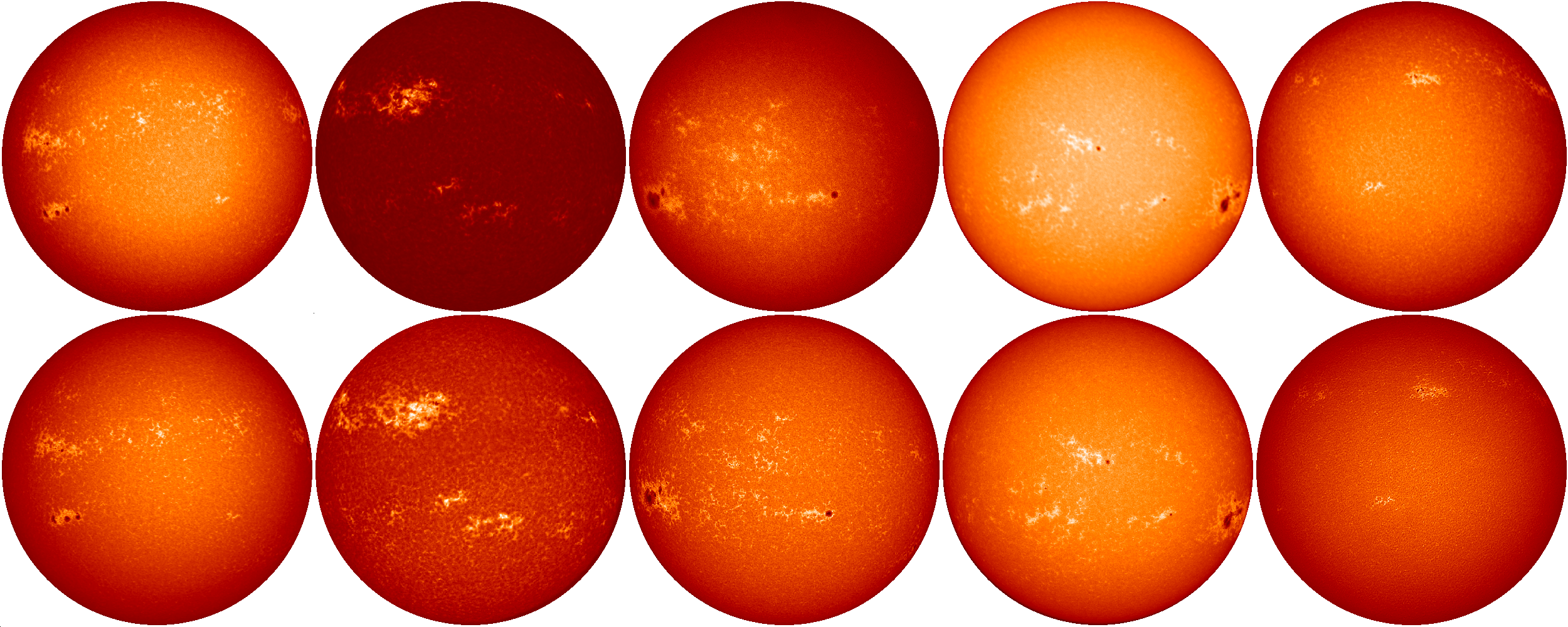}
\end{center}
\caption{Examples of amateur \ca observations (bottom row) along with co-temporal images from professional observatories (top row). From left to right, the images were taken on 24/12/2015, 21/04/2022, 19/10/2014, 28/10/2014, and 13/02/2010
by the amateur observers (top):
Michael Borman, Didier Favre, André Gabriël, Lastrofieffe, Stefano Sello, and at professional observatories (bottom): Kanzelhöhe, Meudon, Valašské Meziříčí, Calern, and Pic du Midi.}
\label{fig:amateur}
\end{figure*}

All these factors considerably reduce the applicability of such data for scientific studies, especially for studies such as reconstructions of past irradiance variations (see Sect. \ref{sec:irradiance}).
However, that is not to say that such data might not be usable at all. 
In Fig. \ref{fig:amateur} we present raw images from 5 observers from IOSD along with a co-temporal image from a professional observatory showing that the data can be of comparable quality. 
Thus, potentially, such data from amateur observers might be used for studies of plage areas to fill temporal gaps in professional archives.
Should amateur observers account for the issues mentioned above, they would potentially complement the professional archives and provide valuable information for solar and stellar physics studies.
For instance, some of the issues mentioned above can be alleviated if the observer preserved the original raw observation in a lossless file format.

\section{Processing methods}
\label{sec:processing}
In this section, we briefly describe and discuss the processing methods that are applied on \ca data.
The typical order of the processing steps is to apply the CCD calibration (including the data digitised with a CCD camera from photographic archives), then the photometric calibration for the photographic plates, followed by a number of pre-processing steps aiming at the compensation of the limb-darkening. 
To make the description of the methods easier to understand, here we will, however, deviate from this order. 
After an introduction into the various pre-processing steps, we will first describe the limb-darkening compensation process and finally the photometric calibration process. 

\subsection{Pre-processing}
As pre-processing we consider all steps that need to be applied to the data before the compensation for the limb darkening. 
As a first step, all images that are not yet stored as Flexible Image Transport System \citep[FITS;][]{wells_fits_1981} files, are converted into this format.
CCD-based data (again, including those digitised with a CCD camera) require the standard dark, bias, and flat-field calibration.

The most important step is to identify the solar disc on the images. 
This is usually done in an automatic way with edge detection techniques, such as Sobel filter, a Canny edge detector, or Hough transform \citep{veronig_automatic_2000,zharkova_full-disk_2003,curto_automatic_2008,ermolli_digitized_2009,chatterjee_butterfly_2016,suo_full-disk_2020,zhu_new_2020, potzi_kanzelhohe_2021}, while in some cases this was done manually by clicking with the computer mouse at 3-5 points at the limb \citep[e.g.][]{priyal_long_2014}.
The assumed shape of the recorded solar disc is usually either circular or elliptical. 
Assuming the recorded disc as an ellipse is more accurate in view of the fact that due to instrumental issues the recorded disc typically has an elliptical shape. 
The ellipticity is usually relatively weak for filtergrams but can be quite high for spectroheliograms \citep[ellipticities up to 0.82 have been found among the data analysed by][]{chatzistergos_analysis_2020}. %Kyoto Ca_19300911_02 
The coordinates of the centre of the disc along with the radius or the semi-major and semi-minor axes are typically added in the header of the FITS files. 
An exception to that was by \cite{priyal_long_2014} who determined a circular disc and then resized and cropped the images.
If the disc is identified as an ellipse it has to be re-sampled so to be circularised \citep{tlatov_new_2009,chatzistergos_historical_2020}.
This is not always straightforward either.
Sometimes stretching is inhomogeneous over individual parts of the disc. 
In such cases circularisation introduces new distortions.
Also other processing artefacts sometimes occur \citep[see][for more details]{chatzistergos_historical_2020}

Another pre-processing step is the determination of the orientation of the solar disc.
During this step, the images are typically (re-)oriented such as to have the north pole at the top and East at the left side.
The optical components of the telescope (e.g. existence and number of mirrors) determine whether the image needs to be inverted. 
Typically, compensation for ephemeris is sufficient to orient the images.
If the plates were placed appropriately and consistently during their digitisation, this would be the case for the historical photographic data as well.
However, since this is not always the case, pole markings were placed on the photographic plates of some archives, either during the observation or right before their digitisation, which should allow the determination of the orientation of the solar disc after the digitisation.
Such pole markings exist at least for samples of the Meudon, Kodaikanal, Mt Wilson, and Sacramento Peak data. 
The pole markings were used by \cite{priyal_long_2014} and \cite{tlatov_polar_2019} to orient the Kodaikanal and Kodaikanal, Mt Wilson, Sacramento Peak, and Meudon images, respectively. 
\cite{sheeley_carrington_2011} and \cite{bertello_70_2020} used a cross-correlation approach applied on subsequent images in order to determine the orientation of the images.

Finally, the photographic data are typically given in negatives (values of transparency). 
These need to be converted into densities (see Sect. \ref{sec:imagecapturing}).

Additional processing steps, often individual for specific archives, might be implemented at this point. 
For instance, this can include assessment of the quality of images or simply excluding very problematic observations.
As an example, \cite{jarolim_image-quality_2020} used generative adversarial networks to assess the quality of the images within the Kanzelhöhe data (they used H$\alpha$ but in principle it is applicable to \ca too) and categorise them.
They also used their approach to correct for various observational artefacts, such as large scale intensity variations or patterns introduced by the passage of clouds in the sky. 

\subsection{Limb-darkening and artefact compensation}
\label{sec:limbdarkening}
Limb darkening refers to the gradual decrease of the intensity of the disc from centre to its edge, when observed at near ultraviolet, visual, and  near infrared bands, due to the decrease of the temperature of the solar plasma with height from the solar surface to the bottom of the chromosphere.
It is important to compensate for the limb darkening in order to render the intensity values across the solar disc consistent and directly comparable with each other.
Furthermore, the recorded images suffer from numerous artefacts (see Sect. \ref{sec:spectroheliograph}), which need to be accounted for too.
In the following we describe processing techniques employed to compensate the limb darkening and correct various image artefacts. 
Some of them have been applied on different lines, such as H$\alpha$ or 1600\AA, but in principle they are applicable on \ca data too.
All methods described below are typically used up to 0.98$R$, in a few cases up to 0.99$R$ \citep{chatzistergos_analysis_2018}.
The available processing techniques can be put in the following 5 categories:
\begin{itemize}
    \item 1D polynomial fit to the average radial intensity profile;
    \item 2D polynomial fit to the entire image;
    \item 1D polynomial fits along columns and rows of the image;
    \item 2D running window median filter;
    \item combination of the above approaches.
\end{itemize}

The first group of methods determine a radially symmetric limb darkening background \citep{brandt_determination_1998, walton_processing_1998,johannesson_10-year_1998,denker_synoptic_1999,zharkova_full-disk_2003,diercke_chromospheric_2019,potzi_kanzelhohe_2021}. 
This is done by fitting polynomial functions (typically 4th order) along the average radial profile of the solar disc in terms of $\mu$ (cosine of the heliocentric angle). 
A caveat is that the recorded solar disc is rarely radially symmetric due to various observational circumstances (see Sect.~\ref{sec:spectroheliograph}), including essentially all photographic data and spectroheliograms.  
Obviously, this will not be accounted for with such methods. 
Sometimes, this approach is used as a first approximation followed by another approach, such as Zernike polynomials, removing the residual patterns \citep[e.g.][]{diercke_solar_2022,dineva_characterization_2022}. 

Several studies applied a 2D (4th order) polynomial fitting to the entire image \citep{caccin_variations_1997,caccin_variation_1998,worden_plage_1998}. 
This approach returned a background that was not radially symmetric, but it could not properly handle artefacts in the images.

More suitable for processing of spectroheliograms proved to be 1D polynomial fittings along columns and rows of the images \citep{worden_evolution_1998}. 
\cite{worden_evolution_1998} and \cite{priyal_long_2014} used 5th and 3rd degree polynomials, respectively. 
The downside is that different linear cuts through the solar disc cover quite different $\mu$-ranges, while they are all fitted with polynomials of a fixed degree.
This reduces the accuracy of the fits, especially closer to the limb. 
To account for this, \cite{worden_evolution_1998} repeated the fits after rotating the image by 45$^\circ$.

\begin{figure*}[t!]
\begin{center}
\includegraphics[width=0.95\linewidth]{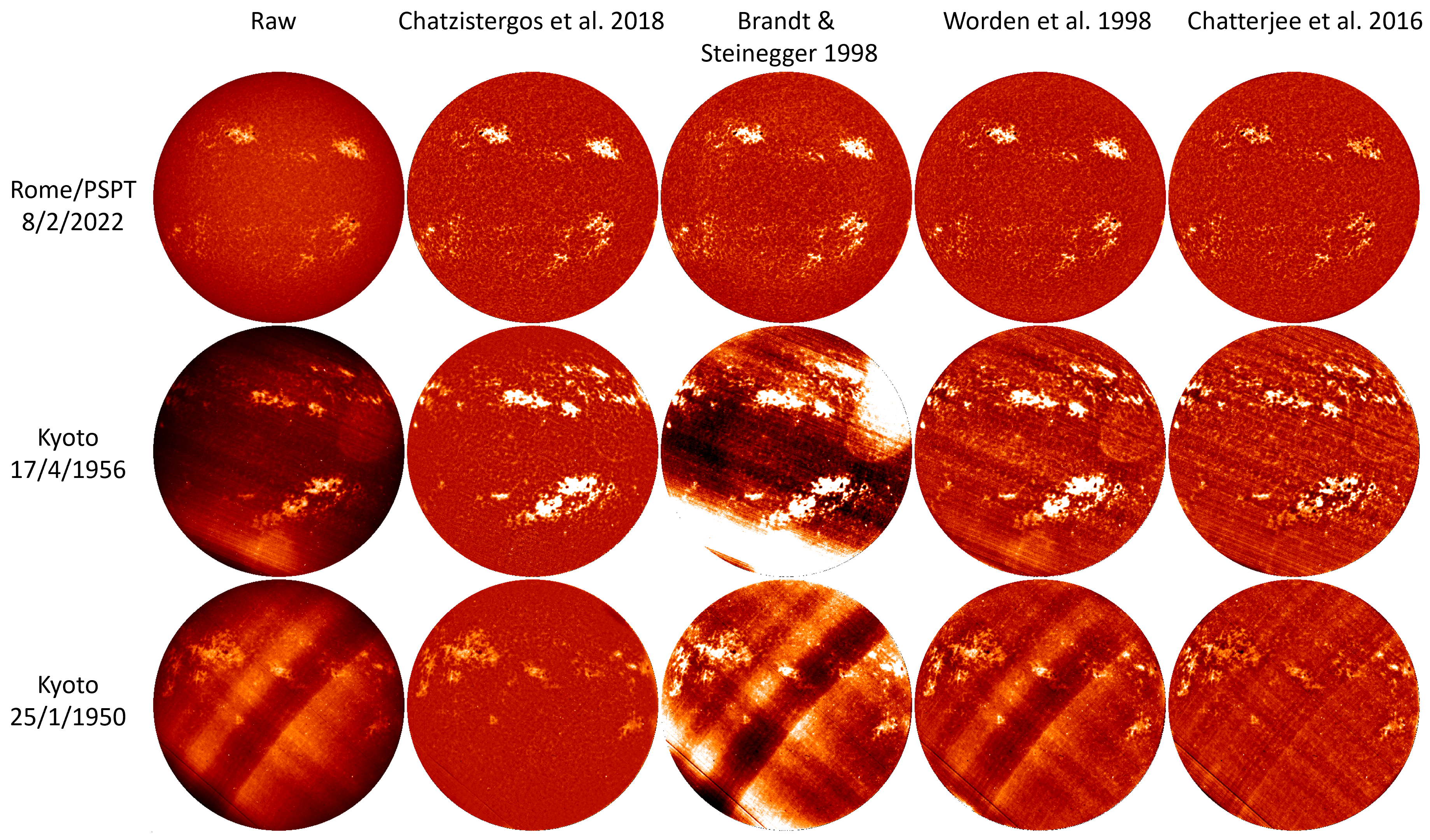}
\end{center}
\caption{Examples of the performance of selected methods to compensate for the limb darkening on observations from Rome/PSPT (8 February 2022; top row) and Kyoto observatory (17 April 1956 in the middle row and 25 January 1950 in the bottom row). Raw images (i.e. including the limb darkening and image artefacts) are shown in the left column. Columns 2--5 show images compensated for the limb-darkening with the methods by \cite{chatzistergos_analysis_2020}, \cite{brandt_determination_1998}, \cite{worden_evolution_1998}, and \cite{chatterjee_butterfly_2016}, respectively. Raw images are shown to their full intensity range, while the limb darkening-compensated images are shown in the range [-0.5--0.5] in contrast.}\label{fig:processing}
\end{figure*}

A more commonly used approach to determine the limb darkening employs a 2D running window median filter \citep{lefebvre_solar_2005,bertello_mount_2010,bertello_70_2020,chatterjee_butterfly_2016,bose_variability_2018,tahtinen_reconstructing_2022}. 
This means that for each pixel, the median of the values within a square box with a predefined width is assigned. 
This approach makes no assumption on the shape of the limb darkening, thus it can easily be used for data with very different observational characteristics (for instance, data taken in different spectral lines), while it also accounts for artefacts in the images, which are typically larger than the window width. 
The width of the window is thus an important factor determining the performance of this approach. 
\cite{bose_variability_2018} used $\sim R/2$, while \cite{chatterjee_butterfly_2016}  used $\sim R/6$. 
\cite{chatzistergos_analysis_2018} studied how the window width affects the results and found that the optimum value for \ca data is within the range $R/8-R/6$, while \cite{tahtinen_reconstructing_2022} found the same optimum values for 1600\AA~AIA data.
Using a median filter to compensate for the limb darkening has three major caveats: (1) how to treat regions closer than half-width of the median filter to the limb, (2) how to avoid biasing the background by the presence of bright regions, and (3) how to account for artefacts that are not smooth (such as the intensity variations across linear segments caused by the changed exposure of different rasters with a spectroheliograph; see Sect. \ref{sec:imagecapturing}). 
The performance of the method declines near the limb, because the median filter typically considers pixels outside of the solar disc and thus artificially underestimates the background. 
Active regions make parts of the disc in \ca line appear brighter, which means that the median filter might overestimate the limb darkening over active regions. 
Depending on the window width this might spill over to the surrounding pixels. 
Thus when correcting the image with the determined background, besides lowering the intensity of the bright regions, it might also create a dark ring with width depending on the window of the median filter, around the active regions (see last column in Fig. \ref{fig:processing}). 

As already mentioned, all of the proposed techniques have caveats and cannot account for all issues affecting the images. 
For this reason, a number of studies have employed diverse combinations of the above approaches \citep{centrone_image_2005,ermolli_comparison_2009,singh_determination_2012,singh_application_2022,priyal_long_2014,priyal_long-term_2017,priyal_periodic_2019,chatzistergos_exploiting_2016,chatzistergos_analysis_2018,chatzistergos_analysis_2020,tahtinen_reconstructing_2022}.

Here we will briefly discuss the method used by \citet{chatzistergos_analysis_2018,chatzistergos_analysis_2019,chatzistergos_historical_2019,chatzistergos_delving_2019,chatzistergos_historical_2020,chatzistergos_analysis_2020} because it is the only method whose performance was tested, and it was shown to fare better than other techniques.
The method combines some of the methods mentioned above, partly modified and extended. 
The processing starts with a first quick and rough estimate of the background using 1D 5th order polynomial fit on radial profiles \citep[which is a variation of the method by][]{brandt_determination_1998}, but applied to azimuthal slices of 30$^\circ$ in steps of 5$^\circ$ rather than the whole disc at once.
This first background is used to preliminary exclude bright and dark regions.
Then a map is constructed by applying the method by \citet[][]{worden_evolution_1998}, that is 1D polynomial fittings along columns and rows of the images, here without the 45$^\circ$ rotation (because the latter raises the noise in the result). 
An important aspect of this process is that it allows the fits on curved segments so to account for such artefacts present in some historical archives, such as Kyoto \citep{chatzistergos_historical_2020}.
This first map is merely used to replace active regions in the original image, and then the 2D running window median filter is applied to the result.
The final background is the sum of a map resulting from the application of a 2D running-window median filter and a map produced by stitching together 1D 5th order polynomial fits across columns and rows applied to the residual image between the original observation and the result of the 2D running-window median filter.
This process is repeated iteratively until the active region exclusion converges.
In this way, the method by \cite{chatzistergos_analysis_2018} overcomes all three caveats mentioned above.

Figure \ref{fig:processing} shows examples of application of four limb-darkening compensation methods on \ca observations. 
In particular, we show results obtained with the methods by \cite{chatzistergos_analysis_2018}, \cite{brandt_determination_1998}, \cite{worden_evolution_1998}, and \cite{chatterjee_butterfly_2016}.
The figure illustrates the limitations of the various approaches. 
All methods perform rather well with artefact-free CCD-based images, although there are still some mild artefacts evident on the image processed with the method by \cite{brandt_determination_1998}. 
However, all methods, except the one by \cite{chatzistergos_analysis_2018}, perform poorly on the historical data suffering from severe artefacts. 
Furthermore, the figure highlights the effect of inaccurate active region exclusion on processing, which manifests as a decrease of plage contrast as can be seen in the last column of Fig. \ref{fig:processing}.

\subsection{Photometric calibration}
As already mentioned in Sect.~\ref{sec:imagecapturing}, due to non-linear response to the incident radiation, data stored on photographic plates need to be photometrically calibrated.
Some observatories complemented the images with so-called calibration wedges. 
These were typically obtained by exposing part of the same plate outside the solar disc (although not necessarily at an unexposed part of the plate which would be ideal) with a narrow-field view of the Sun or a controlled light box, using  different known exposures.
Three examples from Arcetri observations are shown in Fig. \ref{fig:characteristiccurve}.
Unfortunately, utilisation of the calibration wedges has serious limitations \citep[see also][]{foukal_century_2009}. 
Firstly, the exposure range covered by the wedges often does not cover the entire range needed to accurately determine the characteristic curve (see panel c of Fig. \ref{fig:characteristiccurve}). 
Wedges taken with the same exposure sometimes exhibit big differences, thus not allowing accurate assessment of the sigmoid curve.
Finally and most importantly, the need for photometric calibration was not recognised at those times, and the majority of available data do not include calibration wedges.
For instance, wedges exist for Arcetri data after 22 February 1938 \citep{ermolli_digitized_2009}, for Mt Wilson data after 9 October 1961, and Kodaikanal data between 1958 and 2006 \citep{priyal_long_2014}.

Despite these shortcomings, various studies employed the calibration wedges to perform the photometric calibration \citep{fredga_comparison_1971,kariyappa_variability_1994,worden_evolution_1998,giorgi_calibration_2005,ermolli_digitized_2009}, but they were limited by the availability of the wedges.

For photometric calibration of data lacking calibration wedges (that is the bulk of the data),
four main approaches have been proposed:

\begin{enumerate}

\item
\cite{steinegger_sunspot_1996} and \cite{priyal_long_2014} applied an average characteristic curve, computed from all wedges of Sacramento Peak and Kodaikanal data,  respectively.
However, as mentioned in Sect.~\ref{sec:imagecapturing} there are significant deviations between characteristic curves of different observations, rendering this approach not recommendable.

\item
\cite{ermolli_comparison_2009} applied the method by \cite{mickaelian_digitized_2007}, originally used to photometrically calibrate star survey plates, on \ca data. 
\cite{mickaelian_digitized_2007} proposed the following formula to recover the characteristic curve:
\begin{equation}
    I_i=\frac{V-B}{T_i-B},
\end{equation}
where $I_i$ is the intensity of the $i$th pixel, $V$ the average value in an unexposed section of the photographic plate, $B$ is the mean value of the darkest exposed regions, and $T_i$ the transparency value of the $i$th pixel.

\item
\cite{tlatov_new_2009} calibrated \ca data with linear scaling. 
The scaling law was determined by relating the measured plate density at two $\mu$ positions (0 and 0.9) to that of standard centre-to-limb variation  (CLV, hereafter) profiles measured by \cite{pierce_solar_1977}.

\item
\cite{chatzistergos_analysis_2018} proposed a novel approach to perform the photometric calibration. 
This approach relates the centre-to-limb variation (CLV, hereafter) of the quiet Sun regions measured in the photographic data to a reference CLV of quiet-Sun regions determined from high-quality modern CCD-based data. 
The method is based on the assumption that the darker parts of quiet Sun regions do not vary significantly with time, which is in agreement with the results by \cite{white_solar_1978,white_solar_1981,livingston_suns_2003,livingston_sun-as--star_2007,buhler_quiet_2013,lites_solar_2014}. 
\cite{kakuwa_investigation_2021} used a rather similar method to that by \cite{chatzistergos_analysis_2018}.
Importantly, \cite{chatzistergos_analysis_2018,chatzistergos_analysis_2019} tested the accuracy of their approach with synthetic data created to imitate most issues affecting historical data (e.g. large scale inhomegeneities, non-linear and varying characteristic curves). 
They found that their approach allowed to recover the intensity with a mean error of $<$1\%.
Furthermore, they also tested some selected methods from the literature and showed that their method performed significantly better.
\end{enumerate}

One common drawback of all the previously described methods is that they derive one characteristic curve for each plate. 
This is reasonable for filtergrams stored on photographic plates, but not very accurate for spectroheliograms. 
That is because spectroheliograms are slowly scanned, one strip after the other (see Sect.~\ref{sec:spectroheliograph}).
In practice, it can happen that different image strips have different exposures, which introduces uncertainties in recovering the characteristic curve.
Also, the observational conditions sometimes changed over the course of the scan leading to variations of the characteristic curve over the image.
We note, however, that these uncertainties are still lower than when using photometrically uncalibrated data, while severe cases of variable exposure over an image are very likely to be excluded by the researchers from further analyses during data processing.

\clearpage
\section{Carrington maps}
\label{sec:carringtonmaps}
For most purposes, no further processing of the \ca images after the photometric calibration and limb-darkening compensation is required. 
However, there are applications for which information on the entire surface of the Sun is needed, for example coronal field extrapolations with potential field source-surface models \citep{wang_potential_1992,wiegelmann_magnetic_2014,asvestari_modelling_2021}.
In such cases, Carrington maps are produced \citep{sheeley_carrington_2011,chatterjee_butterfly_2016,bertello_70_2020}.

Carrington maps are Mercator projections of the Sun, showing the entire solar surface in one map. 
Due to the fact that we essentially have \ca observations (as in fact most other solar observations) from only one vantage point, that of the Earth, Carrington maps are not instantaneous snapshots, but a collage of observations taken over the course of one solar rotation (27.2753 days).
That means that values in Carrington maps at certain longitudes result from averaging observations taken a few days apart, which can smear and smooth out features. 
To reduce this effect, usually only a window in longitudes for each observation is used instead of the entire solar disc, for example \cite{chatterjee_butterfly_2016} used $60^\circ$ longitudinal bands.  
Maps are produced by summing up all the slabs and dividing this map by a streak map,  which is essentially a map counting how many images have been used for each pixel.

\begin{figure*}[t!]
\begin{center}
\includegraphics[width=0.95\linewidth]{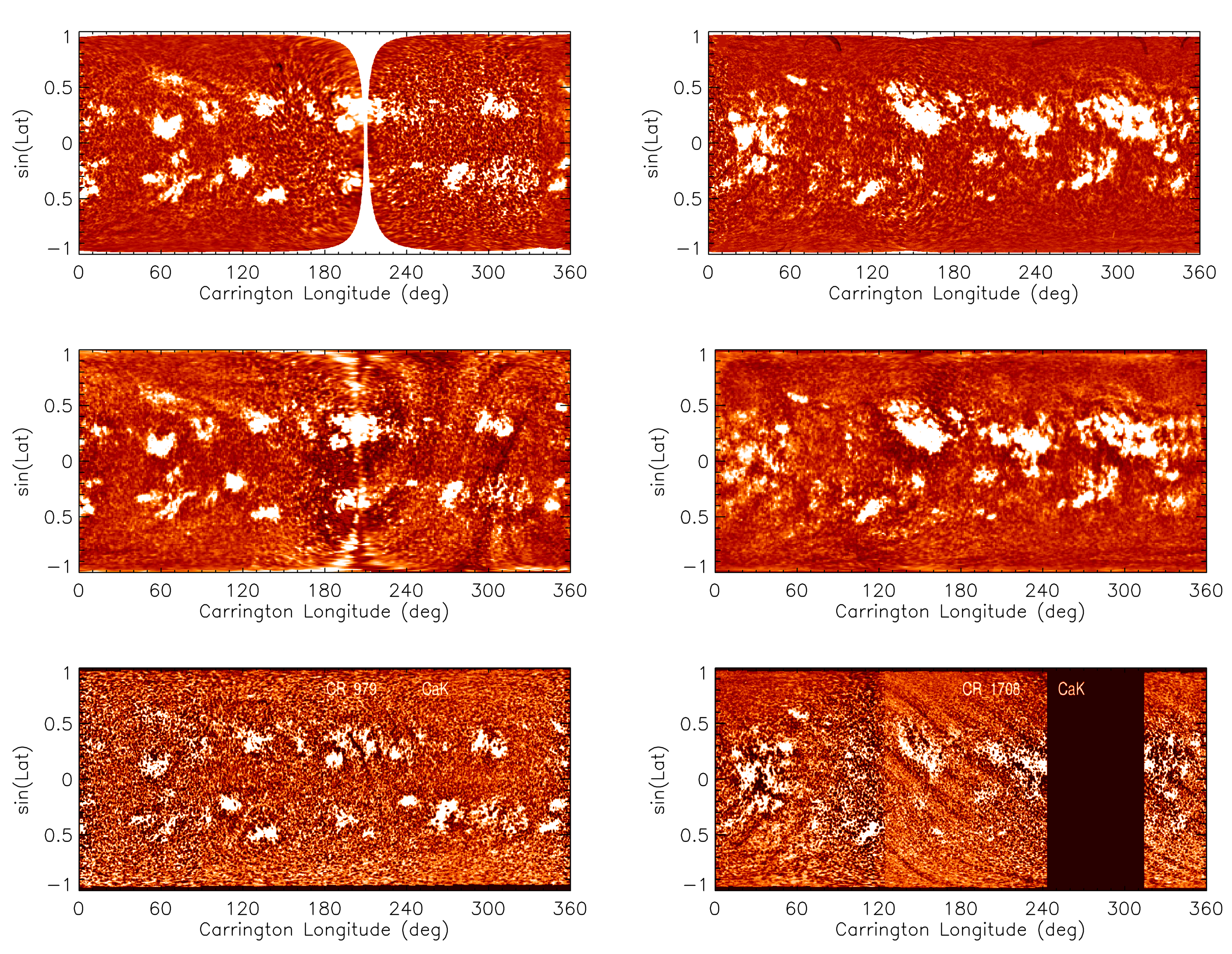}
\end{center}
\caption{Examples of Carrington maps constructed from \ca data over rotations 979 (23/11--19/12/1926, left) and 1708 (2/5--29/5/1981, right). Shown are Mt Wilson data processed by \citet[][top panel]{chatzistergos_analysis_2020} and by \citet[][middle panel]{bertello_70_2020} as well as Kodaikanal data processed by \citet[][bottom panel]{chatterjee_butterfly_2016}. Due to differences in the image processing the images are saturated at different levels in order to show plage at roughly similar levels. In particular, the contrast ranges are [-0.2--0.2] and [-0.5--0.5] for the first and second rows, respectively, while the 3rd row shows images to their entire range. The data by \cite{chatterjee_butterfly_2016} are provided in PNG file format so the exact contrast values are not known to us. }\label{fig:carringtonmaps}
\end{figure*}

Figure \ref{fig:carringtonmaps} shows Carrington maps produced by \cite{chatzistergos_analysis_2020}, \cite{bertello_70_2020}, and \cite{chatterjee_butterfly_2016} from Mt Wilson and Kodaikanal data.
The figure reveals differences between the various Carrington maps even when derived from the same data. 
The majority of the differences arise from differences in processing applied to compensate for the limb darkening. 
Quite evident are darkened rings around plage areas due to inaccurate exclusion of plage areas when computing the background as well as residual artefacts which were not properly accounted for by the respective processing method (see Sect. \ref{sec:limbdarkening}). 
However, there are also differences caused by the procedures through which the Carrington maps were created. 
For instance, maps from \cite{chatzistergos_analysis_2020} and \cite{chatterjee_butterfly_2016} have clear gaps over the areas not covered by observations, while the maps from \cite{bertello_70_2020} do not show gaps, but stretched parts, which actually belong to regions outside of the disc, instead.

\section{Plage areas}
\label{sec:plageareas}
\begin{figure*}[t!]
\begin{center}
\includegraphics[width=0.95\linewidth]{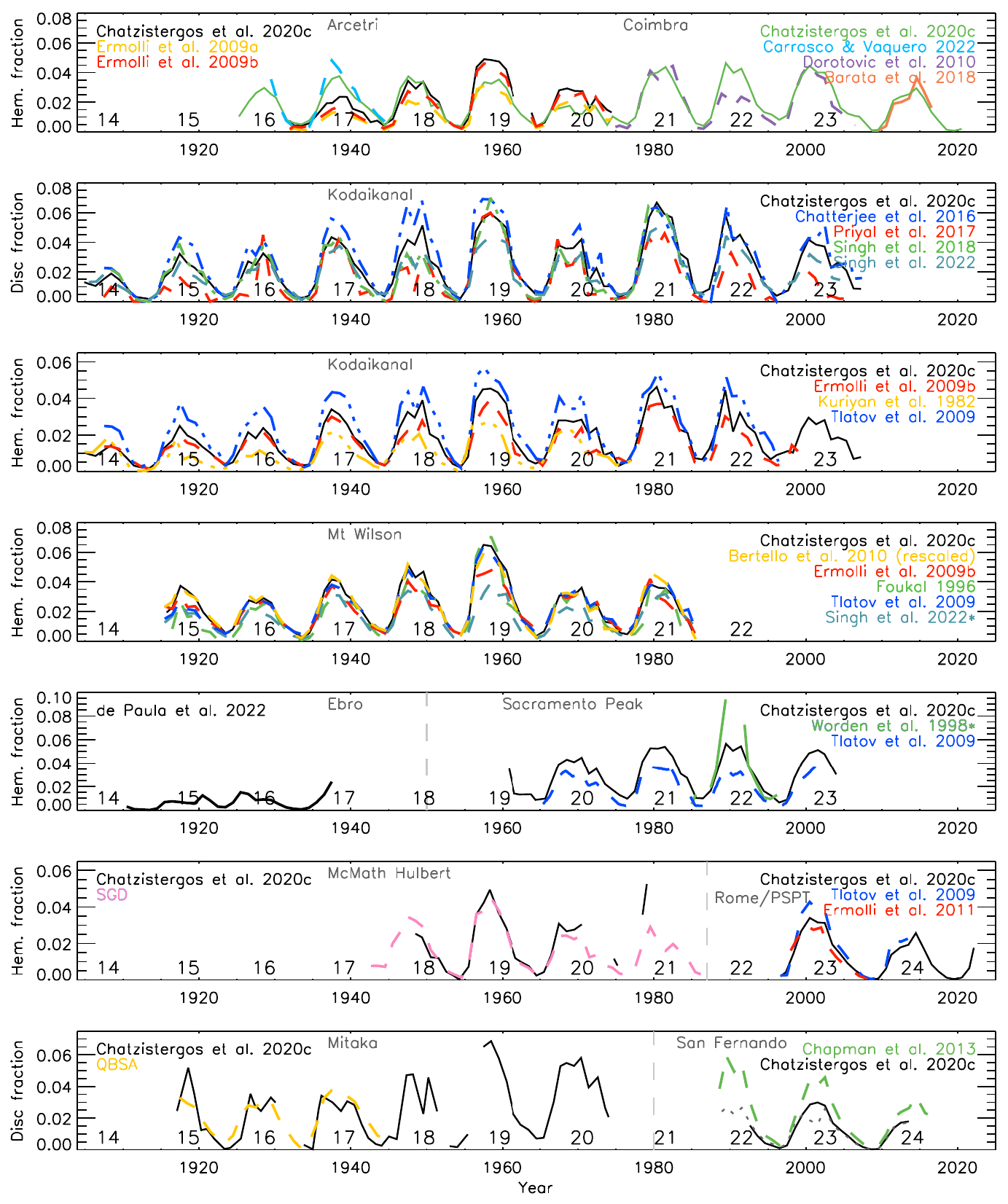}
\end{center}
\caption{Plage areas derived from various \ca archives (as indicated in grey) and different studies.
For the San Fernando series by \cite{chatzistergos_analysis_2020} we show separately the results for CFDT1 (dotted grey) and CFDT2 (black) data.
Shown are annual median values in hemispheric fraction, except for the series in 2nd and 7th rows as well as the one by \cite{singh_application_2022} in the 4th row and \cite{worden_evolution_1998} in the 5th row which are given in disc fraction.  
The numbers in the lower part of each panel denote the conventional solar cycle numbering.
The vertical lines roughly mark the separation of archives in panels including timeseries from more than one archive.
}\label{fig:plageareas}
\end{figure*}

\begin{figure*}[t!]
\begin{center}
\includegraphics[width=0.95\linewidth]{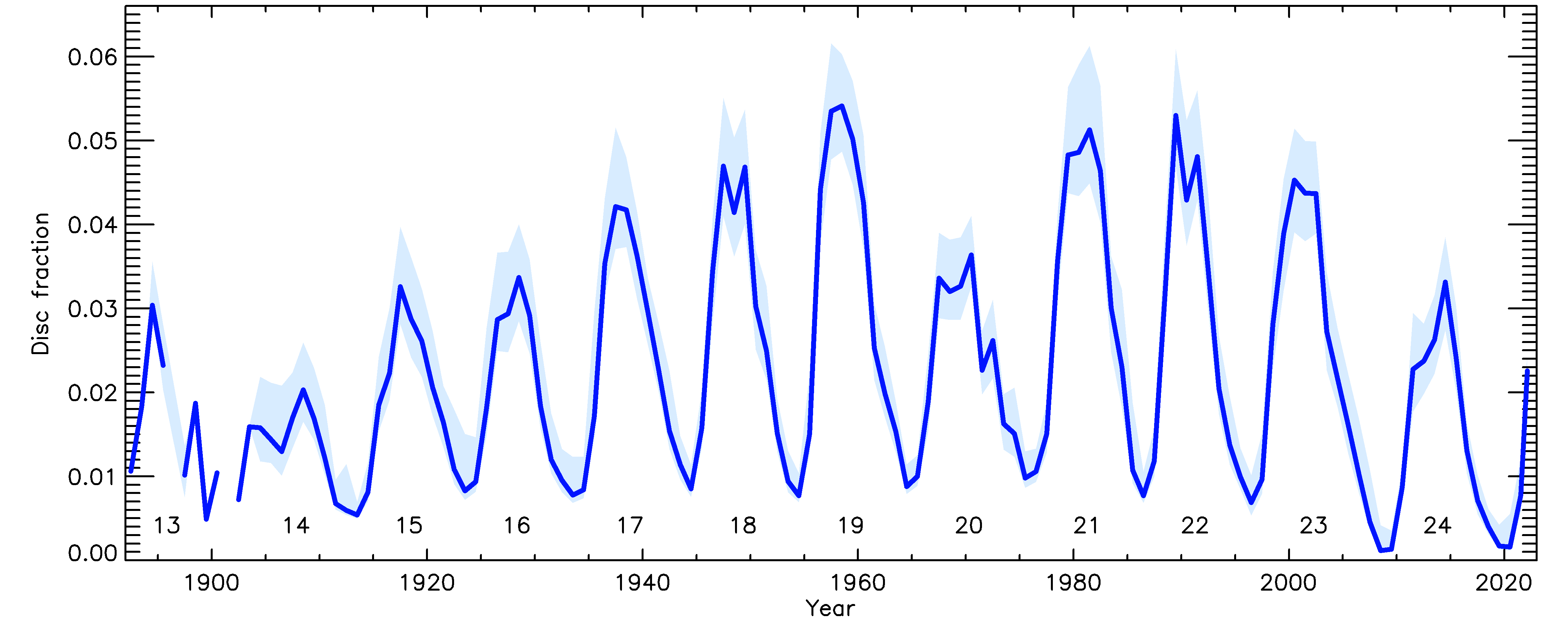}
\includegraphics[width=0.95\linewidth]{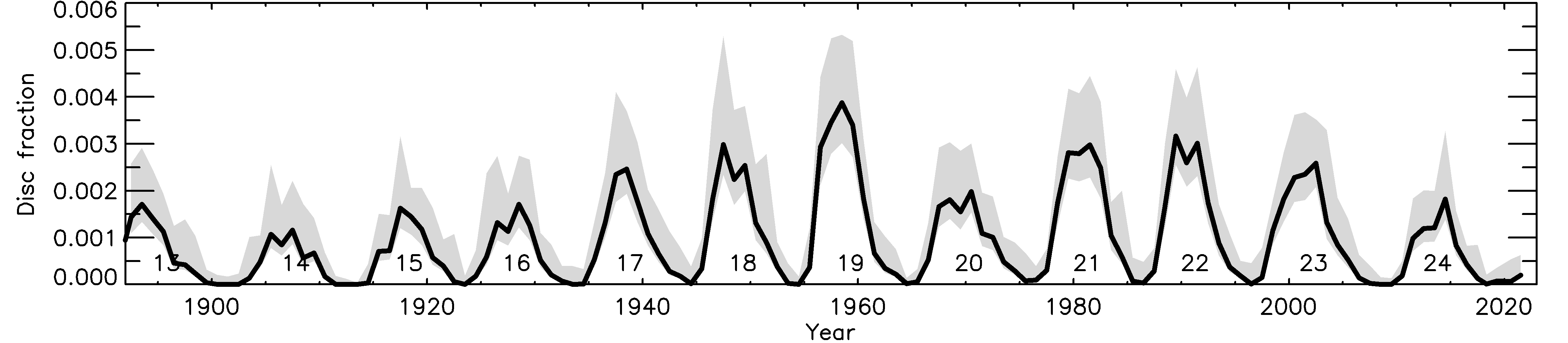}
\end{center}
\caption{\textit{Top:} Plage area composite series by \cite{chatzistergos_analysis_2020} extended to 10 March 2022 with Rome/PSPT data processed in the same way as the rest of the data used for the composite series. \textit{Bottom:} Sunspot area series by \cite{mandal_sunspot_2020}. Shaded areas mark the asymmetric 1$\sigma$ intervals, while the numbers at the lower part of each panel denote the conventional solar cycle numbering.}
\label{fig:plageareacomposite}
\end{figure*}

One of the most commonly studied quantities derived from \ca data is the total plage area on the visible solar hemisphere and its evolution with time.
Determining plage areas, in principle, does not require any image processing and can be done on images without photometric calibration or limb-darkening compensation. 
This allowed various studies to determine plage areas without following the processing steps described in Sect. \ref{sec:processing} \citep[e.g.][]{kuriyan_long-term_1983,segui_temporal_2019,de_paula_evolution_2020,de_paula_application_2021,de_paula_cyclic_2022,carrasco_catalog_2022}.
Plage areas in these studies were counted manually from the physical photographs, typically by overlaying transparencies with circles of known areas.
The early plage area timeseries were thus produced in a more qualitative way, applying classification schemes based on characteristics of plage regions.
One of the earliest such series is from the Ebro observatory, which kept records of plage areas between 1910--1937 \citep[][shown in Fig. \ref{fig:plageareas}]{segui_temporal_2019,de_paula_application_2021}. 
They used a classification scheme, where they presented areas for four classes (one class had three further subdivisions) of plage regions identified based on their appearance (whether they were compact, scattered, a combination of both, or plage they could not associate with any of these groups). 
Similarly, the Quarterly Bulletin on Solar Activity (QBSA)\footnote{Available at \url{https://solarwww.mtk.nao.ac.jp/en/wdc/qbsa.html}} series included a plage index from Mitaka \ca data where plage regions were manually sorted depending on their size and brightness.
SGD\footnote{Available at \url{https://www.ngdc.noaa.gov/stp/solar/calciumplages.html}} also maintained a series of plage areas derived from physical photographs from McMath-Hulbert (06/1942--09/1979), Mt Wilson (10/1979--09/1981), and Big Bear (10/1981--11/1987) observatories.

Digitisation of the data over the last decades (see Sect. \ref{sect:digitisation}) allowed more quantitative analyses of \ca data. 
Limb-darkening compensation and automatic approaches to identify plage regions have been developed and applied. 
Various series were produced in this way both from historical photographic and modern CCD-based data \citep[e.g.][]{chapman_solar_1997,chapman_improved_2001,ermolli_digitized_2009,ermolli_comparison_2009,tlatov_new_2009,bertello_mount_2010,bertello_70_2020,dorotovic_north-south_2010,priyal_long_2014,priyal_long-term_2017,priyal_periodic_2019,chatterjee_butterfly_2016,singh_variations_2018,singh_application_2022}.
The segmentation of plage regions was typically done with a contrast threshold, which was either set to a constant value for all images, or was allowed to vary depending on the standard deviation of contrast values over the disc or the quiet Sun regions \citep[see][]{ermolli_photometric_2007,ermolli_comparison_2009,chatzistergos_analysis_2017}.
We note, however, that one of the earliest series based on digitised data from Mt Wilson by  \cite{foukal_behavior_1996} still selected plage regions manually.
Further, \cite{barata_software_2018} used morphological operators to single out plage areas from CCD-based Coimbra data, without the need to compensate the images for the limb-darkening.

Even though plage areas can be derived from uncalibrated data, there are nevertheless disadvantages of using photometrically uncalibrated historical data (see Sect. \ref{sec:imagecapturing}).
Photometric calibration returns data with homogeneous intensity in plage regions, which in turn allows more accurate identification of plage areas.
Also the accuracy of the limb-darkening compensation affects the resulting plage areas (see Sect. \ref{sec:limbdarkening}), for instance through unaccounted or introduced artefacts, or due to potentially varying accuracy of the processing, especially if it depends on activity.

Figure \ref{fig:plageareas} shows some of the available plage area series, focusing on those from historical long-term datasets. 
It is immediately evident that the various published series show (sometimes significant) differences. 
This is not only the case for plage areas derived from different \ca archives, but also for those derived from the same archive but with different processing techniques \citep[see also][]{ermolli_potential_2018}.
Most of the discrepancies come from the applied processing, 
including the limb darkening compensation, photometrical calibration, the segmentation, and even divergence in the definition of plage.
Furthermore, the samples of the data used for the analysis are typically not identical either.
However, part of the differences is also due to the intrinsic differences between the archives, in particular in the employed bandwidth (see Sect. \ref{sec:spectroheliograph}). 
This was shown by \cite{chatzistergos_analysis_2020} who analysed 43 \ca datasets (38 centred at the core of the line and 5 centred at different wavelengths across the wings) processed with exactly the same methods. 

In addition to purely plage area series, some authors constructed various composite records of facular indices or proxies, which also included plage areas.
For instance, \cite{fligge_long-term_1998} combined the series of the international sunspot number, sunspot areas, F10.7 flux, white-light facular areas, and Mt Wilson plage areas by \cite{foukal_behavior_1996}. 
\cite{bertello_correlation_2016,bertello_ca_2017} produced a composite emission index by combining disc-integrated 1\AA~\ca indices from Sacramento Peak and Synoptic Optical Long-term Investigations of the Sun (SOLIS) Integrated Sunlight Spectrometer (ISS) with the plage areas from Kodaikanal by \cite{tlatov_new_2009}. 
The first composite of plage areas based solely on \ca data was presented by \cite{chatzistergos_analysis_2019}, who combined the data from nine \ca datasets (Arcetri, Kodaikanal, McMath-Hulbert, Meudon, Mitaka, Mt Wilson, Rome/PSPT, Schauinsland, and Wendelstein). 
Later, the composite was updated to include data from 38 \ca datasets  \citep{chatzistergos_analysis_2020}. 
It provides daily values between 1892 and 2019 and is shown in Fig.~\ref{fig:plageareacomposite} along with the sunspot area series by \cite{mandal_sunspot_2020}.

\citet{chatterjee_butterfly_2016,priyal_long-term_2017} and \citet{tlatov_polar_2019} have also analysed the distribution of plage areas over latitudes and time, the so-called 
butterfly diagrams, analogous to the well-known sunspot butterfly diagram.
At the beginning of a cycle, active regions typically emerge at mid latitudes, while as the cycle progresses they move towards lower latitudes, such that the shape of these graphs resembles wings of a butterfly.

The North--South asymmetry of solar cycles has usually been studied with sunspot data \citep[e.g.][]{veronig_hemispheric_2021,ravindra_solar-cycle_2021}, but
 \cite{dorotovic_north-south_2007,dorotovic_north-south_2010,segui_temporal_2019,el-borie_influence_2020,de_paula_evolution_2020,de_paula_cyclic_2022} and \cite{chowdhury_analysis_2022} used \ca data for this purpose. 
Most of these studies covered relatively short intervals up to two solar cycles, with the exception of \cite{el-borie_influence_2020} and \cite{chowdhury_analysis_2022}.
The result of the study by \cite{el-borie_influence_2020} is, however, simply mirroring that from sunspot data, as these authors 
used the composite of full-disc plage areas by \cite{chatzistergos_analysis_2019}, which they separated into the north and south components by imposing the same asymmetry level as found in Royal Greenwich observatory sunspot area data. 
Using Kodaikanal plage areas from \citet{chatterjee_butterfly_2016} over cycles 14--21 derived for each hemisphere separately 
\cite{chowdhury_analysis_2022} found only three cycles (14, 15, and 21), for which the activity peak roughly coincided in both hemispheres, while higher activity was found in the northern hemisphere for all cycles except 14, 17, and 21.
This differs from the results based on sunspot data \citep[][]{veronig_hemispheric_2021} which imply that cycles 16, 18, 22, 23, and 24 had higher activity in the southern hemisphere.
The North--South asymmetry was found to be highest over cycle 19, in agreement with \cite[][]{veronig_hemispheric_2021}.

The relation between plage areas and sunspot areas or numbers has also been studied \citep[see][]{foukal_effect_1979,schatten_importance_1985,lawrence_ratio_1987,steinegger_energy_1996,chapman_solar_1997,chapman_facular_2011,fligge_long-term_1998,bertello_correlation_2016,mandal_association_2017,yeo_how_2020,chatzistergos_ca_2018,chatzistergos_scrutinising_2022}. 
Past studies, in general, suggest a quadratic relation between plage and sunspot areas for daily values.
plage and sunspot areas for daily values. A linear relationship is typically favored for annual sunspot values, however \citet{chatzistergos_scrutinising_2022} reported a, in general, non linear relationship even for annual values.
For sunspot numbers the reported relation is typically linear when annual values are considered, but non-linearities have been reported for daily values \citep[][]{chatzistergos_scrutinising_2022}.
No qualitative difference has been reported when alternative sunspot number series, such as those by \cite{svalgaard_reconstruction_2016,usoskin_new_2016,chatzistergos_new_2017,willamo_updated_2017}, were used for the analysis.
\citet{chatzistergos_scrutinising_2022} also reported that the bandwidth of the \ca observations affects the relationship between plage areas and sunspot data.

\begin{figure}[t!]
\begin{center}
\includegraphics[width=0.95\linewidth]{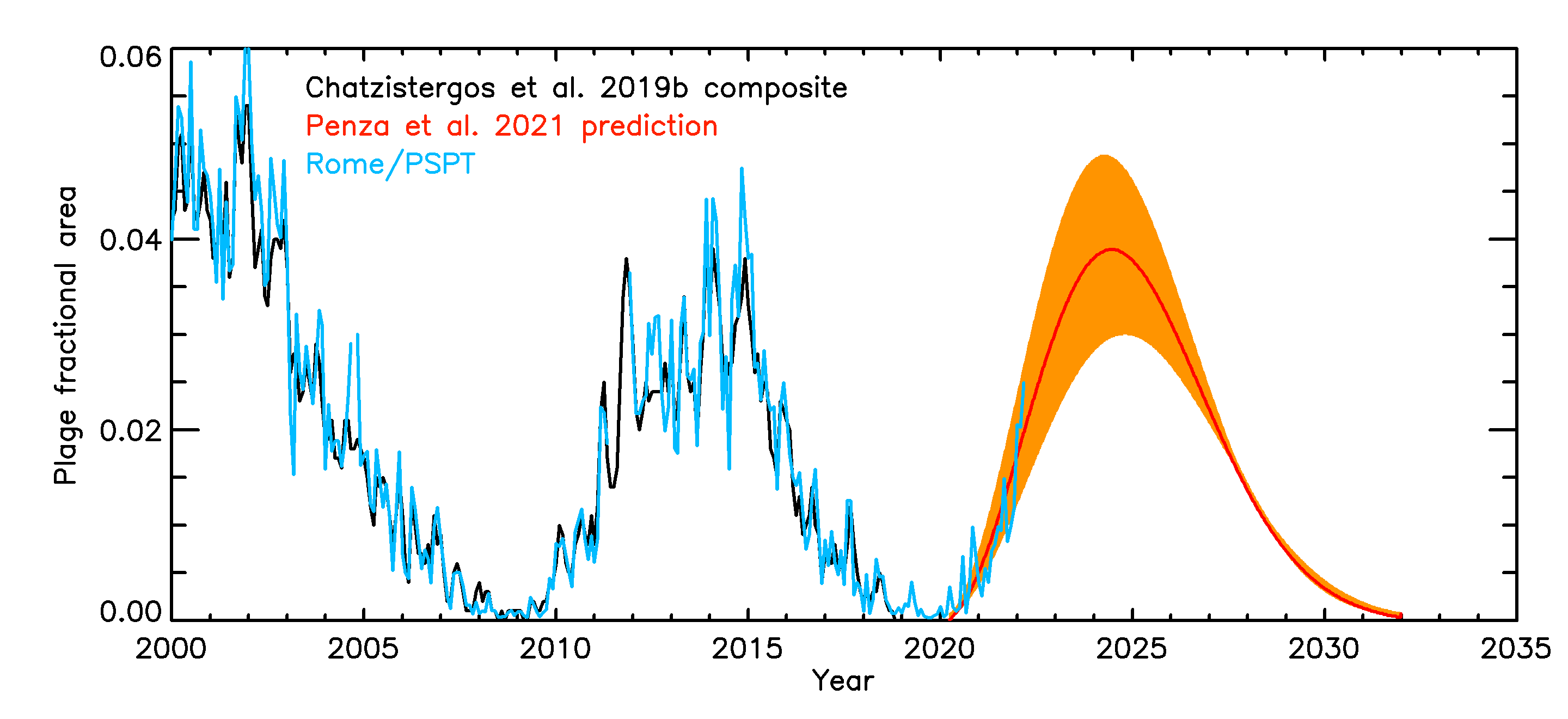}
\end{center}
\caption{Prediction of solar cycle 25 amplitude in plage fractional areas by \citet[][red, with confidence intervals in shaded orange surface]{penza_prediction_2021}. Shown also is the \cite{chatzistergos_analysis_2019} plage area composite (black), on which the prediction was based, as well as more recent data from Rome/PSPT (light blue), which was the reference observatory in the \cite{chatzistergos_analysis_2019} plage area composite since 1996.}\label{fig:plageareaprediction}
\end{figure}

Finally, \cite{penza_prediction_2021} used the composite by \cite{chatzistergos_analysis_2019} going up to the end of 2018 to make, to our knowledge, the first prediction of the amplitude of solar cycle 25 in terms of plage areas (Fig.~\ref{fig:plageareaprediction})
They used an empirical parametrisation of the solar cycle in plage and sunspot areas as suggested by \cite{2009SoPh..258..319V}.
This model has one free parameter varying from cycle to cycle.
By determining an empirical relation between the free parameter of the model for subsequent odd and even cycles, \cite{penza_prediction_2021} estimated the amplitude of cycle 25 using the value of the parameter for cycle 24.
Their prediction suggests the amplitude of cycle 25 to be rather similar to or slightly higher than that of cycle 24. 
Extending the composite by \cite{chatzistergos_analysis_2019} with more recent data from Rome/PSPT (up to 10 March 2022), agrees with the prediction until then within the uncertainty level.
However, data covering the next few years will be required to assess the performance of the prediction more meaningfully.

\section{Network}
\label{sec:network}
\subsection{Network area evolution}

Network regions are small magnetic ﬂux concentrations appearing as bright web-like structures outlining dark cells, which are the chromospheric counterparts of the supergranulation pattern. Studying the network evolution is particularly important, because they are thought to be responsible for the secular variation of solar irradiance \citep[e.g.][]{solanki_secular_2002}.

Network regions are usually subdivided into active, enhanced, and quiet network in order to distinguish between products of decaying active regions and the ``quiet-Sun'' network outside of active regions. 
However, often this distinction is rather arbitrary and is based merely on the contrast of network regions. 
In the following we will refer to all of them collectively as network unless specified otherwise.

The mean level of network fractional areas 
varies significantly among the various studies, lying on average between 0.02 \citep[][using Mt Wilson data]{foukal_extension_1998} and 0.35 \citep[][using Rome/PSPT data]{ermolli_measure_2003}.
The total network area varies in-phase with the solar cycle \citep{caccin_variations_1997,caccin_variations_1998,foukal_extension_1998,worden_evolution_1998,worden_plage_1998,ermolli_measure_2003,priyal_long_2014,priyal_long-term_2017,ermolli_romepspt_2022}.
A latitudinal dependence of network areas on the solar cycle has also been reported \citep[][using Kodaikanal data]{devi_variation_2021}, with the amplitude of variations decreasing from the equator towards $\pm50^\circ$ latitude and increasing again towards the poles. 
\cite{foukal_measurement_2001} argued that there is no evidence for a long-term trend in network areas by analysing Mt Wilson and Sacramento Peak data around cycle minima periods between 1914 and 1996.
However, the scatter of their derived fractional areas is rather large (roughly between 0.1 and 0.19) with too few data points (3 images per cycle minimum) to accurately assess this.
Furthermore, they used photometrically uncalibrated data, which is an additional important limiting factor.

The controversies described above are largely due to different definitions of network regions, but also due to the employed data and the processing.
In particular, the accuracy of the processing becomes more important for determining network regions than for plage ones, because of the lower contrast of network regions, which is frequently lower than that of artefacts found in historical data.
Furthermore, the segmentation approach can also significantly affect the determined network regions.
Analysing modern high-quality Rome/PSPT observations \cite{chatzistergos_analysis_2017} showed that segmentation with a constant threshold resulted in an in-phase relation between network areas and the solar cycle. 
Thresholding with a multiplicative factor to the standard deviation of the solar disc contrast resulted in anti-phase variability, instead. 
They also noticed that the latter effect was more pronounced when the contrast of the entire disc (that is including also active regions) was used to compute the standard deviation (thus being affected by activity variations), which is in fact the more commonly used method. 
This effect is marginal when active regions are excluded for the computation of the standard deviation  \citep[as used by][]{chatzistergos_analysis_2020}, suggesting a very small to no variation of network areas over the solar cycle.
This is in good agreement with the conclusion of \citet{harvey_magnetic_1993,harvey_solar_1994}, who found that the amplitude of the solar cycle variation decreased markedly with size of magnetic regions, being several times stronger for active regions than for smaller ephemeral regions.

Due to the absence of information on network regions before the beginning of \ca observations, there have been studies aiming to establish a relationship between network areas and other solar indices which go further back in time than \ca observations.
For instance, \cite{singh_determining_2021} compared the network areas from Kodaikanal to the international sunspot number series \citep{clette_new_2016-1} and reported linear correlation factors of 0.87 and 0.77 for the active and enhanced network, respectively.
\cite{berrilli_long-term_2020} reconstructed monthly network areas back to 1749 with a power law relation applied to the international sunspot number series.

\begin{figure*}[t!]
\begin{center}
\includegraphics[width=0.95\linewidth]{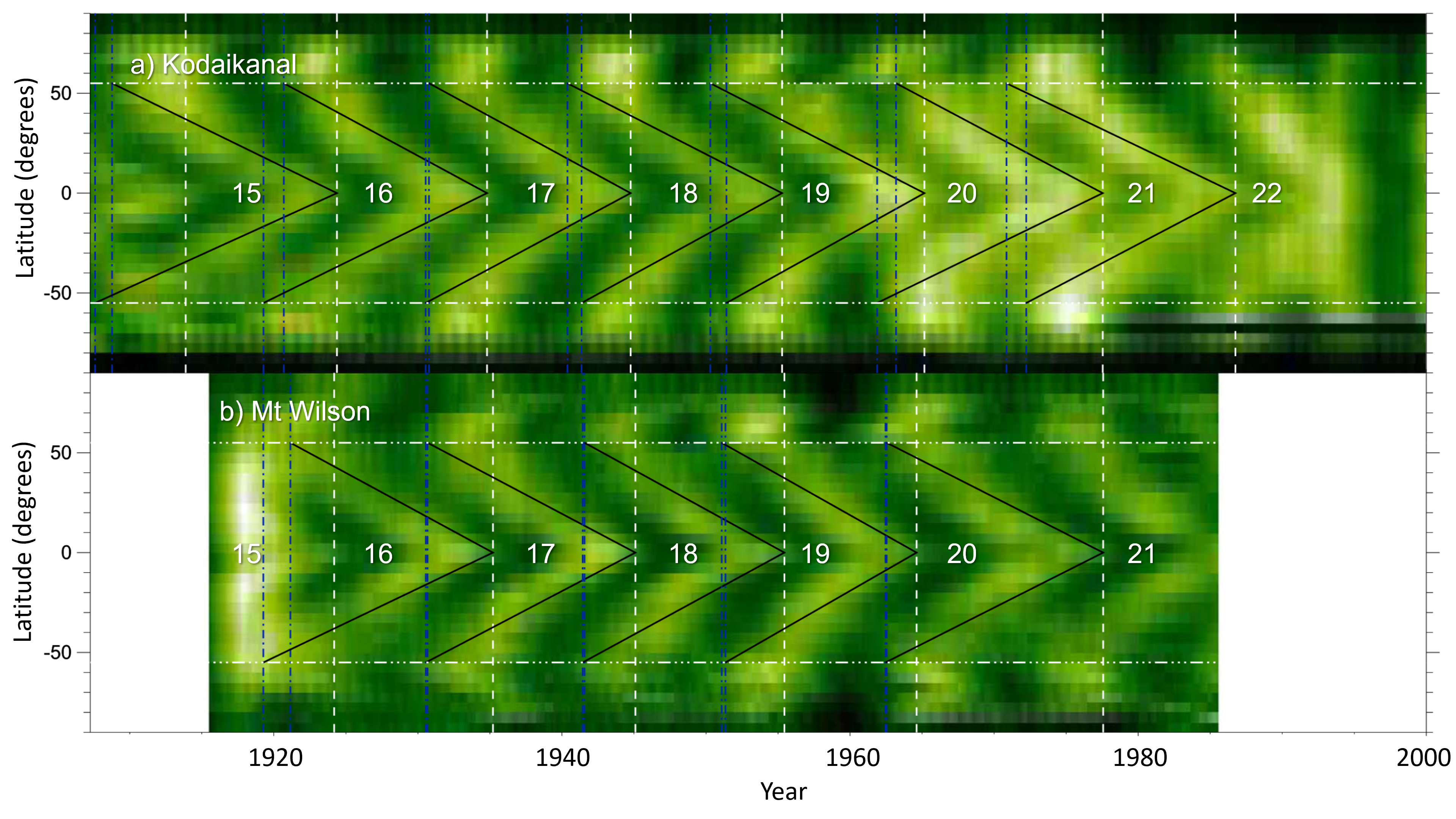}
\end{center}
\caption{Time-latitude maps of network areas from Kodaikanal (top panel) and Mt Wilson (bottom) observatories. The black lines are fits to the extended equatorward branches.   
The vertical dashed blue lines denote the onset of the extended equatorward branches, while the white vertical dashed lines denote the time when those branches reach the equator. The horizontal white dashed lines mark latitudes at $\pm55^\circ$. The numbers at the centre of each panel denote the conventional solar cycle numbering. The figure is adapted from \cite{chatterjee_signature_2019}}\label{fig:nbe_chatterjee}
\end{figure*}

Finally, \cite{chatterjee_signature_2019} produced time-latitude maps of network areas (Fig. \ref{fig:nbe_chatterjee}). 
They used Carringtom maps from Kodaikanal produced by \cite{chatterjee_butterfly_2016} and Mt Wilson by \cite{sheeley_carrington_2011}. 
For each Carrington map they counted the network areas within latitudinal strips of 5$^\circ$. 
They also applied a smoothing to each latitudinal strip with a kernel of 200 Carrington rotations (roughly 15 years). 
Their results show branches of equatorward migration of network areas that start at approximately $\pm55^\circ$ latitude and take approximately $15\pm1$ years to reach the equator.

\subsection{Network cell characteristics}
\cite{simon_velocity_1964} noticed a spatial correspondence between the chromospheric network in \ca observations and boundaries of supergranular cells seen in the photosphere.
This triggered a series of studies of supergranular parameters by using \ca observations \citep{sotnikova_statistical_1978,singh_dependence_1981,raghavan_quantitative_1983,munzer_pole-equator-difference_1989,kariyappa_variability_1994,hagenaar_distribution_1997,berrilli_geometrical_1998,berrilli_average_1999,ermolli_chromospheric_1998,ermolli_chromospheric_1998-1,pietropaolo_chromospheric_1998,raju_dependence_1998,goldbaum_intensity_2009,raju_network_2014,raju_asymmetry_2020,mcintosh_observing_2011,chatterjee_variation_2017,mandal_association_2017-1,rajani_solar_2022}.
The cell sizes determined from \ca data were in general smaller than the diameter of the supergranular cells in the photosphere reported by \citet[][about 32 Mm]{simon_velocity_1964}.
For instance, \cite{hagenaar_distribution_1997}, \cite{mcintosh_observing_2011}, and \cite{singh_dependence_1981} with South Pole, Mauna Loa Solar observatory (MLSO) PSPT, and Kodaikanal \ca data found the mean cell diameter of 13--18 Mm, 22--32 Mm, and 22 Mm, respectively. 
However, the estimate by \cite{singh_dependence_1981} was obtained by selecting network regions manually, whereas if an auto-correlation technique was applied, the cell diameter increased to about 32 Mm, in accordance with \cite{simon_velocity_1964}.
We note that procedures employed to determine the network cells as well as the data used differ considerably among different studies, causing discrepancies in the reported values \citep{rincon_suns_2018}.

In general, results from the literature suggest that cell sizes decrease with increasing activity level.
This was reported by \cite{singh_dependence_1981}, \cite{raghavan_quantitative_1983}, \cite{kariyappa_variability_1994}, and \cite{rajani_solar_2022} using Kodaikanal data, and by \cite{mcintosh_observing_2011} using Mt Wilson data covering 1944--1976.
In contrast, \cite{chatterjee_variation_2017} used Kodaikanal, Rome/PSPT, and MLSO/PSPT data and found an increase of cell sizes with increasing solar activity when considering the entire solar disc without distinguishing between active and quiet regions.  
However, they also found an anti-correlation between cell sizes and activity when only quiet regions were considered.
Furthermore, based on Arcetri spectroheliograms covering 1950--1970, \cite{caccin_variations_1998} found that the histogram of contrast values of non-plage regions exhibited an asymmetry depending on the activity level.

\section{Connection between Ca II K brightness and magnetic field strength}
\label{sec:bvscak}
\ca data have been recognised very early as good tracers of solar surface magnetic fields \cite[][]{babcock_suns_1955,howard_observations_1959,leighton_observations_1959}. 
Various studies assessed the relationship between Ca II brightness and magnetic field strength \citep{frazier_multi-channel_1971,skumanich_statistical_1975,schrijver_relations_1989, nindos_relation_1998,harvey_magnetic_1999,rast_scales_2003,ortiz_how_2005,rezaei_relation_2007,loukitcheva_relationship_2009,kahil_brightness_2017,kahil_intensity_2019,chatzistergos_recovering_2019}.
Most of them favoured a power law relationship between the Ca II brightness and the magnetic field strength, while \cite{kahil_brightness_2017,kahil_intensity_2019} found a logarithmic function to fit the data best.
These previous studies used very diverse data, in terms of the periods covered, the bandwidth of observations, and types of considered regions. 
Most of them considered only small parts of the disc in either the quiet Sun or active regions separately. 
All of these factors let to a rather high spread in the derived power-law exponents, typically between 0.3 and 0.6.
The analysis by \cite{chatzistergos_recovering_2019} is the most comprehensive to date, considering high-quality full-disc data covering half a solar cycle.
They saw no dependence (within the uncertainty of the fits) of the relation on the $\mu$ position or activity level.

The connection between the \ca brightness and the magnetic field strength suggests that \ca data can be used to reconstruct magnetograms.
This is very important considering that high-quality magnetograms of the Sun exist since the 1970's \citep{livingston_kitt_1976} and \ca data can be used to extend the magnetic field data back to 1892. 
To our knowledge, there have been only four such reconstructions until now.

\cite{shin_generation_2020} employed a machine learning image-to-image translation approach to convert CCD-based Rome/PSPT \ca images to SDO/HMI-like magnetograms. 
They based their work on the "pix2pixHD" model by \cite{wang_high-resolution_2018}. 
They reported a linear correlation factor between the total unsigned magnetic flux measured in the generated magnetograms to the actual one of 0.99. 
They reported that pixel-by-pixel correlation is on average 0.74, but it depends on the selected regions, being higher for active regions (0.81) and considerably lower for quiet Sun regions (0.24).
However, it should be noted that they employed data without compensation for limb-darkening, which might be a factor affecting the performance of the training.

\cite{pevtsov_reconstructing_2016} was, to our knowledge, the first to reconstruct magnetograms from historical photographic \ca data. 
They used low resolution Carrington maps produced from photometrically uncalibrated Mt Wilson \ca data processed by \cite{bertello_70_2020} (see Sect. \ref{sec:carringtonmaps}). 
Their magnetogram reconstruction is based on the empirical relation between the magnetic flux and the integrated intensity of a plage region.
They singled out plage regions with a contrast threshold of 2$\sigma$ from each synoptic map and assigned to each region a single value of magnetic field strength. 
They also used the Mt Wilson sunspot dataset, which includes information on the polarity of sunspots, to recover the polarity of plage regions in the vicinity of sunspots.
The processing applied by these authors has, unfortunately, introduced clear artefacts affecting the brightness of plage regions and thus the recovered magnetic field strength (see Carrington maps in Fig. \ref{fig:carringtonmaps} and Sect. \ref{sec:limbdarkening}).
Other limitations of this study are the low resolution of the Carrington maps, the single value of the magnetic field strength assigned to plage regions, and the absence of magnetic field and polarity information outside of bright plage regions.

Similarly, \cite{mordvinov_long-term_2020} reconstructed magnetograms over 1907--1965 from Kodaikanal \ca Carrington maps produced by \cite{chatterjee_butterfly_2016}. 
They used a polynomial relationship between the magnetic flux density and \ca contrast. 
The degree of the polynomial was chosen to vary depending on the activity level. 
In particular, a 3rd degree polynomial was used for low activity levels and 4th degree for high activity levels.
This cycle-dependent variation of the relation might be an artefact introduced by the processing of the images (see Sect. \ref{sec:limbdarkening}).

Another approach for reconstructing unsigned magnetograms was proposed by \cite{chatzistergos_recovering_2019}. 
The pixel-by-pixel relation between \ca brightness and magnetic field strength was directly used to reconstruct solar magnetograms. 
This approach produces unsigned magnetograms and does not recover the magnetic field strength in sunspot regions. 
\cite{chatzistergos_reconstructing_2021-1} used this approach to reconstruct unsigned magnetograms from 13 \ca archives, including the historical photographic datasets from Mt Wilson and Meudon.
We note that the spatial resolution is an important factor when recovering magnetograms, since low spatial resolution results in high probability of ``missing'' part of the magnetic flux due to sub-pixel cancellation of opposite-polarity elements  \citep{krivova_effect_2004}.

\section{Irradiance reconstruction}
\label{sec:irradiance}
One of the most important applications of \ca data is reconstruction of past irradiance variations.
Measurements of TSI variations from space exist only since 1978, while longer records are required for example for climate studies.
The driver of the irradiance variations on timescales of days and longer is the evolution of the solar surface magnetic field \citep{shapiro_nature_2017,yeo_solar_2017}, as a result of a competition between plage and sunspot regions enhancing and suppressing TSI, respectively.
Thus, appropriate facular and sunspot data are needed by models to reconstruct past irradiance variations.
While sunspot data are readily available back to early 1600s \citep{vaquero_revised_2016}, facular data are significantly more scarce. 
In fact, most of the available facular data cover roughly the same period as direct TSI measurements, while \ca data are the only direct facular dataset extending back to 1892 \citep[potentially also white-light faculae;][]{foukal_curious_1993,carrasco_note_2021}. 
Most of the existing irradiance reconstructions extending back to 1600's use sunspot data to infer the characteristics of faculae. This results in a roughly an order of magnitude difference among the various estimates of the long-term evolution of TSI since 1700's \citep{solanki_solar_2013-1}, although more recently \cite{yeo_dimmest_2020} set an upper limit of 2.0$\pm0.7$ W/m$^2$ on the possible difference between the TSI at modern activity minima and grand minima.

Until now,
most studies used CCD-based \ca data for irradiance reconstructions, covering shorter periods than the direct TSI measurements \citep{chapman_variations_1996,chapman_modeling_2013,walton_contribution_2003,ermolli_modeling_2003,ermolli_recent_2011,penza_modeling_2003,fontenla_bright_2018,puiu_modeling_2019,choudhary_variability_2020,chatzistergos_modelling_2020}.
Notwithstanding the short interval, such data allowed an assessment of the accuracy of the irradiance reconstructions.
Furthermore, they provided an independent estimate on the TSI variations between activity minima over the last three cycles.
For instance, by reconstructing TSI variations with an empirical model using Rome/PSPT \ca and continuum observations, \cite{chatzistergos_modelling_2020} found that TSI  in general declined over the minima of the last three cycles.

\begin{figure*}[t!]
\begin{center}
\includegraphics[width=0.95\linewidth]{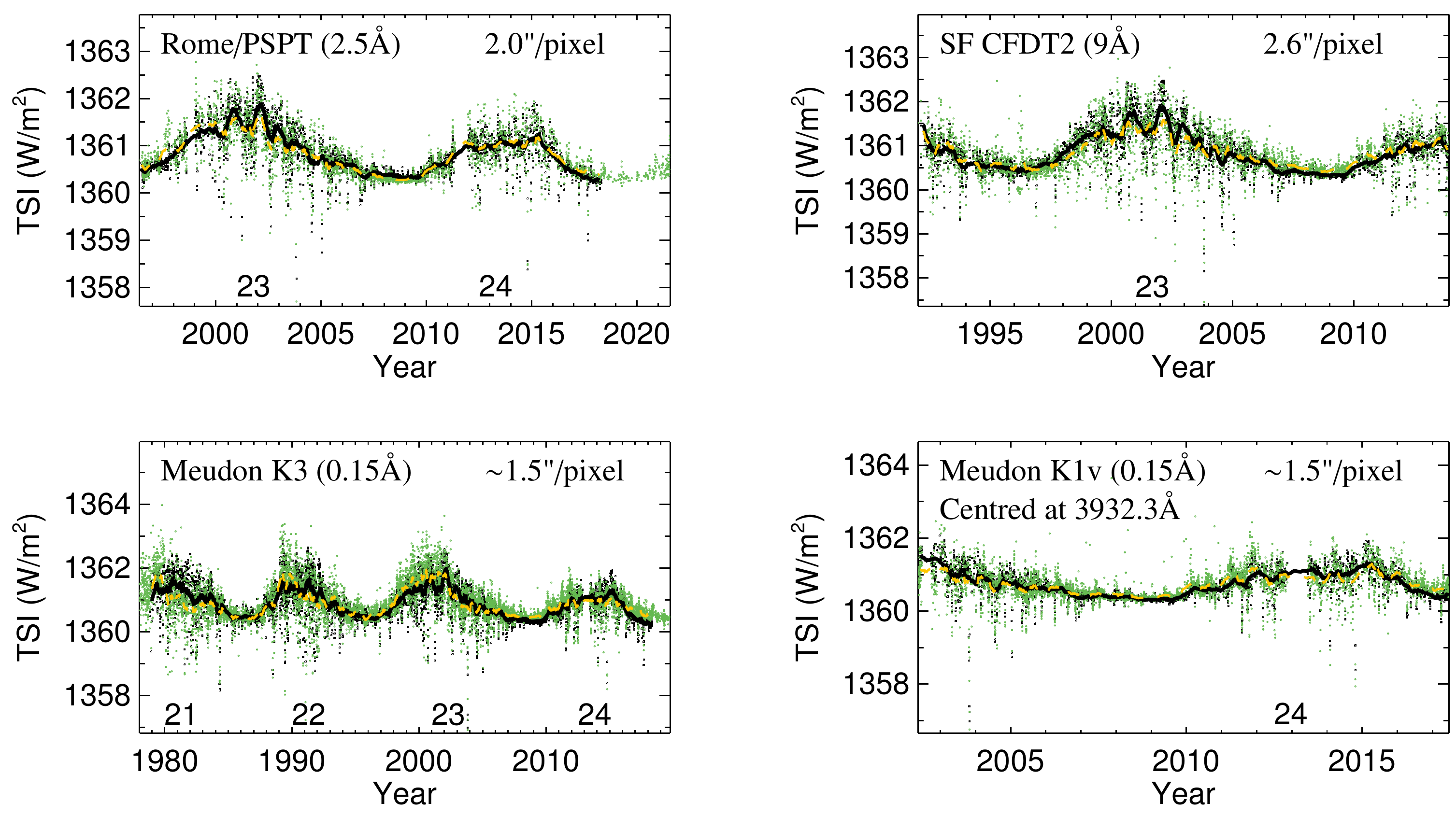}
\end{center}
\caption{Reconstructed TSI series by \cite{chatzistergos_reconstructing_2021-1} with \ca data from Rome/PSPT, San Fernando CFDT2, Meudon K3, and Meudon K1v. The reconstructions are shown in green dots for daily values and dashed yellow for 81-day running means. Also shown is the PMOD TSI composite (black dots for daily values and black line for 81-day running mean values). The numbers at the lower part of the figures denote the conventional solar cycle numbering. Also listed within the figures are the bandwidth used for the observations as well as the pixel scale (for Meudon K1v we also list the central wavelength).}\label{fig:tsireconstructions}
\end{figure*}

Use of historical photographic data for irradiance reconstructions has so far been rather limited, with only five TSI reconstructions from historical \ca observations known to us,
namely,  by \cite{ambelu_estimation_2011}, \cite{foukal_new_2012},  \cite{penza_total_2022}, \cite{xu_reconstruction_2021}, and \cite{chatzistergos_reconstructing_2021-1}.
\cite{ambelu_estimation_2011}, \cite{foukal_new_2012}, and \cite{xu_reconstruction_2021} used the photometrically uncalibrated Mt Wilson, Mt Wilson, and Kodaikanal \ca data, respectively to reconstruct TSI variations with a linear regression model. 
These studies also did not account for the various instrumental changes in Mt Wilson, thus rendering their results highly uncertain.
\cite{penza_total_2022} used the plage area composite series by \cite{chatzistergos_analysis_2019} to reconstruct TSI over the last five centuries with an empirical model.
\cite{chatzistergos_reconstructing_2021-1,chatzistergos_reconstructing_2021} used 13 \ca archives, including historical photographic archives from Meudon and Mt Wilson, to reconstruct past irradiance variations with the Spectral and Total Iiradiance REconstruction \citep[SATIRE][]{krivova_reconstruction_2003} model. 
Figure \ref{fig:tsireconstructions} shows the reconstructed TSI series from four \ca archives by \cite{chatzistergos_reconstructing_2021-1}, in particular those from Rome/PSPT, San Fernando CFDT2, Meudon K3, and Meudon K1v (see Fig. \ref{fig:passbandarchives} for the correspondence of the K3 and K1v parts of the \ca line).
The reconstructions done by \cite{chatzistergos_reconstructing_2021-1} were limited to the periods with overlap to the direct TSI measurements, since their aim was to assess the quality of the reconstruction.
Their results showed that \ca data can be used to recover past irradiance variations almost as accurately as with SATIRE-S \citep{yeo_reconstruction_2014} which employs actual magnetograms. 
This was a very important step, which demonstrated the ability of \ca data to produce highly accurate irradiance reconstructions, provided the data have been consistently and accurately processed, while also accounting for the various inhomogeneities within the datasets.

\section{Conclusions}
\label{sec:summary}

Full-disc \ca observations of the Sun have a long legacy, which is unrivaled by most direct solar data.
Such observations have been performed routinely since 1892 at numerous sites around the world and continue to this day.

In this review, we have highlighted the fact that \ca archives are far from a coherent and consistent dataset, rather they are collections made with quite diverse  instruments and observational settings. 
Deriving accurate and reliable results from \ca data requires robust processing as well as knowledge of and accounting for the various differences and inconsistencies of the available archives.
A very important aspect, usually neglected, is the need for the photometric calibration to account for the non-linear response of the photographic plates to the incident radiation.

The increased interest of the recent decades and digitisation of historical photographic \ca datasets have opened up the possibility for precise quantitative studies.
This led to the development of various modern processing techniques allowing one to overcome many of the issues affecting the images, including the photometric calibration of the images \citep[e.g.][]{chatzistergos_analysis_2018,chatzistergos_analysis_2019}.
Comprehensive analyses of plage and network regions have been performed on diverse \ca data, while plage areas determined from 43 archives have been combined into a composite series covering 1892--2019 \citep{chatzistergos_analysis_2020} with an almost full temporal coverage since 1907.

\ca observations are also invaluable for recovering information on the solar surface magnetic field, especially in the past. 
Due to the tight relation between the magnetic field strength and \ca brightness they can be used to reconstruct magnetograms back to 1892.
One of the most important applications of \ca observations is their potential to help us improve our understanding of the long-term variations in solar irradiance. 
\cite{chatzistergos_reconstructing_2021-1} showed that, if accurately processed and after accounting for the problems and artefacts, \ca data can be used to reconstruct past irradiance variations almost as accurately as with actual high-quality magnetograms.
This highlights the significance of \ca data for irradiance studies as well as for studies of the solar influence on Earth's climate.

We stress, however, that there are still a lot of unexplored \ca observations around, either because they have not been digitised yet (such as the data from Kenwood and Kandilli) or because their whereabouts are unknown (like the data from Hamburg).
Recovering and digitising the remaining \ca observations would be highly beneficial.

\section*{Conflict of Interest Statement}
%All financial, commercial or other relationships that might be perceived by the academic community as representing a potential conflict of interest must be disclosed. If no such relationship exists, authors will be asked to confirm the following statement: 

The authors declare that the research was conducted in the absence of any commercial or financial relationships that could be construed as a potential conflict of interest.

\section*{Author Contributions}
TC drafted the manuscript, while all co-authors provided parts of the text and proofread it.
%The Author Contributions section is mandatory for all articles, including articles by sole authors. If an appropriate statement is not provided on submission, a standard one will be inserted during the production process. The Author Contributions statement must describe the contributions of individual authors referred to by their initials and, in doing so, all authors agree to be accountable for the content of the work. Please see  \href{https://www.frontiersin.org/about/policies-and-publication-ethics#AuthorshipAuthorResponsibilities}{here} for full authorship criteria.

\section*{Funding}
This work was supported by the Italian MIUR-PRIN grant 2017 ''Circumterrestrial Environment: Impact of Sun--Earth Interaction'', by the German Federal Ministry of Education and Research (Project No. 01LG1909C), and by the
European Union's Horizon 2020 research and Innovation program under grant agreement No 824135 (SOLARNET) and No 739500 (PRE-EST).

\section*{Acknowledgments}
	The authors thank the observers at the Arcetri, Calern, Coimbra, Ebro, Kanzelhöhe, Kharkiv, Kodaikanal, Kyoto, Meudon, Mitaka, Mt Wilson, Pic du Midi, Rome, Sac Peak, San Fernando, and Valašské Meziříčí sites as well all other \ca observatories helping to create such an important archive of solar data. 
	We thank Isabelle Buale for her continued efforts at digitising the archive of Meudon observatory.
	We thank Teresa Barata, Luca Bertello, Victor M. S. Carrasco, Subhamoy Chatterjee, Angie Cookson, Juan José Curto, Ivan Dorotovič, Muthu Priyal, Jagdev Singh, and Andrey Tlatov for providing us with their produced timeseries.
	This publication uses data from the solar section of the solar observatory in Roquetes, Spain, owned and operated by the Fundació Observatori de l'Ebre.
	We thank Michael Borman, Didier Favre, André Gabriël, Lastrofieffe, Stefano Sello, and the other amateur observers at the international online solar database for their efforts at observing the Sun and allowing us to use their observations in this publication.
	We thank the referees who helped to improve this paper.
	This research has made use of NASA's Astrophysics Data System.

%\section*{Supplemental Data}
% \href{http://home.frontiersin.org/about/author-guidelines#SupplementaryMaterial}{Supplementary Material} should be uploaded separately on submission, if there are Supplementary Figures, please include the caption in the same file as the figure. LaTeX Supplementary Material templates can be found in the Frontiers LaTeX folder.

%\section*{Data Availability Statement}
%The datasets [GENERATED/ANALYZED] for this study can be found in the [NAME OF REPOSITORY] [LINK].
% Please see the availability of data guidelines for more information, at https://www.frontiersin.org/about/author-guidelines#AvailabilityofData

\bibliographystyle{Frontiers-Harvard} %  Many Frontiers journals use the Harvard referencing system (Author-date), to find the style and resources for the journal you are submitting to: https://zendesk.frontiersin.org/hc/en-us/articles/360017860337-Frontiers-Reference-Styles-by-Journal. For Humanities and Social Sciences articles please include page numbers in the in-text citations 
\bibliography{_biblio1}

\begin{thebibliography}{274}
\providecommand{\natexlab}[1]{#1}
\expandafter\ifx\csname urlstyle\endcsname\relax
  \providecommand{\doi}[1]{doi:\discretionary{}{}{}#1}\else
  \providecommand{\doi}{doi:\discretionary{}{}{}\begingroup
  \urlstyle{rm}\Url}\fi
\providecommand{\selectlanguage}[1]{\relax}
\providecommand{\bibAnnoteFile}[1]{%
  \IfFileExists{#1}{\begin{quotation}\noindent\textsc{Key:} #1\\
  \textsc{Annotation:}\ \input{#1}\end{quotation}}{}}
\providecommand{\bibAnnote}[2]{%
  \begin{quotation}\noindent\textsc{Key:} #1\\
  \textsc{Annotation:}\ #2\end{quotation}}

\bibitem[{Ambelu et~al.(2011)Ambelu, Falayi, Elemo, and
  Oladosu}]{ambelu_estimation_2011}
Ambelu, T., Falayi, E.~O., Elemo, E.~O., and Oladosu, O. (2011).
\newblock Estimation of total solar irradiance from sunspot number.
\newblock \emph{Latin-American Journal of Physics Education} 5
\bibAnnoteFile{ambelu_estimation_2011}

\bibitem[{Antonucci et~al.(1977)Antonucci, Azzarelli, Casalini, and
  Cerri}]{antonucci_chromospheric_1977}
Antonucci, E., Azzarelli, L., Casalini, P., and Cerri, S. (1977).
\newblock Chromospheric rotation during 1972-73, years of declining activity.
\newblock \emph{Solar Physics} 53, 519--529.
\newblock \doi{10.1007/BF00160294}
\bibAnnoteFile{antonucci_chromospheric_1977}

\bibitem[{Asvestari et~al.(2021)Asvestari, Pomoell, Kilpua, Good,
  Chatzistergos, Temmer et~al.}]{asvestari_modelling_2021}
Asvestari, E., Pomoell, J., Kilpua, E., Good, S., Chatzistergos, T., Temmer,
  M., et~al. (2021).
\newblock Modelling a multi-spacecraft coronal mass ejection encounter with
  {EUHFORIA}.
\newblock \emph{Astronomy \& Astrophysics} 652, A27.
\newblock \doi{10.1051/0004-6361/202140315}
\bibAnnoteFile{asvestari_modelling_2021}

\bibitem[{Babcock and Babcock(1955)}]{babcock_suns_1955}
Babcock, H.~W. and Babcock, H.~D. (1955).
\newblock The {Sun}'s {Magnetic} {Field}, 1952-1954.
\newblock \emph{The Astrophysical Journal} 121, 349.
\newblock \doi{10.1086/145994}
\bibAnnoteFile{babcock_suns_1955}

\bibitem[{Barata et~al.(2018)Barata, Carvalho, Dorotovič, Pinheiro, Garcia,
  Fernandes et~al.}]{barata_software_2018}
Barata, T., Carvalho, S., Dorotovič, I., Pinheiro, F. J.~G., Garcia, A.,
  Fernandes, J., et~al. (2018).
\newblock Software tool for automatic detection of solar plages in the
  {Coimbra} {Observatory} spectroheliograms.
\newblock \emph{Astronomy and Computing} 24, 70--83.
\newblock \doi{10.1016/j.ascom.2018.06.003}
\bibAnnoteFile{barata_software_2018}

\bibitem[{Bates and McDowell(1972)}]{bates_compact_1972}
Bates, B. and McDowell, M.~W. (1972).
\newblock A compact grating spectroheliograph for the {Mg} ii resonance lines.
\newblock \emph{Solar Physics} 23, 26--29.
\newblock \doi{10.1007/BF00153889}.
\newblock ADS Bibcode: 1972SoPh...23...26B
\bibAnnoteFile{bates_compact_1972}

\bibitem[{Belkina et~al.(1996)Belkina, Beletskij, Gretskij, and
  Marchenko}]{belkina_ccd_1996}
Belkina, I.~L., Beletskij, S.~A., Gretskij, A.~M., and Marchenko, G.~P. (1996).
\newblock {CCD} observations of the {Sun} in the lines {He} {I} 1083 nm,
  {H$\alpha$}, and {K} {Ca} {II}.
\newblock \emph{Kinematics and Physics of Celestial Bodies} 12, 55
\bibAnnoteFile{belkina_ccd_1996}

\bibitem[{Berrilli et~al.(2020)Berrilli, Criscuoli, Penza, and
  Lovric}]{berrilli_long-term_2020}
Berrilli, F., Criscuoli, S., Penza, V., and Lovric, M. (2020).
\newblock Long-term (1749–2015) {Variations} of {Solar} {UV} {Spectral}
  {Indices}.
\newblock \emph{Solar Physics} 295, 38.
\newblock \doi{10.1007/s11207-020-01603-5}
\bibAnnoteFile{berrilli_long-term_2020}

\bibitem[{Berrilli et~al.(1998)Berrilli, Ermolli, Florio, and
  Pietropaolo}]{berrilli_geometrical_1998}
Berrilli, F., Ermolli, I., Florio, A., and Pietropaolo, E. (1998).
\newblock Geometrical properties of the chromospheric network cells from
  {OAR}/{PSPT} images.
\newblock \emph{Memorie della Societa Astronomica Italiana} 69, 635
\bibAnnoteFile{berrilli_geometrical_1998}

\bibitem[{Berrilli et~al.(1999)Berrilli, Ermolli, Florio, and
  Pietropaolo}]{berrilli_average_1999}
Berrilli, F., Ermolli, I., Florio, A., and Pietropaolo, E. (1999).
\newblock Average properties and temporal variations of the geometry of solar
  network cells.
\newblock \emph{Astronomy and Astrophysics} 344, 965--972
\bibAnnoteFile{berrilli_average_1999}

\bibitem[{Bertello et~al.(2017)Bertello, Marble, and
  Pevtsov}]{bertello_ca_2017}
Bertello, L., Marble, A.~R., and Pevtsov, A.~A. (2017).
\newblock Ca {II} {K} 1-{A} {Emission} {Index} {Composites}.
\newblock \emph{ArXiv e-prints} 1702, arXiv:1702.00838
\bibAnnoteFile{bertello_ca_2017}

\bibitem[{Bertello et~al.(2016)Bertello, Pevtsov, Tlatov, and
  Singh}]{bertello_correlation_2016}
Bertello, L., Pevtsov, A., Tlatov, A., and Singh, J. (2016).
\newblock Correlation {Between} {Sunspot} {Number} and {Ca} {II} {K} {Emission}
  {Index}.
\newblock \emph{Solar Physics} \doi{10.1007/s11207-016-0927-9}
\bibAnnoteFile{bertello_correlation_2016}

\bibitem[{Bertello et~al.(2020)Bertello, Pevtsov, and
  Ulrich}]{bertello_70_2020}
Bertello, L., Pevtsov, A.~A., and Ulrich, R.~K. (2020).
\newblock 70 {Years} of {Chromospheric} {Solar} {Activity} and {Dynamics}.
\newblock \emph{The Astrophysical Journal} 897, 181.
\newblock \doi{10.3847/1538-4357/ab9746}
\bibAnnoteFile{bertello_70_2020}

\bibitem[{Bertello et~al.(2010)Bertello, Ulrich, and
  Boyden}]{bertello_mount_2010}
Bertello, L., Ulrich, R.~K., and Boyden, J.~E. (2010).
\newblock The {Mount} {Wilson} {Ca} ii {K} {Plage} {Index} {Time} {Series}.
\newblock \emph{Solar Physics} 264, 31--44.
\newblock \doi{10.1007/s11207-010-9570-z}
\bibAnnoteFile{bertello_mount_2010}

\bibitem[{Bethge et~al.(2011)Bethge, Peter, Kentischer, Halbgewachs, Elmore,
  and Beck}]{bethge_chromospheric_2011}
Bethge, C., Peter, H., Kentischer, T.~J., Halbgewachs, C., Elmore, D.~F., and
  Beck, C. (2011).
\newblock The {Chromospheric} {Telescope}.
\newblock \emph{Astronomy and Astrophysics} 534, A105.
\newblock \doi{10.1051/0004-6361/201117456}
\bibAnnoteFile{bethge_chromospheric_2011}

\bibitem[{Bühler et~al.(2013)Bühler, Lagg, and Solanki}]{buhler_quiet_2013}
Bühler, D., Lagg, A., and Solanki, S.~K. (2013).
\newblock Quiet {Sun} magnetic fields observed by {Hinode}: {Support} for a
  local dynamo.
\newblock \emph{Astronomy and Astrophysics} 555, A33.
\newblock \doi{10.1051/0004-6361/201321152}
\bibAnnoteFile{buhler_quiet_2013}

\bibitem[{Bose and Nagaraju(2018)}]{bose_variability_2018}
Bose, S. and Nagaraju, K. (2018).
\newblock On the {Variability} of the {Solar} {Mean} {Magnetic} {Field}:
  {Contributions} from {Various} {Magnetic} {Features} on the {Surface} of the
  {Sun}.
\newblock \emph{The Astrophysical Journal} 862, 35.
\newblock \doi{10.3847/1538-4357/aaccf1}
\bibAnnoteFile{bose_variability_2018}

\bibitem[{Brandt and Steinegger(1998)}]{brandt_determination_1998}
Brandt, P.~N. and Steinegger, M. (1998).
\newblock On the {Determination} of the {Quiet}-{Sun} {Center}-to-{Limb}
  {Variation} in {Ca} {II} {K} {Spectroheliograms}.
\newblock \emph{Solar Physics} 177, 287--294.
\newblock \doi{10.1023/A:1004953032251}.
\newblock Number: 1-2
\bibAnnoteFile{brandt_determination_1998}

\bibitem[{Caccin et~al.(1997)Caccin, Ermolli, Fofi, and
  Sambuco}]{caccin_variations_1997}
Caccin, B., Ermolli, I., Fofi, M., and Sambuco, A.~M. (1997).
\newblock Variations of the network contribution to the solar irradiance.
\newblock \emph{Memorie della Societa Astronomica Italiana} 68, 459
\bibAnnoteFile{caccin_variations_1997}

\bibitem[{Caccin et~al.(1998{\natexlab{a}})Caccin, Ermolli, Fofi, and
  Sambuco}]{caccin_variations_1998}
Caccin, B., Ermolli, I., Fofi, M., and Sambuco, A.~M. (1998{\natexlab{a}}).
\newblock Variations of the {Chromospheric} {Network} with the {Solar} {Cycle}.
\newblock \emph{Solar Physics} 177, 295--303.
\newblock \doi{10.1023/A:1004938412420}
\bibAnnoteFile{caccin_variations_1998}

\bibitem[{Caccin et~al.(1998{\natexlab{b}})Caccin, Staro, and
  Gomez}]{caccin_variation_1998}
Caccin, B., Staro, F., and Gomez, M.~T. (1998{\natexlab{b}}).
\newblock Variation of the effective temperature with the solar cycle.
\newblock \emph{Memorie della Societa Astronomica Italiana} 69, 595
\bibAnnoteFile{caccin_variation_1998}

\bibitem[{Carrasco et~al.(2021)Carrasco, Nogales, Vaquero, Chatzistergos, and
  Ermolli}]{carrasco_note_2021}
Carrasco, V. M.~S., Nogales, J.~M., Vaquero, J.~M., Chatzistergos, T., and
  Ermolli, I. (2021).
\newblock A note on the sunspot and prominence records made by {Angelo}
  {Secchi} during the period 1871–1875.
\newblock \emph{Journal of Space Weather and Space Climate} 11, 51.
\newblock \doi{10.1051/swsc/2021033}
\bibAnnoteFile{carrasco_note_2021}

\bibitem[{Carrasco and Vaquero(2022)}]{carrasco_catalog_2022}
Carrasco, V. M.~S. and Vaquero, J.~M. (2022).
\newblock A {Catalog} of {Faculae}, {Prominences}, and {Filaments} for the
  {Period} 1929–1944 from the {Astronomical} {Observatory} of the
  {University} of {Coimbra}.
\newblock \emph{The Astrophysical Journal Supplement Series} 262, 44.
\newblock \doi{10.3847/1538-4365/ac85dd}
\bibAnnoteFile{carrasco_catalog_2022}

\bibitem[{Centrone et~al.(2005)Centrone, Ermolli, and
  Giorgi}]{centrone_image_2005}
Centrone, M., Ermolli, I., and Giorgi, F. (2005).
\newblock Image processing for the {Arcetri} {Solar} {Archive}.
\newblock \emph{Memorie della Societa Astronomica Italiana} 76, 941
\bibAnnoteFile{centrone_image_2005}

\bibitem[{Chapman et~al.(1996)Chapman, Cookson, and
  Dobias}]{chapman_variations_1996}
Chapman, G.~A., Cookson, A.~M., and Dobias, J.~J. (1996).
\newblock Variations in total solar irradiance during solar cycle 22.
\newblock \emph{Journal of Geophysical Research} 101, 13541--13548.
\newblock \doi{10.1029/96JA00683}
\bibAnnoteFile{chapman_variations_1996}

\bibitem[{Chapman et~al.(1997)Chapman, Cookson, and
  Dobias}]{chapman_solar_1997}
Chapman, G.~A., Cookson, A.~M., and Dobias, J.~J. (1997).
\newblock Solar {Variability} and the {Relation} of {Facular} to {Sunspot}
  {Areas} during {Solar} {Cycle} 22.
\newblock \emph{The Astrophysical Journal} 482, 541--545.
\newblock \doi{10.1086/304138}
\bibAnnoteFile{chapman_solar_1997}

\bibitem[{Chapman et~al.(2001)Chapman, Cookson, Dobias, and
  Walton}]{chapman_improved_2001}
Chapman, G.~A., Cookson, A.~M., Dobias, J.~J., and Walton, S.~R. (2001).
\newblock An {Improved} {Determination} of the {Area} {Ratio} of {Faculae} to
  {Sunspots}.
\newblock \emph{The Astrophysical Journal} 555, 462.
\newblock \doi{10.1086/321466}
\bibAnnoteFile{chapman_improved_2001}

\bibitem[{Chapman et~al.(2013)Chapman, Cookson, and
  Preminger}]{chapman_modeling_2013}
Chapman, G.~A., Cookson, A.~M., and Preminger, D.~G. (2013).
\newblock Modeling {Total} {Solar} {Irradiance} with {San} {Fernando}
  {Observatory} {Ground}-{Based} {Photometry}: {Comparison} with {ACRIM},
  {PMOD}, and {RMIB} {Composites}.
\newblock \emph{Solar Physics} 283, 295--305.
\newblock \doi{10.1007/s11207-013-0233-8}
\bibAnnoteFile{chapman_modeling_2013}

\bibitem[{Chapman et~al.(2011)Chapman, Dobias, and
  Arias}]{chapman_facular_2011}
Chapman, G.~A., Dobias, J.~J., and Arias, T. (2011).
\newblock Facular and {Sunspot} {Areas} {During} {Solar} {Cycles} 22 and 23.
\newblock \emph{The Astrophysical Journal} 728, 150.
\newblock \doi{10.1088/0004-637X/728/2/150}
\bibAnnoteFile{chapman_facular_2011}

\bibitem[{Chatterjee et~al.(2019)Chatterjee, Banerjee, McIntosh, Leamon,
  Dikpati, Srivastava et~al.}]{chatterjee_signature_2019}
Chatterjee, S., Banerjee, D., McIntosh, S.~W., Leamon, R.~J., Dikpati, M.,
  Srivastava, A.~K., et~al. (2019).
\newblock Signature of {Extended} {Solar} {Cycles} as {Detected} from {Ca} ii
  {K} {Synoptic} {Maps} of {Kodaikanal} and {Mount} {Wilson} {Observatory}.
\newblock \emph{The Astrophysical Journal} 874, L4.
\newblock \doi{10.3847/2041-8213/ab0e0e}
\bibAnnoteFile{chatterjee_signature_2019}

\bibitem[{Chatterjee et~al.(2016)Chatterjee, Banerjee, and
  Ravindra}]{chatterjee_butterfly_2016}
Chatterjee, S., Banerjee, D., and Ravindra, B. (2016).
\newblock A butterfly diagram and carrington maps for century-long {Ca} {II}
  {K} spectroheliograms from the {Kodaikanal} observatory.
\newblock \emph{The Astrophysical Journal} 827, 87.
\newblock \doi{10.3847/0004-637X/827/1/87}
\bibAnnoteFile{chatterjee_butterfly_2016}

\bibitem[{Chatterjee et~al.(2017{\natexlab{a}})Chatterjee, Hegde, Banerjee, and
  Ravindra}]{chatterjee_long-term_2017}
Chatterjee, S., Hegde, M., Banerjee, D., and Ravindra, B. (2017{\natexlab{a}}).
\newblock Long-term {Study} of the {Solar} {Filaments} from the {Synoptic}
  {Maps} as {Derived} from {H}\$\_{\textbackslash}alpha\$ {Spectroheliograms}
  of the {Kodaikanal} {Observatory}.
\newblock \emph{The Astrophysical Journal} 849, 44.
\newblock \doi{10.3847/1538-4357/aa8ad9}
\bibAnnoteFile{chatterjee_long-term_2017}

\bibitem[{Chatterjee et~al.(2017{\natexlab{b}})Chatterjee, Mandal, and
  Banerjee}]{chatterjee_variation_2017}
Chatterjee, S., Mandal, S., and Banerjee, D. (2017{\natexlab{b}}).
\newblock Variation of {Supergranule} {Parameters} with {Solar} {Cycles}:
  {Results} from {Century}-long {Kodaikanal} {Digitized} {Ca} {II} {K} {Data}.
\newblock \emph{The Astrophysical Journal} 841, 70.
\newblock \doi{10.3847/1538-4357/aa709d}
\bibAnnoteFile{chatterjee_variation_2017}

\bibitem[{Chatzistergos(2017)}]{chatzistergos_analysis_2017}
Chatzistergos, T. (2017).
\newblock \emph{Analysis of historical solar observations and long-term changes
  in solar irradiance}.
\newblock {PhD} thesis (Uni-edition)
\bibAnnoteFile{chatzistergos_analysis_2017}

\bibitem[{Chatzistergos et~al.(2019{\natexlab{a}})Chatzistergos, Ermolli,
  Falco, Giorgi, Guglielmino, Krivova et~al.}]{chatzistergos_historical_2019}
Chatzistergos, T., Ermolli, I., Falco, M., Giorgi, F., Guglielmino, S.~L.,
  Krivova, N.~A., et~al. (2019{\natexlab{a}}).
\newblock Historical solar {Ca} {II} {K} observations at the {Rome} and
  {Catania} observatories.
\newblock In \emph{Il {Nuovo} {Cimento}}. vol. 42C, 5.
\newblock \doi{10.1393/ncc/i2019-19005-2}
\bibAnnoteFile{chatzistergos_historical_2019}

\bibitem[{Chatzistergos et~al.(2020{\natexlab{a}})Chatzistergos, Ermolli,
  Giorgi, Krivova, and Puiu}]{chatzistergos_modelling_2020}
Chatzistergos, T., Ermolli, I., Giorgi, F., Krivova, N.~A., and Puiu, C.~C.
  (2020{\natexlab{a}}).
\newblock Modelling solar irradiance from ground-based photometric
  observations.
\newblock \emph{Journal of Space Weather and Space Climate} 10, 45.
\newblock \doi{10.1051/swsc/2020047}
\bibAnnoteFile{chatzistergos_modelling_2020}

\bibitem[{Chatzistergos et~al.(2022)Chatzistergos, Ermolli, Krivova, Barata,
  Carvalho, and Malherbe}]{chatzistergos_scrutinising_2022}
Chatzistergos, T., Ermolli, I., Krivova, N.~A., Barata, T., Carvalho, S., and
  Malherbe, J.-M. (2022).
\newblock Scrutinising the relationship between plage areas and sunspot areas
  and numbers.
\newblock \emph{Astronomy \& Astrophysics} \doi{10.1051/0004-6361/202244913}
\bibAnnoteFile{chatzistergos_scrutinising_2022}

\bibitem[{Chatzistergos et~al.(2018{\natexlab{a}})Chatzistergos, Ermolli,
  Krivova, and Solanki}]{chatzistergos_ca_2018}
Chatzistergos, T., Ermolli, I., Krivova, N.~A., and Solanki, S.~K.
  (2018{\natexlab{a}}).
\newblock Ca {II} {K} spectroheliograms for studies of long-term changes in
  solar irradiance.
\newblock In \emph{Long-term {Datasets} for the {Understanding} of {Solar} and
  {Stellar} {Magnetic} {Cycles}}, eds. D.~Banerjee, J.~Jiang, K.~Kusano, and
  S.~Solanki (Cambridge, UK: Cambridge University Press), vol. 340 of
  \emph{{IAU} {Symposium}}, 125--128.
\newblock \doi{10.1017/S1743921318001825}
\bibAnnoteFile{chatzistergos_ca_2018}

\bibitem[{Chatzistergos et~al.(2019{\natexlab{b}})Chatzistergos, Ermolli,
  Krivova, and Solanki}]{chatzistergos_analysis_2019}
Chatzistergos, T., Ermolli, I., Krivova, N.~A., and Solanki, S.~K.
  (2019{\natexlab{b}}).
\newblock Analysis of full disc {Ca} {II} {K} spectroheliograms - {II}.
  {Towards} an accurate assessment of long-term variations in plage areas.
\newblock \emph{Astronomy \& Astrophysics} 625, A69.
\newblock \doi{10.1051/0004-6361/201834402}
\bibAnnoteFile{chatzistergos_analysis_2019}

\bibitem[{Chatzistergos et~al.(2020{\natexlab{b}})Chatzistergos, Ermolli,
  Krivova, and Solanki}]{chatzistergos_historical_2020}
Chatzistergos, T., Ermolli, I., Krivova, N.~A., and Solanki, S.~K.
  (2020{\natexlab{b}}).
\newblock Historical solar {Ca} {II} {K} observations at the {Kyoto} and
  {Sacramento} {Peak} observatories.
\newblock \emph{Journal of Physics: Conference Series} 1548, 012007.
\newblock \doi{10.1088/1742-6596/1548/1/012007}
\bibAnnoteFile{chatzistergos_historical_2020}

\bibitem[{Chatzistergos et~al.(2020{\natexlab{c}})Chatzistergos, Ermolli,
  Krivova, Solanki, Banerjee, Barata et~al.}]{chatzistergos_analysis_2020}
Chatzistergos, T., Ermolli, I., Krivova, N.~A., Solanki, S.~K., Banerjee, D.,
  Barata, T., et~al. (2020{\natexlab{c}}).
\newblock Analysis of full-disc {Ca} {II} {K} spectroheliograms - {III}.
  {Plage} area composite series covering 1892–2019.
\newblock \emph{Astronomy \& Astrophysics} 639, A88.
\newblock \doi{10.1051/0004-6361/202037746}
\bibAnnoteFile{chatzistergos_analysis_2020}

\bibitem[{Chatzistergos et~al.(2016)Chatzistergos, Ermolli, Solanki, and
  Krivova}]{chatzistergos_exploiting_2016}
Chatzistergos, T., Ermolli, I., Solanki, S.~K., and Krivova, N.~A. (2016).
\newblock Exploiting {Four} {Historical} {Ca} {II} {K} {Spectroheliogram}
  {Archives}.
\newblock In \emph{Coimbra {Solar} {Physics} {Meeting}: {Ground}-based {Solar}
  {Observations} in the {Space} {Instrumentation} {Era}}, eds. I.~Dorotovic,
  C.~E. Fischer, and M.~Temmer (San Francisco: Astronomical Society of the
  Pacific), vol. 504 of \emph{Astronomical {Society} of the {Pacific}
  {Conference} {Series}}, 227--231
\bibAnnoteFile{chatzistergos_exploiting_2016}

\bibitem[{Chatzistergos et~al.(2018{\natexlab{b}})Chatzistergos, Ermolli,
  Solanki, and Krivova}]{chatzistergos_analysis_2018}
Chatzistergos, T., Ermolli, I., Solanki, S.~K., and Krivova, N.~A.
  (2018{\natexlab{b}}).
\newblock Analysis of full disc {Ca} {II} {K} spectroheliograms - {I}.
  {Photometric} calibration and centre-to-limb variation compensation.
\newblock \emph{Astronomy \& Astrophysics} 609, A92.
\newblock \doi{10.1051/0004-6361/201731511}
\bibAnnoteFile{chatzistergos_analysis_2018}

\bibitem[{Chatzistergos et~al.(2019{\natexlab{c}})Chatzistergos, Ermolli,
  Solanki, Krivova, Banerjee, Jha et~al.}]{chatzistergos_delving_2019}
Chatzistergos, T., Ermolli, I., Solanki, S.~K., Krivova, N.~A., Banerjee, D.,
  Jha, B.~K., et~al. (2019{\natexlab{c}}).
\newblock Delving into the {Historical} {Ca} ii {K} {Archive} from the
  {Kodaikanal} {Observatory}: {The} {Potential} of the {Most} {Recent}
  {Digitized} {Series}.
\newblock \emph{Solar Physics} 294, 145.
\newblock \doi{10.1007/s11207-019-1532-5}
\bibAnnoteFile{chatzistergos_delving_2019}

\bibitem[{Chatzistergos et~al.(2019{\natexlab{d}})Chatzistergos, {Ermolli,
  Ilaria}, {Solanki, Sami K.}, {Krivova, Natalie A.}, {Giorgi, Fabrizio}, and
  {Yeo, Kok Leng}}]{chatzistergos_recovering_2019}
Chatzistergos, T., {Ermolli, Ilaria}, {Solanki, Sami K.}, {Krivova, Natalie
  A.}, {Giorgi, Fabrizio}, and {Yeo, Kok Leng} (2019{\natexlab{d}}).
\newblock Recovering the unsigned photospheric magnetic field from {Ca} {II}
  {K} observations.
\newblock \emph{Astronomy \& Astrophysics} 626, A114.
\newblock \doi{10.1051/0004-6361/201935131}
\bibAnnoteFile{chatzistergos_recovering_2019}

\bibitem[{Chatzistergos et~al.(2021{\natexlab{a}})Chatzistergos, Krivova,
  Ermolli, Yeo, Mandal, Solanki et~al.}]{chatzistergos_reconstructing_2021-1}
Chatzistergos, T., Krivova, N.~A., Ermolli, I., Yeo, K.~L., Mandal, S.,
  Solanki, S.~K., et~al. (2021{\natexlab{a}}).
\newblock Reconstructing solar irradiance from historical {Ca} {II} {K}
  observations - {I}. {Method} and its validation.
\newblock \emph{Astronomy \& Astrophysics} 656, A104.
\newblock \doi{10.1051/0004-6361/202141516}
\bibAnnoteFile{chatzistergos_reconstructing_2021-1}

\bibitem[{Chatzistergos et~al.(2021{\natexlab{b}})Chatzistergos, Krivova,
  Ermolli, Yeo, Solanki, Puiu et~al.}]{chatzistergos_reconstructing_2021}
Chatzistergos, T., Krivova, N.~A., Ermolli, I., Yeo, K.~L., Solanki, S.~K.,
  Puiu, C.~C., et~al. (2021{\natexlab{b}}).
\newblock Reconstructing solar irradiance from {Ca} {II} {K} observations.
\newblock In \emph{{ESSOAr}} (San Francisco), 3.
\newblock \doi{10.1002/essoar.10505862.1}
\bibAnnoteFile{chatzistergos_reconstructing_2021}

\bibitem[{Chatzistergos et~al.(2017)Chatzistergos, Usoskin, Kovaltsov, Krivova,
  and Solanki}]{chatzistergos_new_2017}
Chatzistergos, T., Usoskin, I.~G., Kovaltsov, G.~A., Krivova, N.~A., and
  Solanki, S.~K. (2017).
\newblock New reconstruction of the sunspot group numbers since 1739 using
  direct calibration and "backbone" methods.
\newblock \emph{Astronomy \& Astrophysics} 602, A69.
\newblock \doi{10.1051/0004-6361/201630045}
\bibAnnoteFile{chatzistergos_new_2017}

\bibitem[{Chinnici and Consolmagno(2021)}]{chinnici_angelo_2021}
Chinnici, I. and Consolmagno, G. (eds.) (2021).
\newblock \emph{Angelo {Secchi} and {Nineteenth} {Century} {Science}: {The}
  {Multidisciplinary} {Contributions} of a {Pioneer} and {Innovator}}.
\newblock Historical \& {Cultural} {Astronomy} (Cham: Springer International
  Publishing)
\bibAnnoteFile{chinnici_angelo_2021}

\bibitem[{Choudhary et~al.(2020)Choudhary, Cadavid, Cookson, and
  Chapman}]{choudhary_variability_2020}
Choudhary, D.~P., Cadavid, A.~C., Cookson, A., and Chapman, G.~A. (2020).
\newblock Variability in {Irradiance} and {Photometric} {Indices} {During} the
  {Last} {Two} {Solar} {Cycles}.
\newblock \emph{Solar Physics} 295, 15.
\newblock \doi{10.1007/s11207-019-1559-7}
\bibAnnoteFile{choudhary_variability_2020}

\bibitem[{Chowdhury et~al.(2022)Chowdhury, Belur, Bertello, and
  Pevtsov}]{chowdhury_analysis_2022}
Chowdhury, P., Belur, R., Bertello, L., and Pevtsov, A.~A. (2022).
\newblock Analysis of {Solar} {Hemispheric} {Chromosphere} {Properties} using
  the {Kodaikanal} {Observatory} {Ca}–{K} {Index}.
\newblock \emph{The Astrophysical Journal} 925, 81.
\newblock \doi{10.3847/1538-4357/ac3983}
\bibAnnoteFile{chowdhury_analysis_2022}

\bibitem[{Clette and Lefèvre(2016)}]{clette_new_2016-1}
Clette, F. and Lefèvre, L. (2016).
\newblock The {New} {Sunspot} {Number}: {Assembling} {All} {Corrections}.
\newblock \emph{Solar Physics} 291, 2629--2651.
\newblock \doi{10.1007/s11207-016-1014-y}
\bibAnnoteFile{clette_new_2016-1}

\bibitem[{Curto et~al.(2008)Curto, Blanca, and
  Martínez}]{curto_automatic_2008}
Curto, J.~J., Blanca, M., and Martínez, E. (2008).
\newblock Automatic {Sunspots} {Detection} on {Full}-{Disk} {Solar} {Images}
  using {Mathematical} {Morphology}.
\newblock \emph{Solar Physics} 250, 411--429.
\newblock \doi{10.1007/s11207-008-9224-6}
\bibAnnoteFile{curto_automatic_2008}

\bibitem[{Curto et~al.(2016)Curto, Solé, Genescà, Blanca, and
  Vaquero}]{curto_historical_2016}
Curto, J.~J., Solé, J.~G., Genescà, M., Blanca, M.~J., and Vaquero, J.~M.
  (2016).
\newblock Historical {Heliophysical} {Series} of the {Ebro} {Observatory}.
\newblock \emph{Solar Physics} \doi{10.1007/s11207-016-0896-z}
\bibAnnoteFile{curto_historical_2016}

\bibitem[{Dainty and Shaw(1974)}]{dainty_image_1974}
Dainty, J.~C. and Shaw, R. (1974).
\newblock Image science. {Principles}, analysis and evaluation of
  photographic-type imaging processes.
\newblock \emph{London: Academic Press}
\bibAnnoteFile{dainty_image_1974}

\bibitem[{D'Azambuja(1930)}]{dazambuja_annales_1930}
D'Azambuja, L. (1930).
\newblock \emph{Annales de l'{Observatoire} de {Paris}, {Section} de {Meudon}}
  (Gauthier-VillarsGauthier-Villars (Paris))
\bibAnnoteFile{dazambuja_annales_1930}

\bibitem[{de~Paula and Curto(2020)}]{de_paula_evolution_2020}
de~Paula, V. and Curto, J.~J. (2020).
\newblock The {Evolution} over {Time} and {North}–{South} {Asymmetry} of
  {Sunspots} and {Solar} {Plages} for the {Period} 1910 to 1937 {Using} {Data}
  from {Ebro} {Catalogues}.
\newblock \emph{Solar Physics} 295, 99.
\newblock \doi{10.1007/s11207-020-01648-6}
\bibAnnoteFile{de_paula_evolution_2020}

\bibitem[{de~Paula et~al.(2022)de~Paula, Curto, and
  Oliver}]{de_paula_cyclic_2022}
de~Paula, V., Curto, J.~J., and Oliver, R. (2022).
\newblock The cyclic behaviour in the {N}-{S} asymmetry of sunspots and solar
  plages for the period 1910 to 1937 using data from {Ebro} catalogues.
\newblock \emph{Monthly Notices of the Royal Astronomical Society} 512,
  5726--5742.
\newblock \doi{10.1093/mnras/stac424}.
\newblock ADS Bibcode: 2022MNRAS.512.5726D
\bibAnnoteFile{de_paula_cyclic_2022}

\bibitem[{de~Paula et~al.(2021)de~Paula, Curto, and
  Sole}]{de_paula_application_2021}
de~Paula, V., Curto, J.~J., and Sole, T. (2021).
\newblock Application of the {Markov} {Chain} {Model} to {Sunspots} and {Solar}
  {Plages} for the {Period} 1910 to 1937 {Using} {Data} from {Ebro}
  {Catalogues}.
\newblock \emph{Solar Physics} 296, 92.
\newblock \doi{10.1007/s11207-021-01838-w}
\bibAnnoteFile{de_paula_application_2021}

\bibitem[{Deng et~al.(1997)Deng, Ai, Wang, Song, Zhang, and
  Ye}]{deng_reports_1997}
Deng, Y., Ai, G., Wang, J., Song, G., Zhang, B., and Ye, X. (1997).
\newblock Reports on {Test} {Observations} with the {Multi}-{Channel} {Solar}
  {Telescope}.
\newblock \emph{Solar Physics} 173, 207--221.
\newblock \doi{10.1023/A:1004960617982}.
\newblock ADS Bibcode: 1997SoPh..173..207D
\bibAnnoteFile{deng_reports_1997}

\bibitem[{Denker et~al.(1999)Denker, Johannesson, Marquette, Goode, Wang, and
  Zirin}]{denker_synoptic_1999}
Denker, C., Johannesson, A., Marquette, W., Goode, P.~R., Wang, H., and Zirin,
  H. (1999).
\newblock Synoptic {H$\alpha$} {Full}-{Disk} {Observations} of the {Sun} from
  {BigBear} {Solar} {Observatory} - {I}. {Instrumentation}, {Image}
  {Processing}, {Data} {Products}, and {First} {Results}.
\newblock \emph{Solar Physics} 184, 87--102.
\newblock \doi{10.1023/A:1005047906097}
\bibAnnoteFile{denker_synoptic_1999}

\bibitem[{Devi et~al.(2021)Devi, Singh, Chandra, Priyal, and
  Joshi}]{devi_variation_2021}
Devi, P., Singh, J., Chandra, R., Priyal, M., and Joshi, R. (2021).
\newblock Variation of {Chromospheric} {Features} as a {Function} of {Latitude}
  and {Time} {Using} {Ca}-{K} {Spectroheliograms} for {Solar} {Cycles}
  15 – 23: {Implications} for {Meridional} {Flow}.
\newblock \emph{Solar Physics} 296, 49.
\newblock \doi{10.1007/s11207-021-01798-1}
\bibAnnoteFile{devi_variation_2021}

\bibitem[{Diercke and Denker(2019)}]{diercke_chromospheric_2019}
Diercke, A. and Denker, C. (2019).
\newblock Chromospheric {Synoptic} {Maps} of {Polar} {Crown} {Filaments}.
\newblock \emph{Solar Physics} 294, 152.
\newblock \doi{10.1007/s11207-019-1538-z}
\bibAnnoteFile{diercke_chromospheric_2019}

\bibitem[{Diercke et~al.(2022)Diercke, Kuckein, Cauley, Poppenhäger,
  Alvarado-Gómez, Dineva et~al.}]{diercke_solar_2022}
Diercke, A., Kuckein, C., Cauley, P.~W., Poppenhäger, K., Alvarado-Gómez,
  J.~D., Dineva, E., et~al. (2022).
\newblock Solar {H$\alpha$} excess during {Solar} {Cycle} 24 from full-disk
  filtergrams of the {Chromospheric} {Telescope}.
\newblock \emph{Astronomy and Astrophysics} 661, A107.
\newblock \doi{10.1051/0004-6361/202040091}
\bibAnnoteFile{diercke_solar_2022}

\bibitem[{Dineva et~al.(2022)Dineva, Pearson, Ilyin, Verma, Diercke,
  Strassmeier et~al.}]{dineva_characterization_2022}
Dineva, E., Pearson, J., Ilyin, I., Verma, M., Diercke, A., Strassmeier, K.~G.,
  et~al. (2022).
\newblock Characterization of chromospheric activity based on {Sun}-as-a-star
  spectral and disk-resolved activity indices.
\newblock \emph{Astronomische Nachrichten} 343, e223996.
\newblock \doi{10.1002/asna.20223996}
\bibAnnoteFile{dineva_characterization_2022}

\bibitem[{Dizer(1968)}]{dizer_kandilli_1968}
Dizer, M. (1968).
\newblock {Kandİllİ} observatory, {Istanbul}.
\newblock \emph{Solar Physics} 3, 491--492.
\newblock \doi{10.1007/BF00171622}
\bibAnnoteFile{dizer_kandilli_1968}

\bibitem[{Doerr et~al.(2016)Doerr, Vitas, and Fabbian}]{doerr_how_2016}
Doerr, H.-P., Vitas, N., and Fabbian, D. (2016).
\newblock How different are the {Liège} and {Hamburg} atlases of the solar
  spectrum?
\newblock \emph{Astronomy and Astrophysics} 590, A118.
\newblock \doi{10.1051/0004-6361/201628570}
\bibAnnoteFile{doerr_how_2016}

\bibitem[{Domingo et~al.(1995)Domingo, Fleck, and Poland}]{domingo_soho_1995}
Domingo, V., Fleck, B., and Poland, A.~I. (1995).
\newblock The {SOHO} {Mission}: an {Overview}.
\newblock \emph{Solar Physics} 162, 1--37.
\newblock \doi{10.1007/BF00733425}.
\newblock ADS Bibcode: 1995SoPh..162....1D
\bibAnnoteFile{domingo_soho_1995}

\bibitem[{Donnelly et~al.(1977)Donnelly, Grubb, and
  Cowley}]{donnelly_solar_1977}
Donnelly, R.~F., Grubb, R.~N., and Cowley, F.~C. (1977).
\newblock Solar {X}-ray measurements from {SMS}-1, {SMS}-2, and {GOES}-1,
  information for data users.
\newblock \emph{NASA STI/Recon Technical Report N} 78, 13992
\bibAnnoteFile{donnelly_solar_1977}

\bibitem[{Dorotovič et~al.(2007)Dorotovič, Journoud, Rybák, and
  Sýkora}]{dorotovic_north-south_2007}
Dorotovič, I., Journoud, P., Rybák, J., and Sýkora, J. (2007).
\newblock North-{South} {Asymmetry} of {Ca} {II} {K} {Plages}.
\newblock \emph{The Physics of Chromospheric Plasmas} 368, 527
\bibAnnoteFile{dorotovic_north-south_2007}

\bibitem[{Dorotovič et~al.(2010)Dorotovič, Rybák, Garcia, and
  Journoud}]{dorotovic_north-south_2010}
Dorotovič, I., Rybák, J., Garcia, A., and Journoud, P. (2010).
\newblock North-south asymmetry of {Ca} {II} {K} regions determined from {OAUC}
  spectroheliograms: 1996 - 2006.
\newblock \emph{Proceedings of the 20th National Solar Physics Meeting} 20, 58
\bibAnnoteFile{dorotovic_north-south_2010}

\bibitem[{El-Borie et~al.(2020)El-Borie, El-Taher, Thabet, Ibrahim, Aly, and
  Bishara}]{el-borie_influence_2020}
El-Borie, M.~A., El-Taher, A.~M., Thabet, A.~A., Ibrahim, S.~F., Aly, N.~S.,
  and Bishara, A.~A. (2020).
\newblock The {Influence} of {Asymmetrical} {Distribution} of {Hemispheric}
  {Sunspot} {Areas} on {Some} {Solar} {Parameters}' {Periodicities} during the
  {Period} 1945–2017: {Wavelet} {Analysis}.
\newblock \emph{The Astrophysical Journal} 898, 73.
\newblock \doi{10.3847/1538-4357/ab9d21}
\bibAnnoteFile{el-borie_influence_2020}

\bibitem[{Ermolli et~al.(2003{\natexlab{a}})Ermolli, Berrilli, and
  Florio}]{ermolli_measure_2003}
Ermolli, I., Berrilli, F., and Florio, A. (2003{\natexlab{a}}).
\newblock A measure of the network radiative properties over the solar activity
  cycle.
\newblock \emph{Astronomy and Astrophysics} 412, 857--864.
\newblock \doi{10.1051/0004-6361:20031479}
\bibAnnoteFile{ermolli_measure_2003}

\bibitem[{Ermolli et~al.(1998{\natexlab{a}})Ermolli, Berrilli, Florio, and
  Pietropaolo}]{ermolli_chromospheric_1998}
Ermolli, I., Berrilli, F., Florio, A., and Pietropaolo, E.
  (1998{\natexlab{a}}).
\newblock Chromospheric {Network} {Properties} {Derived} {From} {One} {Year} of
  {PSPT} {Images}.
\newblock In \emph{Synoptic {Solar} {Physics}} (Astronomical Society of the
  Pacific), vol. 140 of \emph{Astronomical {Society} of the {Pacific}
  {Conference} {Series}}, 223
\bibAnnoteFile{ermolli_chromospheric_1998}

\bibitem[{Ermolli et~al.(2003{\natexlab{b}})Ermolli, Caccin, Centrone, and
  Penza}]{ermolli_modeling_2003}
Ermolli, I., Caccin, B., Centrone, M., and Penza, V. (2003{\natexlab{b}}).
\newblock Modeling solar irradiance variations through full-disk images and
  semi-empirical atmospheric models.
\newblock \emph{Memorie della Societa Astronomica Italiana} 74, 603
\bibAnnoteFile{ermolli_modeling_2003}

\bibitem[{Ermolli et~al.(2018)Ermolli, Chatzistergos, Krivova, and
  Solanki}]{ermolli_potential_2018}
Ermolli, I., Chatzistergos, T., Krivova, N.~A., and Solanki, S.~K. (2018).
\newblock The potential of {Ca} {II} {K} observations for solar activity and
  variability studies.
\newblock In \emph{Long-term {Datasets} for the {Understanding} of {Solar} and
  {Stellar} {Magnetic} {Cycles}}, eds. D.~Banerjee, J.~Jiang, K.~Kusano, and
  S.~Solanki (Cambridge, UK: Cambridge University Press), vol. 340 of
  \emph{{IAU} {Symposium}}, 115--120.
\newblock \doi{10.1017/S1743921318001849}
\bibAnnoteFile{ermolli_potential_2018}

\bibitem[{Ermolli et~al.(2007)Ermolli, Criscuoli, Centrone, Giorgi, and
  Penza}]{ermolli_photometric_2007}
Ermolli, I., Criscuoli, S., Centrone, M., Giorgi, F., and Penza, V. (2007).
\newblock Photometric properties of facular features over the activity cycle.
\newblock \emph{Astronomy and Astrophysics} 465, 305--314.
\newblock \doi{10.1051/0004-6361:20065995}
\bibAnnoteFile{ermolli_photometric_2007}

\bibitem[{Ermolli et~al.(2011)Ermolli, Criscuoli, and
  Giorgi}]{ermolli_recent_2011}
Ermolli, I., Criscuoli, S., and Giorgi, F. (2011).
\newblock Recent results from optical synoptic observations of the solar
  atmosphere with ground-based instruments.
\newblock \emph{Contributions of the Astronomical Observatory Skalnate Pleso}
  41, 73--84
\bibAnnoteFile{ermolli_recent_2011}

\bibitem[{Ermolli et~al.(2010)Ermolli, Criscuoli, Uitenbroek, Giorgi, Rast, and
  Solanki}]{ermolli_radiative_2010}
Ermolli, I., Criscuoli, S., Uitenbroek, H., Giorgi, F., Rast, M.~P., and
  Solanki, S.~K. (2010).
\newblock Radiative emission of solar features in the {Ca} {II} {K} line:
  comparison of measurements and models.
\newblock \emph{Astronomy and Astrophysics} 523, 55.
\newblock \doi{10.1051/0004-6361/201014762;}
\bibAnnoteFile{ermolli_radiative_2010}

\bibitem[{Ermolli and Ferrucci(2021)}]{ermolli_legacy_2021}
Ermolli, I. and Ferrucci, M. (2021).
\newblock The {Legacy} of {Angelo} {Secchi} at the {Forefront} of {Solar}
  {Physics} {Research}.
\newblock In \emph{Angelo {Secchi} and {Nineteenth} {Century} {Science}: {The}
  {Multidisciplinary} {Contributions} of a {Pioneer} and {Innovator}}, eds.
  I.~Chinnici and G.~Consolmagno (Cham: Springer International Publishing),
  Historical \& {Cultural} {Astronomy}. 123--136
\bibAnnoteFile{ermolli_legacy_2021}

\bibitem[{Ermolli et~al.(2022)Ermolli, Giorgi, and
  Chatzistergos}]{ermolli_romepspt_2022}
Ermolli, I., Giorgi, F., and Chatzistergos, T. (2022).
\newblock Rome/{PSPT}: precision solar full-disk photometry during solar cycles
  23-25.
\newblock \emph{Frontiers in Astronomy and Space Sciences}
\bibAnnoteFile{ermolli_romepspt_2022}

\bibitem[{Ermolli et~al.(2009{\natexlab{a}})Ermolli, Marchei, Centrone,
  Criscuoli, Giorgi, and Perna}]{ermolli_digitized_2009}
Ermolli, I., Marchei, E., Centrone, M., Criscuoli, S., Giorgi, F., and Perna,
  C. (2009{\natexlab{a}}).
\newblock The digitized archive of the {Arcetri} spectroheliograms.
  {Preliminary} results from the analysis of {Ca} {II} {K} images.
\newblock \emph{Astronomy and Astrophysics} 499, 627--632.
\newblock \doi{10.1051/0004-6361/200811406}
\bibAnnoteFile{ermolli_digitized_2009}

\bibitem[{Ermolli et~al.(1998{\natexlab{b}})Ermolli, Pietropaolo, Florio, and
  Berrilli}]{ermolli_chromospheric_1998-1}
Ermolli, I., Pietropaolo, E., Florio, A., and Berrilli, F.
  (1998{\natexlab{b}}).
\newblock Chromospheric {Network} {Properties} {On} {Short} {Time} {Scales}
  {From} {PSPT} {Images}.
\newblock In \emph{Synoptic {Solar} {Physics}} (Astronomical Society of the
  Pacific), vol. 140 of \emph{Astronomical {Society} of the {Pacific}
  {Conference} {Series}}, 231
\bibAnnoteFile{ermolli_chromospheric_1998-1}

\bibitem[{Ermolli et~al.(2015)Ermolli, Shibasaki, Tlatov, and van
  Driel-Gesztelyi}]{ermolli_solar_2015}
Ermolli, I., Shibasaki, K., Tlatov, A., and van Driel-Gesztelyi, L. (2015).
\newblock Solar {Cycle} {Indices} from the {Photosphere} to the {Corona}:
  {Measurements} and {Underlying} {Physics}.
\newblock In \emph{The {Solar} {Activity} {Cycle}: {Physical} {Causes} and
  {Consequences}}, eds. A.~Balogh, H.~Hudson, K.~Petrovay, and R.~von Steiger
  (New York, NY: Springer), Space {Sciences} {Series} of {ISSI}. 105--135.
\newblock \doi{10.1007/978-1-4939-2584-1_4}
\bibAnnoteFile{ermolli_solar_2015}

\bibitem[{Ermolli et~al.(2009{\natexlab{b}})Ermolli, Solanki, Tlatov, Krivova,
  Ulrich, and Singh}]{ermolli_comparison_2009}
Ermolli, I., Solanki, S.~K., Tlatov, A.~G., Krivova, N.~A., Ulrich, R.~K., and
  Singh, J. (2009{\natexlab{b}}).
\newblock Comparison {Among} {Ca} {II} {K} {Spectroheliogram} {Time} {Series}
  with an {Application} to {Solar} {Activity} {Studies}.
\newblock \emph{The Astrophysical Journal} 698, 1000--1009.
\newblock \doi{10.1088/0004-637X/698/2/1000}
\bibAnnoteFile{ermolli_comparison_2009}

\bibitem[{Feller et~al.(2020)Feller, Gandorfer, Iglesias, Lagg, Riethmüller,
  Solanki et~al.}]{feller_sunrise_2020}
Feller, A., Gandorfer, A., Iglesias, F.~A., Lagg, A., Riethmüller, T.~L.,
  Solanki, S.~K., et~al. (2020).
\newblock The {SUNRISE} {UV} {Spectropolarimeter} and imager for {SUNRISE}
  {III}.
\newblock In \emph{Proc. {SPIE}}. vol. 1447, 11447AK.
\newblock \doi{10.1117/12.2562666}.
\newblock Conference Name: Society of Photo-Optical Instrumentation Engineers
  (SPIE) Conference Series ISBN: 9781510636811
\bibAnnoteFile{feller_sunrise_2020}

\bibitem[{Fligge and Solanki(1998)}]{fligge_long-term_1998}
Fligge, M. and Solanki, S.~K. (1998).
\newblock Long-term behavior of emission from solar faculae: steps towards a
  robust index.
\newblock \emph{Astronomy and Astrophysics} 332, 1082--1086
\bibAnnoteFile{fligge_long-term_1998}

\bibitem[{Fontenla and Landi(2018)}]{fontenla_bright_2018}
Fontenla, J.~M. and Landi, E. (2018).
\newblock Bright {Network}, {UVA}, and the {Physical} {Modeling} of {Solar}
  {Spectral} and {Total} {Irradiance} in {Recent} {Solar} {Cycles}.
\newblock \emph{The Astrophysical Journal} 861, 120.
\newblock \doi{10.3847/1538-4357/aac388}
\bibAnnoteFile{fontenla_bright_2018}

\bibitem[{Fossum(1993)}]{fossum_active_1993}
Fossum, E.~R. (1993).
\newblock Active pixel sensors: are {CCDs} dinosaurs?
\newblock In \emph{Proceedings of the {SPIE}}. vol. 1900, 2--14.
\newblock \doi{10.1117/12.148585}
\bibAnnoteFile{fossum_active_1993}

\bibitem[{Foukal(1993)}]{foukal_curious_1993}
Foukal, P. (1993).
\newblock The {Curious} {Case} of the {Greenwich} {Faculae}.
\newblock \emph{Solar Physics} 148, 219--232.
\newblock \doi{10.1007/BF00645087}
\bibAnnoteFile{foukal_curious_1993}

\bibitem[{Foukal(1996)}]{foukal_behavior_1996}
Foukal, P. (1996).
\newblock The {Behavior} of solar magnetic plages measured from {Mt}. {Wilson}
  observations between 1915-1984.
\newblock \emph{Geophysical Research Letters} 23, 2169--2172.
\newblock \doi{10.1029/96GL01356}
\bibAnnoteFile{foukal_behavior_1996}

\bibitem[{Foukal(1998)}]{foukal_extension_1998}
Foukal, P. (1998).
\newblock Extension of the {F10}.7 {Index} to 1905 using {Mt}. {Wilson} {Ca}
  {K} {Spectroheliograms}.
\newblock \emph{Geophysical Research Letters} 25, 2909--2912.
\newblock \doi{10.1029/98GL02057}
\bibAnnoteFile{foukal_extension_1998}

\bibitem[{Foukal(2012)}]{foukal_new_2012}
Foukal, P. (2012).
\newblock A {New} {Look} at {Solar} {Irradiance} {Variation}.
\newblock \emph{Solar Physics} 279, 365--381.
\newblock \doi{10.1007/s11207-012-0017-6}
\bibAnnoteFile{foukal_new_2012}

\bibitem[{Foukal et~al.(2009)Foukal, Bertello, Livingston, Pevtsov, Singh,
  Tlatov et~al.}]{foukal_century_2009}
Foukal, P., Bertello, L., Livingston, W.~C., Pevtsov, A.~A., Singh, J., Tlatov,
  A.~G., et~al. (2009).
\newblock A {Century} of {Solar} {Ca} ii {Measurements} and {Their}
  {Implication} for {Solar} {UV} {Driving} of {Climate}.
\newblock \emph{Solar Physics} 255, 229--238.
\newblock \doi{10.1007/s11207-009-9330-0}
\bibAnnoteFile{foukal_century_2009}

\bibitem[{Foukal and Milano(2001)}]{foukal_measurement_2001}
Foukal, P. and Milano, L. (2001).
\newblock A measurement of the quiet network contribution to solar irradiance
  variation.
\newblock \emph{Geophysical Research Letters} 28, 883--886.
\newblock \doi{10.1029/2000GL012072}
\bibAnnoteFile{foukal_measurement_2001}

\bibitem[{Foukal and Vernazza(1979)}]{foukal_effect_1979}
Foukal, P. and Vernazza, J. (1979).
\newblock The effect of magnetic fields on solar luminosity.
\newblock \emph{The Astrophysical Journal} 234, 707--715.
\newblock \doi{10.1086/157547}
\bibAnnoteFile{foukal_effect_1979}

\bibitem[{Fox et~al.(2016)Fox, Velli, Bale, Decker, Driesman, Howard
  et~al.}]{fox_solar_2016}
Fox, N.~J., Velli, M.~C., Bale, S.~D., Decker, R., Driesman, A., Howard, R.~A.,
  et~al. (2016).
\newblock The {Solar} {Probe} {Plus} {Mission}: {Humanity}’s {First} {Visit}
  to {Our} {Star}.
\newblock \emph{Space Science Reviews} 204, 7--48.
\newblock \doi{10.1007/s11214-015-0211-6}
\bibAnnoteFile{fox_solar_2016}

\bibitem[{Frazier(1971)}]{frazier_multi-channel_1971}
Frazier, E.~N. (1971).
\newblock Multi-{Channel} {Magnetograph} {Observations}. {III}: {Faculae}.
\newblock \emph{Solar Physics} 21, 42--53.
\newblock \doi{10.1007/BF00155772}
\bibAnnoteFile{frazier_multi-channel_1971}

\bibitem[{Fredga(1971)}]{fredga_comparison_1971}
Fredga, K. (1971).
\newblock A {Comparison} between {Mg} {II} and {Ca} {II} {Spectroheliograms}.
\newblock \emph{Solar Physics} 21, 60--81.
\newblock \doi{10.1007/BF00155775}
\bibAnnoteFile{fredga_comparison_1971}

\bibitem[{Fröhlich et~al.(1995)Fröhlich, Romero, Roth, Wehrli, Andersen,
  Appourchaux et~al.}]{frohlich_virgo_1995}
Fröhlich, C., Romero, J., Roth, H., Wehrli, C., Andersen, B.~N., Appourchaux,
  T., et~al. (1995).
\newblock {VIRGO}: {Experiment} for {Helioseismology} and {Solar} {Irradiance}
  {Monitoring}.
\newblock \emph{Solar Physics} 162, 101--128.
\newblock \doi{10.1007/BF00733428}
\bibAnnoteFile{frohlich_virgo_1995}

\bibitem[{Garcia et~al.(2011)Garcia, Sobotka, Klvana, and
  Bumba}]{garcia_synoptic_2011}
Garcia, A., Sobotka, M., Klvana, M., and Bumba, V. (2011).
\newblock Synoptic observations with the {Coimbra} spectroheliograph.
\newblock \emph{Contributions of the Astronomical Observatory Skalnate Pleso}
  41, 69--72
\bibAnnoteFile{garcia_synoptic_2011}

\bibitem[{Giorgi et~al.(2005)Giorgi, Ermolli, Centrone, and
  Marchei}]{giorgi_calibration_2005}
Giorgi, F., Ermolli, I., Centrone, M., and Marchei, E. (2005).
\newblock Calibration of the {Arcetri} {Solar} {Archive} {Images}.
\newblock \emph{Memorie della Societa Astronomica Italiana} 76, 977
\bibAnnoteFile{giorgi_calibration_2005}

\bibitem[{Goldbaum et~al.(2009)Goldbaum, Rast, Ermolli, Summer~Sands, and
  Berrilli}]{goldbaum_intensity_2009}
Goldbaum, N., Rast, M.~P., Ermolli, I., Summer~Sands, J., and Berrilli, F.
  (2009).
\newblock The intensity profile of the solar supergranulation.
\newblock \emph{The Astrophysical Journal} 707, 67--73.
\newblock \doi{10.1088/0004-637X/707/1/67}
\bibAnnoteFile{goldbaum_intensity_2009}

\bibitem[{Golovko et~al.(2002)Golovko, Golubeva, Grechnev, Myachin, Trifonov,
  and Khlystova}]{golovko_data_2002}
Golovko, A.~A., Golubeva, E.~M., Grechnev, V.~V., Myachin, D.~Y., Trifonov,
  V.~D., and Khlystova, A.~I. (2002).
\newblock Data base of full solar dilk {H}-alpha images from the {Baikal}
  {Observatory}.
\newblock In \emph{Solar {Variability}: {From} {Core} to {Outer} {Frontiers}}
  (ESA Publications Division), vol. 506, 929--932
\bibAnnoteFile{golovko_data_2002}

\bibitem[{Gray et~al.(2010)Gray, Beer, Geller, Haigh, Lockwood, Matthes
  et~al.}]{gray_solar_2010}
Gray, L.~J., Beer, J., Geller, M., Haigh, J.~D., Lockwood, M., Matthes, K.,
  et~al. (2010).
\newblock Solar {Influences} on {Climate}.
\newblock \emph{Reviews of Geophysics} 48, 4001.
\newblock \doi{10.1029/2009RG000282}
\bibAnnoteFile{gray_solar_2010}

\bibitem[{Hagenaar et~al.(1997)Hagenaar, Schrijver, and
  Title}]{hagenaar_distribution_1997}
Hagenaar, H.~J., Schrijver, C.~J., and Title, A.~M. (1997).
\newblock The {Distribution} of {Cell} {Sizes} of the {Solar} {Chromospheric}
  {Network}.
\newblock \emph{The Astrophysical Journal} 481, 988.
\newblock \doi{10.1086/304066}
\bibAnnoteFile{hagenaar_distribution_1997}

\bibitem[{Hale(1890)}]{hale_note_1890}
Hale, G.~E. (1890).
\newblock Note on {Solar} {Prominence} {Photography}.
\newblock \emph{Astronomische Nachrichten} 126, 81.
\newblock \doi{10.1002/asna.18911260602}
\bibAnnoteFile{hale_note_1890}

\bibitem[{Hale(1891)}]{hale_kenwood_1891}
Hale, G.~E. (1891).
\newblock The {Kenwood} {Physical} {Observatory}.
\newblock \emph{Publications of the Astronomical Society of the Pacific} 3,
  30--34.
\newblock \doi{10.1086/120231}
\bibAnnoteFile{hale_kenwood_1891}

\bibitem[{Hale(1893)}]{hale_solar_1893}
Hale, G.~E. (1893).
\newblock Solar photography at the {Kenwood} {Astro}-physical {Observatory}.
\newblock \emph{Memorie della Societa Degli Spettroscopisti Italiani} 21,
  68--74
\bibAnnoteFile{hale_solar_1893}

\bibitem[{Hale and Ellerman(1903)}]{hale_rumford_1903}
Hale, G.~E. and Ellerman, F. (1903).
\newblock The {Rumford} spectroheliograph of the {Yerkes} {Observatory}.
\newblock \emph{Publications of the Yerkes Observatory} 3, I.1--XV.2
\bibAnnoteFile{hale_rumford_1903}

\bibitem[{Hanaoka(2013)}]{hanaoka_long-term_2013}
Hanaoka, Y. (2013).
\newblock Long-term synoptic observations of the {Sun} at the {National}
  {Astronomical} {Observatory} of {Japan}.
\newblock \emph{Journal of Physics Conference Series} 440, 2041.
\newblock \doi{10.1088/1742-6596/440/1/012041}
\bibAnnoteFile{hanaoka_long-term_2013}

\bibitem[{Hanaoka and {Solar Observatory of NAOJ}(2016)}]{hanaoka_past_2016}
Hanaoka, Y. and {Solar Observatory of NAOJ} (2016).
\newblock Past and {Present} of the {Synoptic} {Observations} of the {Sun} at
  the {National} {Astronomical} {Observatory} of {Japan}.
\newblock In \emph{Coimbra {Solar} {Physics} {Meeting}: {Ground}-based {Solar}
  {Observations} in the {Space} {Instrumentation} {Era}}, eds. I.~Dorotovic,
  C.~E. Fischer, and M.~Temmer (San Francisco), vol. 504 of \emph{Astronomical
  {Society} of the {Pacific} {Conference} {Series}}, 313
\bibAnnoteFile{hanaoka_past_2016}

\bibitem[{Harvey(1993)}]{harvey_magnetic_1993}
Harvey, K.~L. (1993).
\newblock \emph{Magnetic {Bipoles} on the {Sun}}.
\newblock Ph.D. thesis, Utrecht University
\bibAnnoteFile{harvey_magnetic_1993}

\bibitem[{Harvey(1994)}]{harvey_solar_1994}
Harvey, K.~L. (1994).
\newblock The solar magnetic cycle.
\newblock In \emph{Solar surface magnetism}, eds. R.~J. Rutten and C.~J.
  Schrijver (Kluwer academic publishers), vol. 433, 347
\bibAnnoteFile{harvey_solar_1994}

\bibitem[{Harvey and White(1999)}]{harvey_magnetic_1999}
Harvey, K.~L. and White, O.~R. (1999).
\newblock Magnetic and {Radiative} {Variability} of {Solar} {Surface}
  {Structures}. {I}. {Image} {Decomposition} and {Magnetic}-{Intensity}
  {Mapping}.
\newblock \emph{The Astrophysical Journal} 515, 812--831.
\newblock \doi{10.1086/307035}
\bibAnnoteFile{harvey_magnetic_1999}

\bibitem[{Heath and Schlesinger(1986)}]{heath_mg_1986}
Heath, D.~F. and Schlesinger, B.~M. (1986).
\newblock The {Mg} 280-nm doublet as a monitor of changes in solar ultraviolet
  irradiance.
\newblock \emph{Journal of Geophysical Research} 91, 8672--8682.
\newblock \doi{10.1029/JD091iD08p08672}
\bibAnnoteFile{heath_mg_1986}

\bibitem[{Herschel(1800)}]{herschel_investigation_1800}
Herschel, W. (1800).
\newblock Investigation of the {Powers} of the {Prismatic} {Colours} to {Heat}
  and {Illuminate} {Objects}; {With} {Remarks}, {That} {Prove} the {Different}
  {Refrangibility} of {Radiant} {Heat}. {To} {Which} is {Added}, an {Inquiry}
  into the {Method} of {Viewing} the {Sun} {Advantageously}, with {Telescopes}
  of {Large} {Apertures} and {High} {Magnifying} {Powers}. {By} {William}
  {Herschel}, {LL}. {D}. {F}. {R}. {S}.
\newblock \emph{Philosophical Transactions of the Royal Society of London
  Series I} 90, 255--283.
\newblock ADS Bibcode: 1800RSPT...90..255H
\bibAnnoteFile{herschel_investigation_1800}

\bibitem[{Hochedez et~al.(2006)Hochedez, Schmutz, Stockman, Schühle,
  BenMoussa, Koller et~al.}]{hochedez_lyra_2006}
Hochedez, J.~F., Schmutz, W., Stockman, Y., Schühle, U., BenMoussa, A.,
  Koller, S., et~al. (2006).
\newblock {LYRA}, a solar {UV} radiometer on {Proba2}.
\newblock \emph{Advances in Space Research} 37, 303--312.
\newblock \doi{10.1016/j.asr.2005.10.041}
\bibAnnoteFile{hochedez_lyra_2006}

\bibitem[{Howard(1959)}]{howard_observations_1959}
Howard, R. (1959).
\newblock Observations of {Solar} {Magnitic} {Fields}.
\newblock \emph{The Astrophysical Journal} 130, 193.
\newblock \doi{10.1086/146708}
\bibAnnoteFile{howard_observations_1959}

\bibitem[{Hubrecht(1912)}]{hubrecht_sun_1912}
Hubrecht, J.~B. (1912).
\newblock Sun, {Rotation} of, {Spectrographic} observations at {Cambridge}
  {Observatory}.
\newblock \emph{Monthly Notices of the Royal Astronomical Society} 73, 5.
\newblock \doi{10.1093/mnras/73.1.5}
\bibAnnoteFile{hubrecht_sun_1912}

\bibitem[{Hurter and Driffield(1890)}]{hurter_photochemical_1890}
Hurter, F. and Driffield, V.~C. (1890).
\newblock Photochemical investigations and a new method of determination of the
  sensitiveness of photographic plates.
\newblock \emph{Journal of the Society of Chemical Industry} 9, 455--469.
\newblock Number: 5
\bibAnnoteFile{hurter_photochemical_1890}

\bibitem[{{Intergovernmental Panel on Climate
  Change}(2021)}]{intergovernmental_panel_on_climate_change_climate_2021}
{Intergovernmental Panel on Climate Change} (ed.) (2021).
\newblock \emph{Climate {Change} 2021: {The} {Physical} {Science} {Basis}.
  {Contribution} of {Working} {Group} {I} to the {Sixth} {Assessment} {Report}
  of the {Intergovernmental} {Panel} on {Climate} {Change}} (in press)
\bibAnnoteFile{intergovernmental_panel_on_climate_change_climate_2021}

\bibitem[{James and Higgins(1968)}]{james_fundamentals_1968}
James, T.~H. and Higgins, G.~C. (1968).
\newblock \emph{Fundamentals of {Photographic} {Theory}} (Morgan and Morgan),
  2nd edition edn.
\bibAnnoteFile{james_fundamentals_1968}

\bibitem[{Janesick(2001)}]{janesick_scientific_2001}
Janesick, J.~R. (2001).
\newblock \emph{Scientific {Charge}-coupled {Devices}} (SPIE Press).
\newblock Google-Books-ID: 3GyE4SWytn4C
\bibAnnoteFile{janesick_scientific_2001}

\bibitem[{Jarolim et~al.(2020)Jarolim, Veronig, Pötzi, and
  Podladchikova}]{jarolim_image-quality_2020}
Jarolim, R., Veronig, A.~M., Pötzi, W., and Podladchikova, T. (2020).
\newblock Image-quality assessment for full-disk solar observations with
  generative adversarial networks.
\newblock \emph{Astronomy \& Astrophysics} 643, A72.
\newblock \doi{10.1051/0004-6361/202038691}
\bibAnnoteFile{jarolim_image-quality_2020}

\bibitem[{Jefferies et~al.(1988)Jefferies, Pomerantz, Duvall, Harvey, and
  Jaksha}]{jefferies_helioseismology_1988}
Jefferies, S.~M., Pomerantz, M.~A., Duvall, T.~L., Jr., Harvey, J.~W., and
  Jaksha, D.~B. (1988).
\newblock Helioseismology from the {South} {Pole}: comparison of 1987 and 1981
  results.
\newblock In \emph{Seismology of the {Sun} and {Sun}-{Like} {Stars}}. vol. 286,
  279--284
\bibAnnoteFile{jefferies_helioseismology_1988}

\bibitem[{Johannesson et~al.(1998)Johannesson, Marquette, and
  Zirin}]{johannesson_10-year_1998}
Johannesson, A., Marquette, W.~H., and Zirin, H. (1998).
\newblock A 10-{Year} {Set} of {CA} {II} {K}-{Line} {Filtergrams}.
\newblock \emph{Solar Physics} 177, 265--278.
\newblock \doi{10.1023/A:1004940227692}
\bibAnnoteFile{johannesson_10-year_1998}

\bibitem[{Kahil et~al.(2017)Kahil, Riethmüller, and
  Solanki}]{kahil_brightness_2017}
Kahil, F., Riethmüller, T.~L., and Solanki, S.~K. (2017).
\newblock Brightness of {Solar} {Magnetic} {Elements} {As} a {Function} of
  {Magnetic} {Flux} at {High} {Spatial} {Resolution}.
\newblock \emph{The Astrophysical Journal Supplement Series} 229, 12.
\newblock \doi{10.3847/1538-4365/229/1/12}.
\newblock Number: 1
\bibAnnoteFile{kahil_brightness_2017}

\bibitem[{Kahil et~al.(2019)Kahil, Riethmüller, and
  Solanki}]{kahil_intensity_2019}
Kahil, F., Riethmüller, T.~L., and Solanki, S.~K. (2019).
\newblock Intensity contrast of solar plage as a function of magnetic flux at
  high spatial resolution.
\newblock \emph{Astronomy \& Astrophysics} 621, A78.
\newblock \doi{10.1051/0004-6361/201833722}
\bibAnnoteFile{kahil_intensity_2019}

\bibitem[{Kaiser et~al.(2005)Kaiser, Hempel, Schmitt, and
  Reiners}]{kaiser_analysis_2005}
Kaiser, C., Hempel, M., Schmitt, J. H. M.~M., and Reiners, A. (2005).
\newblock Analysis of {Ca} {II} emission lines in active stars.
\newblock In \emph{13th {Cambridge} {Workshop} on {Cool} {Stars}, {Stellar}
  {Systems} and the {Sun}}. vol. 560, 693
\bibAnnoteFile{kaiser_analysis_2005}

\bibitem[{Kakuwa and Ueno(2021)}]{kakuwa_investigation_2021}
Kakuwa, J. and Ueno, S. (2021).
\newblock Investigation of the {Long}-term {Variation} of {Solar} {Ca} ii {K}
  {Intensity}. {I}. {Density}-to-intensity {Calibration} {Formula} for
  {Historical} {Photographic} {Plates}.
\newblock \emph{The Astrophysical Journal Supplement Series} 254, 44.
\newblock \doi{10.3847/1538-4365/abfbe3}
\bibAnnoteFile{kakuwa_investigation_2021}

\bibitem[{Kariyappa and Sivaraman(1994)}]{kariyappa_variability_1994}
Kariyappa, R. and Sivaraman, K.~R. (1994).
\newblock Variability of the solar chromospheric network over the solar cycle.
\newblock \emph{Solar Physics} 152, 139--144.
\newblock \doi{10.1007/BF01473196}
\bibAnnoteFile{kariyappa_variability_1994}

\bibitem[{Khetsuriani(1967)}]{khetsuriani_abastumani_1967}
Khetsuriani, T.~S. (1967).
\newblock Abastumani {Astrophysical} {Observatory} of the {Academy} of
  {Sciences} of the {Georgian} {SSR}.
\newblock \emph{Solar Physics} 2, 237--239.
\newblock \doi{10.1007/BF00155927}
\bibAnnoteFile{khetsuriani_abastumani_1967}

\bibitem[{Kiepenheuer(1969)}]{kiepenheuer_fraunhofer_1969}
Kiepenheuer, K.~O. (1969).
\newblock Fraunhofer {Institut} mit den {Observatorien} {Schauinsland} und
  {Anacapri}. {Report} 1968.
\newblock \emph{Mitteilungen der Astronomischen Gesellschaft Hamburg} 26,
  42--47
\bibAnnoteFile{kiepenheuer_fraunhofer_1969}

\bibitem[{Kiepenheuer(1974)}]{kiepenheuer_fraunhofer-institut_1974}
Kiepenheuer, K.~O. (1974).
\newblock Fraunhofer-{Institut} mit den {Observatorien} {Schauinsland} und
  {Anacapri}. {Report} 1973.
\newblock \emph{Mitteilungen der Astronomischen Gesellschaft Hamburg} 35,
  60--66
\bibAnnoteFile{kiepenheuer_fraunhofer-institut_1974}

\bibitem[{Kitai et~al.(2013)Kitai, Ueno, Maehara, Shirakawa, Katoda, Hada
  et~al.}]{kitai_digital_2013}
Kitai, R., Ueno, S., Maehara, H., Shirakawa, S., Katoda, M., Hada, Y., et~al.
  (2013).
\newblock The {Digital} {Database} of {Long}-{Term} {Solar} {Chromospheric}
  {Variation}.
\newblock \emph{Data Science Journal} 12, WDS213--WDS215.
\newblock \doi{10.2481/dsj.WDS-037}.
\newblock Number: 0
\bibAnnoteFile{kitai_digital_2013}

\bibitem[{Klimeš et~al.(1999)Klimeš, Bělik, Klimeš, and
  Marková}]{klimes_simultaneous_1999}
Klimeš, J., J., Bělik, M., Klimeš, S., J., and Marková, E. (1999).
\newblock Simultaneous {Observation} of the {Sun} in white-light, {H}-alpha,
  {Ca} and radio waves on {Observatory} {Upice}.
\newblock In \emph{8th {SOHO} {Workshop}: {Plasma} {Dynamics} and {Diagnostics}
  in the {Solar} {Transition} {Region} and {Corona}}, eds. J.~C. Vial and
  B.~Kaldeich-Schü. vol. 446 of \emph{{ESA} {Special} {Publication}}, 375
\bibAnnoteFile{klimes_simultaneous_1999}

\bibitem[{Koechlin et~al.(2019)Koechlin, Dettwiller, Audejean, Valais, and
  Ariste}]{koechlin_solar_2019}
Koechlin, L., Dettwiller, L., Audejean, M., Valais, M., and Ariste, A.~L.
  (2019).
\newblock Solar survey at {Pic} du {Midi}: {Calibrated} data and improved
  images.
\newblock \emph{Astronomy \& Astrophysics} 631, A55.
\newblock \doi{10.1051/0004-6361/201732504}
\bibAnnoteFile{koechlin_solar_2019}

\bibitem[{Kosugi et~al.(2007)Kosugi, Matsuzaki, Sakao, Shimizu, Sone, Tachikawa
  et~al.}]{kosugi_hinode_2007}
Kosugi, T., Matsuzaki, K., Sakao, T., Shimizu, T., Sone, Y., Tachikawa, S.,
  et~al. (2007).
\newblock The {Hinode} ({Solar}-{B}) {Mission}: {An} {Overview}.
\newblock \emph{Solar Physics} 243, 3--17.
\newblock \doi{10.1007/s11207-007-9014-6}
\bibAnnoteFile{kosugi_hinode_2007}

\bibitem[{Kren et~al.(2017)Kren, Pilewskie, and Coddington}]{kren_where_2017}
Kren, A.~C., Pilewskie, P., and Coddington, O. (2017).
\newblock Where does {Earth}'s atmosphere get its energy?
\newblock \emph{Journal of Space Weather and Space Climate} 7, A10.
\newblock \doi{10.1051/swsc/2017007}
\bibAnnoteFile{kren_where_2017}

\bibitem[{Krivova(2018)}]{krivova_solar_2018}
Krivova, N.~A. (2018).
\newblock Solar {Irradiance} {Variability} and {Earth}’s {Climate}.
\newblock In \emph{Climate {Changes} in the {Holocene}} (CRC Press). 107--120.
\newblock \doi{10.1201/9781351260244-4}
\bibAnnoteFile{krivova_solar_2018}

\bibitem[{Krivova and Solanki(2004)}]{krivova_effect_2004}
Krivova, N.~A. and Solanki, S.~K. (2004).
\newblock Effect of spatial resolution on estimating the {Sun}'s magnetic flux.
\newblock \emph{Astronomy and Astrophysics} 417, 1125--1132.
\newblock \doi{10.1051/0004-6361:20040022}
\bibAnnoteFile{krivova_effect_2004}

\bibitem[{Krivova et~al.(2003)Krivova, Solanki, Fligge, and
  Unruh}]{krivova_reconstruction_2003}
Krivova, N.~A., Solanki, S.~K., Fligge, M., and Unruh, Y.~C. (2003).
\newblock Reconstruction of solar irradiance variations in cycle 23: {Is} solar
  surface magnetism the cause?
\newblock \emph{Astronomy and Astrophysics} 399, L1--L4.
\newblock \doi{10.1051/0004-6361:20030029}
\bibAnnoteFile{krivova_reconstruction_2003}

\bibitem[{Kuriyan et~al.(1983)Kuriyan, Muralidharan, and
  Sampath}]{kuriyan_long-term_1983}
Kuriyan, P.~P., Muralidharan, V., and Sampath, S. (1983).
\newblock Long-term relationships between sunspots, {Ca}-plages and the
  ionosphere.
\newblock \emph{Journal of Atmospheric and Terrestrial Physics} 45, 285.
\newblock \doi{10.1016/S0021-9169(83)80034-8}
\bibAnnoteFile{kuriyan_long-term_1983}

\bibitem[{Kusano et~al.(2021)Kusano, Ichimoto, Ishii, Miyoshi, Yoden, Akiyoshi
  et~al.}]{kusano_pstep_2021}
Kusano, K., Ichimoto, K., Ishii, M., Miyoshi, Y., Yoden, S., Akiyoshi, H.,
  et~al. (2021).
\newblock {PSTEP}: project for solar–terrestrial environment prediction.
\newblock \emph{Earth, Planets and Space} 73, 159.
\newblock \doi{10.1186/s40623-021-01486-1}
\bibAnnoteFile{kusano_pstep_2021}

\bibitem[{Lawrence(1987)}]{lawrence_ratio_1987}
Lawrence, J.~K. (1987).
\newblock Ratio of calcium plage to sunspot areas of solar active regions.
\newblock \emph{Journal of Geophysical Research: Atmospheres} 92, 813--817.
\newblock \doi{10.1029/JD092iD01p00813}
\bibAnnoteFile{lawrence_ratio_1987}

\bibitem[{Lefebvre et~al.(2005)Lefebvre, Ulrich, Webster, Varadi, Javaraiah,
  Bertello et~al.}]{lefebvre_solar_2005}
Lefebvre, S., Ulrich, R.~K., Webster, L.~S., Varadi, F., Javaraiah, J.,
  Bertello, L., et~al. (2005).
\newblock The solar photograph archive of the {Mount} {Wilson} {Observatory}.
  {A} resource for a century of digital data.
\newblock \emph{Memorie della Societa Astronomica Italiana} 76, 862
\bibAnnoteFile{lefebvre_solar_2005}

\bibitem[{Leighton(1959)}]{leighton_observations_1959}
Leighton, R.~B. (1959).
\newblock Observations of {Solar} {Magnetic} {Fields} in {Plage} {Regions}.
\newblock \emph{The Astrophysical Journal} 130, 366.
\newblock \doi{10.1086/146727}
\bibAnnoteFile{leighton_observations_1959}

\bibitem[{Lenza et~al.(2014)Lenza, Srba, Gregorova, Exnerova, and
  Lenzova}]{lenza_system_2014}
[Dataset] Lenza, L., Srba, J., Gregorova, B., Exnerova, M., and Lenzova, N.
  (2014).
\newblock System for simultaneous observation of solar flares in spectral lines
  of {H}-alpha and {CaII} {K}.
\newblock Presenters: \_:n17573
\bibAnnoteFile{lenza_system_2014}

\bibitem[{Löfdahl et~al.(2011)Löfdahl, Henriques, and
  Kiselman}]{lofdahl_tilted_2011}
Löfdahl, M.~G., Henriques, V. M.~J., and Kiselman, D. (2011).
\newblock A tilted interference filter in a converging beam.
\newblock \emph{Astronomy \& Astrophysics} 533, A82.
\newblock \doi{10.1051/0004-6361/201117305}
\bibAnnoteFile{lofdahl_tilted_2011}

\bibitem[{Lites et~al.(2014)Lites, Centeno, and McIntosh}]{lites_solar_2014}
Lites, B.~W., Centeno, R., and McIntosh, S.~W. (2014).
\newblock The solar cycle dependence of the weak internetwork flux.
\newblock \emph{Publications of the Astronomical Society of Japan} 66, S4.
\newblock \doi{10.1093/pasj/psu082}
\bibAnnoteFile{lites_solar_2014}

\bibitem[{Livingston and Wallace(2003)}]{livingston_suns_2003}
Livingston, W. and Wallace, L. (2003).
\newblock The {Sun}'s immutable basal quiet atmosphere.
\newblock \emph{Solar Physics} 212, 227--237.
\newblock \doi{10.1023/A:1022994002653}
\bibAnnoteFile{livingston_suns_2003}

\bibitem[{Livingston et~al.(2007)Livingston, Wallace, White, and
  Giampapa}]{livingston_sun-as--star_2007}
Livingston, W., Wallace, L., White, O.~R., and Giampapa, M.~S. (2007).
\newblock Sun-as-a-{Star} {Spectrum} {Variations} 1974-2006.
\newblock \emph{The Astrophysical Journal} 657, 1137--1149.
\newblock \doi{10.1086/511127}
\bibAnnoteFile{livingston_sun-as--star_2007}

\bibitem[{Livingston et~al.(1976)Livingston, Harvey, Pierce, Schrage,
  Gillespie, Simmons et~al.}]{livingston_kitt_1976}
Livingston, W.~C., Harvey, J., Pierce, A.~K., Schrage, D., Gillespie, B.,
  Simmons, J., et~al. (1976).
\newblock Kitt {Peak} 60-cm vacuum telescope.
\newblock \emph{Applied Optics} 15, 33--39.
\newblock \doi{10.1364/AO.15.000033}
\bibAnnoteFile{livingston_kitt_1976}

\bibitem[{Lockyer(1909)}]{lockyer_spectroheliograms_1909}
Lockyer, W. J.~S. (1909).
\newblock Spectroheliograms of the solar surface.
\newblock \emph{Monthly Notices of the Royal Astronomical Society} 70, 14.
\newblock \doi{10.1093/mnras/70.1.14}
\bibAnnoteFile{lockyer_spectroheliograms_1909}

\bibitem[{Loukitcheva et~al.(2009)Loukitcheva, Solanki, and
  White}]{loukitcheva_relationship_2009}
Loukitcheva, M., Solanki, S.~K., and White, S.~M. (2009).
\newblock The relationship between chromospheric emissions and magnetic field
  strength.
\newblock \emph{Astronomy and Astrophysics} 497, 273--285.
\newblock \doi{10.1051/0004-6361/200811133}
\bibAnnoteFile{loukitcheva_relationship_2009}

\bibitem[{Makarov et~al.(2004)Makarov, Tlatov, Singh, and
  Gupta}]{makarov_22-years_2004}
Makarov, V.~I., Tlatov, A.~G., Singh, J., and Gupta, S.~S. (2004).
\newblock 22-years magnetic cycle in polar activity of the {Sun}.
\newblock In \emph{Multi-{Wavelength} {Investigations} of {Solar} {Activity}},
  eds. A.~V. Stepanov, E.~Benevolenskaya, and A.~G. Kosovichev (Cambridge, UK:
  Cambridge University Press), vol. 223 of \emph{Proceedings of the
  {International} {Astronomical} {Union}}, 125--126.
\newblock \doi{10.1017/S1743921304005368}
\bibAnnoteFile{makarov_22-years_2004}

\bibitem[{Malherbe et~al.(2022)Malherbe, Corbard, Barbary, Morand, Collin,
  Crussaire et~al.}]{malherbe_monitoring_2022}
Malherbe, J.-M., Corbard, T., Barbary, G., Morand, F., Collin, C., Crussaire,
  D., et~al. (2022).
\newblock Monitoring fast solar chromospheric activity: the {MeteoSpace}
  project.
\newblock \emph{Experimental Astronomy} \doi{10.1007/s10686-022-09848-7}
\bibAnnoteFile{malherbe_monitoring_2022}

\bibitem[{Malherbe and Dalmasse(2019)}]{malherbe_new_2019}
Malherbe, J.-M. and Dalmasse, K. (2019).
\newblock The {New} 2018 {Version} of the {Meudon} {Spectroheliograph}.
\newblock \emph{Solar Physics} 294, 52.
\newblock \doi{10.1007/s11207-019-1441-7}
\bibAnnoteFile{malherbe_new_2019}

\bibitem[{Mandal et~al.(2017{\natexlab{a}})Mandal, Chatterjee, and
  Banerjee}]{mandal_association_2017}
Mandal, S., Chatterjee, S., and Banerjee, D. (2017{\natexlab{a}}).
\newblock Association of {Plages} with {Sunspots}: {A} {Multi}-{Wavelength}
  {Study} {Using} {Kodaikanal} {Ca} ii {K} and {Greenwich} {Sunspot} {Area}
  {Data}.
\newblock \emph{The Astrophysical Journal} 835, 158.
\newblock \doi{10.3847/1538-4357/835/2/158}
\bibAnnoteFile{mandal_association_2017}

\bibitem[{Mandal et~al.(2017{\natexlab{b}})Mandal, Chatterjee, and
  Banerjee}]{mandal_association_2017-1}
Mandal, S., Chatterjee, S., and Banerjee, D. (2017{\natexlab{b}}).
\newblock Association of {Supergranule} {Mean} {Scales} with {Solar} {Cycle}
  {Strengths} and {Total} {Solar} {Irradiance}.
\newblock \emph{The Astrophysical Journal} 844, 24.
\newblock \doi{10.3847/1538-4357/aa76e3}
\bibAnnoteFile{mandal_association_2017-1}

\bibitem[{Mandal et~al.(2020)Mandal, Krivova, Solanki, Sinha, and
  Banerjee}]{mandal_sunspot_2020}
Mandal, S., Krivova, N.~A., Solanki, S.~K., Sinha, N., and Banerjee, D. (2020).
\newblock Sunspot area catalog revisited: {Daily} cross-calibrated areas since
  1874.
\newblock \emph{Astronomy \& Astrophysics} 640, A78.
\newblock \doi{10.1051/0004-6361/202037547}
\bibAnnoteFile{mandal_sunspot_2020}

\bibitem[{Marchei et~al.(2006)Marchei, Ermolli, Centrone, Giorgi, and
  Perna}]{marchei_digitization_2006}
Marchei, E., Ermolli, I., Centrone, M., Giorgi, F., and Perna, C. (2006).
\newblock Digitization of the {Arcetri} {Solar} {Photographic} {Archive}.
\newblock \emph{Memorie della Societa Astronomica Italiana Supplementi} 9, 51
\bibAnnoteFile{marchei_digitization_2006}

\bibitem[{McIntosh et~al.(2011)McIntosh, Leamon, Hock, Rast, and
  Ulrich}]{mcintosh_observing_2011}
McIntosh, S.~W., Leamon, R.~J., Hock, R.~A., Rast, M.~P., and Ulrich, R.~K.
  (2011).
\newblock Observing {Evolution} in the {Supergranular} {Network} {Length}
  {Scale} {During} {Periods} of {Low} {Solar} {Activity}.
\newblock \emph{The Astrophysical Journal Letters} 730, L3.
\newblock \doi{10.1088/2041-8205/730/1/L3}
\bibAnnoteFile{mcintosh_observing_2011}

\bibitem[{Mees(1942)}]{mees_theory_1942}
Mees, C. E.~K. (1942).
\newblock \emph{The theory of the photographic process} (New York: The
  Macmillan Company)
\bibAnnoteFile{mees_theory_1942}

\bibitem[{Meftah et~al.(2018)Meftah, Corbard, Hauchecorne, Morand, Ikhlef,
  Chauvineau et~al.}]{meftah_solar_2018}
Meftah, M., Corbard, T., Hauchecorne, A., Morand, F., Ikhlef, R., Chauvineau,
  B., et~al. (2018).
\newblock Solar radius determined from {PICARD}/{SODISM} observations and
  extremely weak wavelength dependence in the visible and the near-infrared.
\newblock \emph{Astronomy \& Astrophysics} 616, A64.
\newblock \doi{10.1051/0004-6361/201732159}
\bibAnnoteFile{meftah_solar_2018}

\bibitem[{Meftah et~al.(2014)Meftah, Hochedez, Irbah, Hauchecorne, Boumier,
  Corbard et~al.}]{meftah_picard_2014}
Meftah, M., Hochedez, J.-F., Irbah, A., Hauchecorne, A., Boumier, P., Corbard,
  T., et~al. (2014).
\newblock Picard {SODISM}, a {Space} {Telescope} to {Study} the {Sun} from the
  {Middle} {Ultraviolet} to the {Near} {Infrared}.
\newblock \emph{Solar Physics} 289, 1043.
\newblock \doi{10.1007/s11207-013-0373-x}.
\newblock Number: 3
\bibAnnoteFile{meftah_picard_2014}

\bibitem[{Meftah et~al.(2012)Meftah, Irbah, Corbard, Morand, Thuillier,
  Hauchecorne et~al.}]{meftah_picard_2012}
Meftah, M., Irbah, A., Corbard, T., Morand, F., Thuillier, G., Hauchecorne, A.,
  et~al. (2012).
\newblock {PICARD} {SOL} mission, a ground-based facility for long-term solar
  radius measurement.
\newblock \emph{Ground-based and Airborne Instrumentation for Astronomy IV}
  8446, 844676.
\newblock \doi{10.1117/12.925712}
\bibAnnoteFile{meftah_picard_2012}

\bibitem[{Mein and Ribes(1990)}]{mein_spectroheliograms_1990}
Mein, P. and Ribes, E. (1990).
\newblock Spectroheliograms and motions of magnetic tracers.
\newblock \emph{Astronomy and Astrophysics} 227, 577--582
\bibAnnoteFile{mein_spectroheliograms_1990}

\bibitem[{Meurs(1987)}]{meurs_flattening_1987}
Meurs, E. J.~A. (1987).
\newblock Flattening the field - {A} user's view.
\newblock In \emph{European {Southern} {Observatory} {Conference} and
  {Workshop} {Proceedings}}. vol.~25, 105--110
\bibAnnoteFile{meurs_flattening_1987}

\bibitem[{Mickaelian et~al.(2007)Mickaelian, Nesci, Rossi, Weedman, Cirimele,
  Sargsyan et~al.}]{mickaelian_digitized_2007}
Mickaelian, A.~M., Nesci, R., Rossi, C., Weedman, D., Cirimele, G., Sargsyan,
  L.~A., et~al. (2007).
\newblock The digitized first {Byurakan} survey - {DFBS}.
\newblock \emph{Astronomy and Astrophysics} 464, 1177--1180.
\newblock \doi{10.1051/0004-6361:20066241}
\bibAnnoteFile{mickaelian_digitized_2007}

\bibitem[{Miller(1965)}]{miller_new_1965}
Miller, R.~A. (1965).
\newblock New {Spectroheliograph} at {Manila} {Observatory}.
\newblock \emph{Applied Optics} 4, 1085--1089.
\newblock \doi{10.1364/AO.4.001085}.
\newblock Number: 9
\bibAnnoteFile{miller_new_1965}

\bibitem[{Müller et~al.(2020)Müller, St.~Cyr, Zouganelis, Gilbert, Marsden,
  Nieves-Chinchilla et~al.}]{muller_solar_2020}
Müller, D., St.~Cyr, O.~C., Zouganelis, I., Gilbert, H.~R., Marsden, R.,
  Nieves-Chinchilla, T., et~al. (2020).
\newblock The {Solar} {Orbiter} mission: {Science} overview.
\newblock \emph{Astronomy \& Astrophysics} 642, A1.
\newblock \doi{10.1051/0004-6361/202038467}
\bibAnnoteFile{muller_solar_2020}

\bibitem[{Münzer et~al.(1989)Münzer, Hanslmeier, Schröter, and
  Wöhl}]{munzer_pole-equator-difference_1989}
Münzer, H., Hanslmeier, A., Schröter, E.~H., and Wöhl, H. (1989).
\newblock Pole-{Equator}-{Difference} of the {Size} of the {Chromospheric} {Ca}
  {II} ’ {K} ’ {Network} in {Quiet} and {Active} {Solar} {Regions}.
\newblock In \emph{Solar and {Stellar} {Granulation}}, eds. R.~J. Rutten and
  G.~Severino (Springer Netherlands), no. 263 in {NATO} {ASI} {Series}.
  217--218.
\newblock \doi{10.1007/978-94-009-0911-3_25}
\bibAnnoteFile{munzer_pole-equator-difference_1989}

\bibitem[{Mohler and Dodson(1968)}]{mohler_mcmath-hulbert_1968}
Mohler, O.~C. and Dodson, H.~W. (1968).
\newblock {McMath}-{Hulbert} {Observatory} of the {University} of {Michigan}.
\newblock \emph{Solar Physics} 5, 417--422.
\newblock \doi{10.1007/BF00147154}
\bibAnnoteFile{mohler_mcmath-hulbert_1968}

\bibitem[{Mordvinov et~al.(2020)Mordvinov, Karak, Banerjee, Chatterjee,
  Golubeva, and Khlystova}]{mordvinov_long-term_2020}
Mordvinov, A.~V., Karak, B.~B., Banerjee, D., Chatterjee, S., Golubeva, E.~M.,
  and Khlystova, A.~I. (2020).
\newblock Long-term {Evolution} of the {Sun}'s {Magnetic} {Field} during
  {Cycles} 15–19 {Based} on {Their} {Proxies} from {Kodaikanal} {Solar}
  {Observatory}.
\newblock \emph{The Astrophysical Journal} 902, L15.
\newblock \doi{10.3847/2041-8213/abba80}
\bibAnnoteFile{mordvinov_long-term_2020}

\bibitem[{Moss(1942)}]{moss_report_1942}
Moss, W. (1942).
\newblock Report of the proceedings of the (1940 and 1941), {Cambridge}
  {University}, {Solar} {Physics} {Observatory}.
\newblock \emph{Monthly Notices of the Royal Astronomical Society} 102, 86
\bibAnnoteFile{moss_report_1942}

\bibitem[{Naqvi et~al.(2010)Naqvi, Marquette, Tritschler, and
  Denker}]{naqvi_big_2010}
Naqvi, M.~F., Marquette, W.~H., Tritschler, A., and Denker, C. (2010).
\newblock The {Big} {Bear} {Solar} {Observatory} {Ca} {II} {K}-line index for
  solar cycle 23.
\newblock \emph{Astronomische Nachrichten} 331, 696--703.
\newblock \doi{10.1002/asna.201011399}
\bibAnnoteFile{naqvi_big_2010}

\bibitem[{Neckel(1999)}]{neckel_spectral_1999}
Neckel, H. (1999).
\newblock Spectral atlas of solar absolute disk-averaged and disk-center
  intensity from 3290 to 12510 Å ({Brault} and {Neckel}, 1987) now available
  from {Hamburg} observatory {FTP} site.
\newblock \emph{Solar Physics} 184, 421--422.
\newblock \doi{10.1023/A:1017165208013}
\bibAnnoteFile{neckel_spectral_1999}

\bibitem[{Nesme-Ribes et~al.(1996)Nesme-Ribes, Meunier, and
  Collin}]{nesme-ribes_fractal_1996}
Nesme-Ribes, E., Meunier, N., and Collin, B. (1996).
\newblock Fractal analysis of magnetic patterns from {Meudon}
  spectroheliograms.
\newblock \emph{Astronomy and Astrophysics} 308, 213--218
\bibAnnoteFile{nesme-ribes_fractal_1996}

\bibitem[{Nindos and Zirin(1998)}]{nindos_relation_1998}
Nindos, A. and Zirin, H. (1998).
\newblock The {Relation} of {CA} {II} {K} {Features} to {Magnetic} {Field}.
\newblock \emph{Solar Physics} 179, 253--268.
\newblock \doi{10.1023/A:1005046114362}
\bibAnnoteFile{nindos_relation_1998}

\bibitem[{Ohman(1956)}]{ohman_solar_1956}
Ohman, Y. (1956).
\newblock Solar observations in hydrogen and calcium lines.
\newblock \emph{The Observatory} 76, 158--159
\bibAnnoteFile{ohman_solar_1956}

\bibitem[{Ortiz and Rast(2005)}]{ortiz_how_2005}
Ortiz, A. and Rast, M. (2005).
\newblock How good is the {Ca} {II} {K} as a proxy for the magnetic flux?
\newblock \emph{Memorie della Societa Astronomica Italiana} 76, 1018
\bibAnnoteFile{ortiz_how_2005}

\bibitem[{Penza et~al.(2021)Penza, Berrilli, Bertello, Cantoresi, and
  Criscuoli}]{penza_prediction_2021}
Penza, V., Berrilli, F., Bertello, L., Cantoresi, M., and Criscuoli, S. (2021).
\newblock Prediction of {Sunspot} and {Plage} {Coverage} for {Solar} {Cycle}
  25.
\newblock \emph{The Astrophysical Journal Letters} 922, L12.
\newblock \doi{10.3847/2041-8213/ac3663}
\bibAnnoteFile{penza_prediction_2021}

\bibitem[{Penza et~al.(2022)Penza, Berrilli, Bertello, Cantoresi, Criscuoli,
  and Giobbi}]{penza_total_2022}
Penza, V., Berrilli, F., Bertello, L., Cantoresi, M., Criscuoli, S., and
  Giobbi, P. (2022).
\newblock Total {Solar} {Irradiance} during the {Last} {Five} {Centuries}.
\newblock \emph{The Astrophysical Journal} 937, 84.
\newblock \doi{10.3847/1538-4357/ac8a4b}
\bibAnnoteFile{penza_total_2022}

\bibitem[{Penza et~al.(2003)Penza, Caccin, Ermolli, Centrone, and
  Gomez}]{penza_modeling_2003}
Penza, V., Caccin, B., Ermolli, I., Centrone, M., and Gomez, M.~T. (2003).
\newblock Modeling solar irradiance variations through {PSPT} images and
  semiempirical models.
\newblock In \emph{Solar {Variability} as an {Input} to the {Earth}'s
  {Environment}}. vol. 535 of \emph{{ESA} {Special} {Publication}}, 299--302
\bibAnnoteFile{penza_modeling_2003}

\bibitem[{Pesnell et~al.(2012)Pesnell, Thompson, and
  Chamberlin}]{pesnell_solar_2012}
Pesnell, W.~D., Thompson, B.~J., and Chamberlin, P.~C. (2012).
\newblock The {Solar} {Dynamics} {Observatory} ({SDO}).
\newblock \emph{Solar Physics} 275, 3--15.
\newblock \doi{10.1007/s11207-011-9841-3}
\bibAnnoteFile{pesnell_solar_2012}

\bibitem[{Pevtsov et~al.(2019)Pevtsov, Griffin, Grindlay, Kafka, Bartlett,
  Usoskin et~al.}]{pevtsov_historical_2019}
Pevtsov, A., Griffin, E., Grindlay, J., Kafka, S., Bartlett, J., Usoskin, I.,
  et~al. (2019).
\newblock Historical astronomical data: urgent need for preservation,
  digitization enabling scientific exploration.
\newblock \emph{Bulletin of the American Astronomical Society} 51, 190.
\newblock ADS Bibcode: 2019BAAS...51c.190P
\bibAnnoteFile{pevtsov_historical_2019}

\bibitem[{Pevtsov et~al.(2016)Pevtsov, Virtanen, Mursula, Tlatov, and
  Bertello}]{pevtsov_reconstructing_2016}
Pevtsov, A.~A., Virtanen, I., Mursula, K., Tlatov, A., and Bertello, L. (2016).
\newblock Reconstructing solar magnetic fields from historical observations.
  {I}. {Renormalized} {Ca} {K} spectroheliograms and pseudo-magnetograms.
\newblock \emph{Astronomy and Astrophysics} 585, A40.
\newblock \doi{10.1051/0004-6361/201526620}
\bibAnnoteFile{pevtsov_reconstructing_2016}

\bibitem[{Pierce and Slaughter(1977)}]{pierce_solar_1977}
Pierce, A.~K. and Slaughter, C.~D. (1977).
\newblock Solar limb darkening. {I} - {At} wavelengths of 3033-7297.
\newblock \emph{Solar Physics} 51, 25--41.
\newblock \doi{10.1007/BF00240442}
\bibAnnoteFile{pierce_solar_1977}

\bibitem[{Pietropaolo and Ermolli(1998)}]{pietropaolo_chromospheric_1998}
Pietropaolo, E. and Ermolli, I. (1998).
\newblock Chromospheric intensity oscillations from {Ca} {II} {K} {OAR}/{PSPT}
  images.
\newblock \emph{Memorie della Societa Astronomica Italiana} 69, 583
\bibAnnoteFile{pietropaolo_chromospheric_1998}

\bibitem[{Priyal et~al.(2017)Priyal, Singh, Belur, and
  Rathina}]{priyal_long-term_2017}
Priyal, M., Singh, J., Belur, R., and Rathina, S.~K. (2017).
\newblock Long-term {Variations} in the {Intensity} of {Plages} and {Networks}
  as {Observed} in {Kodaikanal} {Ca}-{K} {Digitized} {Data}.
\newblock \emph{Solar Physics} 292, 85.
\newblock \doi{10.1007/s11207-017-1106-3}
\bibAnnoteFile{priyal_long-term_2017}

\bibitem[{Priyal et~al.(2014)Priyal, Singh, Ravindra, Priya, and
  Amareswari}]{priyal_long_2014}
Priyal, M., Singh, J., Ravindra, B., Priya, T.~G., and Amareswari, K. (2014).
\newblock Long {Term} {Variations} in {Chromospheric} {Features} from {Ca}-{K}
  {Images} at {Kodaikanal}.
\newblock \emph{Solar Physics} 289, 137--152.
\newblock \doi{10.1007/s11207-013-0315-7}
\bibAnnoteFile{priyal_long_2014}

\bibitem[{Priyal et~al.(2019)Priyal, Singh, Ravindra, and
  Shekar~B}]{priyal_periodic_2019}
Priyal, M., Singh, J., Ravindra, B., and Shekar~B, C. (2019).
\newblock Periodic and {Quasi}-{Periodic} {Variations} in the {Ca} {K} {Index}
  {During} the 20th {Century} {Using} {Kodaikanal} {Data}.
\newblock \emph{Solar Physics} 294, 131.
\newblock \doi{10.1007/s11207-019-1522-7}
\bibAnnoteFile{priyal_periodic_2019}

\bibitem[{Pruthvi and Ramesh(2015)}]{pruthvi_two-channel_2015}
Pruthvi, H. and Ramesh, K.~B. (2015).
\newblock Two-channel imaging system for the {White} light {Active} {Region}
  {Monitor} ({WARM}) telescope at {Kodaikanal} {Observatory}: design,
  development, and first images.
\newblock \emph{International Conference on Optics and Photonics 2015} 9654,
  96540I.
\newblock \doi{10.1117/12.2182889}
\bibAnnoteFile{pruthvi_two-channel_2015}

\bibitem[{Pötzi et~al.(2021)Pötzi, Veronig, Jarolim, Rodríguez~Gómez,
  Podlachikova, Baumgartner et~al.}]{potzi_kanzelhohe_2021}
Pötzi, W., Veronig, A., Jarolim, R., Rodríguez~Gómez, J.~M., Podlachikova,
  T., Baumgartner, D., et~al. (2021).
\newblock Kanzelhöhe {Observatory}: {Instruments}, {Data} {Processing} and
  {Data} {Products}.
\newblock \emph{Solar Physics} 296, 164.
\newblock \doi{10.1007/s11207-021-01903-4}
\bibAnnoteFile{potzi_kanzelhohe_2021}

\bibitem[{Puiu(2019)}]{puiu_modeling_2019}
Puiu, C.~C. (2019).
\newblock \emph{Modeling solar irradiance variations on timescales from day to
  solar cycle with ground-based observations}.
\newblock Master's thesis, Sapienza – University of Rome, Rome
\bibAnnoteFile{puiu_modeling_2019}

\bibitem[{Raghavan(1983)}]{raghavan_quantitative_1983}
Raghavan, N. (1983).
\newblock A quantitative study of {CA} {II} network geometry.
\newblock \emph{Solar Physics} 89, 35--42.
\newblock \doi{10.1007/BF00211950}
\bibAnnoteFile{raghavan_quantitative_1983}

\bibitem[{Rajani et~al.(2022)Rajani, Sowmya, Paniveni, and
  Srikanth}]{rajani_solar_2022}
Rajani, G., Sowmya, G.~M., Paniveni, U., and Srikanth, R. (2022).
\newblock Solar {Supergranular} {Fractal} {Dimension} {Dependence} on the
  {Solar} {Cycle} {Phase}.
\newblock \emph{Research in Astronomy and Astrophysics} 22, 045006.
\newblock \doi{10.1088/1674-4527/ac5020}.
\newblock Publisher: IOP Publishing
\bibAnnoteFile{rajani_solar_2022}

\bibitem[{Raju(2020)}]{raju_asymmetry_2020}
Raju, K.~P. (2020).
\newblock Asymmetry in the {Length} {Scales} of the {Solar} {Supergranulation}
  {Network}.
\newblock \emph{The Astrophysical Journal} 899, L35.
\newblock \doi{10.3847/2041-8213/abacb7}
\bibAnnoteFile{raju_asymmetry_2020}

\bibitem[{Raju and Singh(2014)}]{raju_network_2014}
Raju, K.~P. and Singh, J. (2014).
\newblock Network and plage indices from {Kodaikanal} {Ca}-{K} data.
\newblock \emph{Research in Astronomy and Astrophysics} 14, 229--232.
\newblock \doi{10.1088/1674-4527/14/2/010}
\bibAnnoteFile{raju_network_2014}

\bibitem[{Raju et~al.(1998)Raju, Srikanth, and Singh}]{raju_dependence_1998}
Raju, K.~P., Srikanth, R., and Singh, J. (1998).
\newblock The {Dependence} of {Chromospheric} {CA} {II} {K} {Network} {Cell}
  {Sizes} on {Solar} {Latitude}.
\newblock \emph{Solar Physics} 180, 47--51.
\newblock \doi{10.1023/A:1005072907000}.
\newblock ADS Bibcode: 1998SoPh..180...47R
\bibAnnoteFile{raju_dependence_1998}

\bibitem[{Rast(2003)}]{rast_scales_2003}
Rast, M.~P. (2003).
\newblock The {Scales} of {Granulation}, {Mesogranulation}, and
  {Supergranulation}.
\newblock \emph{The Astrophysical Journal} 597, 1200--1210.
\newblock \doi{10.1086/381221}
\bibAnnoteFile{rast_scales_2003}

\bibitem[{Rast et~al.(2008)Rast, Ortiz, and Meisner}]{rast_latitudinal_2008}
Rast, M.~P., Ortiz, A., and Meisner, R.~W. (2008).
\newblock Latitudinal {Variation} of the {Solar} {Photospheric} {Intensity}.
\newblock \emph{The Astrophysical Journal} 673, 1209--1217.
\newblock \doi{10.1086/524655;}
\bibAnnoteFile{rast_latitudinal_2008}

\bibitem[{Ravindra et~al.(2021)Ravindra, Chowdhury, and
  Javaraiah}]{ravindra_solar-cycle_2021}
Ravindra, B., Chowdhury, P., and Javaraiah, J. (2021).
\newblock Solar-{Cycle} {Characteristics} in {Kodaikanal} {Sunspot} {Area}:
  {North}–{South} {Asymmetry}, {Phase} {Distribution} and {Gnevyshev} {Gap}.
\newblock \emph{Solar Physics} 296, 2.
\newblock \doi{10.1007/s11207-020-01744-7}
\bibAnnoteFile{ravindra_solar-cycle_2021}

\bibitem[{Rezaei et~al.(2007)Rezaei, Schlichenmaier, Beck, Bruls, and
  Schmidt}]{rezaei_relation_2007}
Rezaei, R., Schlichenmaier, R., Beck, C. A.~R., Bruls, J. H. M.~J., and
  Schmidt, W. (2007).
\newblock Relation between photospheric magnetic field and chromospheric
  emission.
\newblock \emph{Astronomy and Astrophysics} 466, 1131--1144.
\newblock \doi{10.1051/0004-6361:20067017}
\bibAnnoteFile{rezaei_relation_2007}

\bibitem[{Rincon and Rieutord(2018)}]{rincon_suns_2018}
Rincon, F. and Rieutord, M. (2018).
\newblock The {Sun}’s supergranulation.
\newblock \emph{Living Reviews in Solar Physics} 15, 6.
\newblock \doi{10.1007/s41116-018-0013-5}
\bibAnnoteFile{rincon_suns_2018}

\bibitem[{Rottman(2005)}]{rottman_sorce_2005}
Rottman, G. (2005).
\newblock The {SORCE} {Mission}.
\newblock \emph{Solar Physics} 230, 7--25.
\newblock \doi{10.1007/s11207-005-8112-6}
\bibAnnoteFile{rottman_sorce_2005}

\bibitem[{Schatten et~al.(1985)Schatten, Miller, Sofia, Endal, and
  Chapman}]{schatten_importance_1985}
Schatten, K.~H., Miller, N., Sofia, S., Endal, A.~S., and Chapman, G. (1985).
\newblock The importance of improved facular observations in understanding
  solar constant variations.
\newblock \emph{The Astrophysical Journal} 294, 689.
\newblock \doi{10.1086/163339}
\bibAnnoteFile{schatten_importance_1985}

\bibitem[{Schrijver et~al.(1989)Schrijver, Cote, Zwaan, and
  Saar}]{schrijver_relations_1989}
Schrijver, C.~J., Cote, J., Zwaan, C., and Saar, S.~H. (1989).
\newblock Relations between the photospheric magnetic field and the emission
  from the outer atmospheres of cool stars. {I} - {The} solar {CA} {II} {K}
  line core emission.
\newblock \emph{The Astrophysical Journal} 337, 964--976.
\newblock \doi{10.1086/167168}
\bibAnnoteFile{schrijver_relations_1989}

\bibitem[{Seetha and Megala(2017)}]{seetha_aditya-l1_2017}
Seetha, S. and Megala, S. (2017).
\newblock Aditya-{L1} {Mission}.
\newblock \emph{Current Science} 113, 610.
\newblock \doi{10.18520/cs/v113/i04/610-612}
\bibAnnoteFile{seetha_aditya-l1_2017}

\bibitem[{Seguí et~al.(2019)Seguí, Curto, Paula, Rodríguez-Gasén, and
  Vaquero}]{segui_temporal_2019}
Seguí, A., Curto, J.~J., Paula, V.~d., Rodríguez-Gasén, R., and Vaquero,
  J.~M. (2019).
\newblock Temporal variation and asymmetry of sunspot and solar plage types
  from 1930 to 1936.
\newblock \emph{Advances in Space Research}
  \doi{https://doi.org/10.1016/j.asr.2019.02.018}
\bibAnnoteFile{segui_temporal_2019}

\bibitem[{Shapiro et~al.(2017)Shapiro, Solanki, Krivova, Cameron, Yeo, and
  Schmutz}]{shapiro_nature_2017}
Shapiro, A.~I., Solanki, S.~K., Krivova, N.~A., Cameron, R.~H., Yeo, K.~L., and
  Schmutz, W.~K. (2017).
\newblock The nature of solar brightness variations.
\newblock \emph{Nature Astronomy} 1, 612--616.
\newblock \doi{10.1038/s41550-017-0217-y}
\bibAnnoteFile{shapiro_nature_2017}

\bibitem[{Sheeley et~al.(2011)Sheeley, Cooper, and
  Anderson}]{sheeley_carrington_2011}
Sheeley, N.~R., Jr., Cooper, T.~J., and Anderson, J. R.~L. (2011).
\newblock Carrington {Maps} of {Ca} {II} {K}-line {Emission} for the {Years}
  1915-1985.
\newblock \emph{The Astrophysical Journal} 730, 51.
\newblock \doi{10.1088/0004-637X/730/1/51}
\bibAnnoteFile{sheeley_carrington_2011}

\bibitem[{Shimizu et~al.(2020)Shimizu, Imada, Kawate, Suematsu, Hara, Tsuzuki
  et~al.}]{shimizu_solar-c_2020}
Shimizu, T., Imada, S., Kawate, T., Suematsu, Y., Hara, H., Tsuzuki, T., et~al.
  (2020).
\newblock The {Solar}-{C} ({EUVST}) mission: the latest status.
\newblock In \emph{Space {Telescopes} and {Instrumentation} 2020: {Ultraviolet}
  to {Gamma} {Ray}} (Proc. SPIE), vol. 11444, 114440N.
\newblock \doi{10.1117/12.2560887}
\bibAnnoteFile{shimizu_solar-c_2020}

\bibitem[{Shin et~al.(2020)Shin, Moon, Park, Jeong, Lee, and
  Bae}]{shin_generation_2020}
Shin, G., Moon, Y.-J., Park, E., Jeong, H., Lee, H., and Bae, S.-H. (2020).
\newblock Generation of {High}-resolution {Solar} {Pseudo}-magnetograms from
  {Ca} {II} {K} {Images} by {Deep} {Learning}.
\newblock \emph{The Astrophysical Journal Letters} 895, L16.
\newblock \doi{10.3847/2041-8213/ab9085}
\bibAnnoteFile{shin_generation_2020}

\bibitem[{Simon and Leighton(1964)}]{simon_velocity_1964}
Simon, G.~W. and Leighton, R.~B. (1964).
\newblock Velocity {Fields} in the {Solar} {Atmosphere}. {III}. {Large}-{Scale}
  {Motions}, the {Chromospheric} {Network}, and {Magnetic} {Fields}.
\newblock \emph{The Astrophysical Journal} 140, 1120.
\newblock \doi{10.1086/148010}
\bibAnnoteFile{simon_velocity_1964}

\bibitem[{Singh and Bappu(1981)}]{singh_dependence_1981}
Singh, J. and Bappu, M. K.~V. (1981).
\newblock A dependence on solar cycle of the size of the {Ca}/+/ network.
\newblock \emph{Solar Physics} 71, 161--168.
\newblock \doi{10.1007/BF00153615}
\bibAnnoteFile{singh_dependence_1981}

\bibitem[{Singh et~al.(2012)Singh, Belur, Raju, Pichaimani, Priyal,
  Gopalan~Priya et~al.}]{singh_determination_2012}
Singh, J., Belur, R., Raju, S., Pichaimani, K., Priyal, M., Gopalan~Priya, T.,
  et~al. (2012).
\newblock "{Determination} of the chromospheric quiet network element area
  index and its variation between 2008 and 2011" ({RAA}, {Vol}. 12, p.201
  [2012]).
\newblock \emph{Research in Astronomy and Astrophysics} 12, 472.
\newblock \doi{10.1088/1674-4527/12/4/011}
\bibAnnoteFile{singh_determination_2012}

\bibitem[{Singh et~al.(2021)Singh, Priyal, and
  Ravindra}]{singh_determining_2021}
Singh, J., Priyal, M., and Ravindra, B. (2021).
\newblock Determining the {Variations} of {Ca}–{K} {Index} and {Features}
  {Using} {Century}-long {Equal}-contrast {Images} from {Kodaikanal}
  {Observatory}.
\newblock \emph{The Astrophysical Journal} 908, 210.
\newblock \doi{10.3847/1538-4357/abd021}
\bibAnnoteFile{singh_determining_2021}

\bibitem[{Singh et~al.(2022)Singh, Priyal, Ravindra, Bertello, and
  Pevtsov}]{singh_application_2022}
Singh, J., Priyal, M., Ravindra, B., Bertello, L., and Pevtsov, A.~A. (2022).
\newblock On the {Application} of the {Equal}-contrast {Technique} to {Ca}-{K}
  {Data} from {Kodaikanal} and {Other} {Observatories}.
\newblock \emph{The Astrophysical Journal} 927, 154.
\newblock \doi{10.3847/1538-4357/ac4e82}
\bibAnnoteFile{singh_application_2022}

\bibitem[{Singh et~al.(2018)Singh, Priyal, Sindhuja, and
  Ravindra}]{singh_variations_2018}
Singh, J., Priyal, M., Sindhuja, G., and Ravindra, B. (2018).
\newblock Variations in {Ca}-{K} line profiles and {Ca}-{K} line features as a
  function of latitude and solar cycle during the 20th century.
\newblock \emph{Proceedings of the International Astronomical Union} 13,
  23--26.
\newblock \doi{10.1017/S1743921318001540}
\bibAnnoteFile{singh_variations_2018}

\bibitem[{Singh and Ravindra(2012)}]{singh_twin_2012}
Singh, J. and Ravindra, B. (2012).
\newblock Twin {Telescope} observations of the {Sun} at {Kodaikanal}
  {Observatory}.
\newblock \emph{Bulletin of the Astronomical Society of India} 40
\bibAnnoteFile{singh_twin_2012}

\bibitem[{Skumanich et~al.(1975)Skumanich, Smythe, and
  Frazier}]{skumanich_statistical_1975}
Skumanich, A., Smythe, C., and Frazier, E.~N. (1975).
\newblock On the statistical description of inhomogeneities in the quiet solar
  atmosphere. {I} - {Linear} regression analysis and absolute calibration of
  multichannel observations of the {Ca}/+/ emission network.
\newblock \emph{The Astrophysical Journal} 200, 747--764.
\newblock \doi{10.1086/153846}
\bibAnnoteFile{skumanich_statistical_1975}

\bibitem[{Snow et~al.(2014)Snow, Weber, Machol, Viereck, and
  Richard}]{snow_comparison_2014}
Snow, M., Weber, M., Machol, J., Viereck, R., and Richard, E. (2014).
\newblock Comparison of {Magnesium} {II} core-to-wing ratio observations during
  solar minimum 23/24.
\newblock \emph{Journal of Space Weather and Space Climate} 4, A04.
\newblock \doi{10.1051/swsc/2014001}
\bibAnnoteFile{snow_comparison_2014}

\bibitem[{Solanki et~al.(2010)Solanki, Barthol, Danilovic, Feller, Gandorfer,
  Hirzberger et~al.}]{solanki_sunrise_2010}
Solanki, S.~K., Barthol, P., Danilovic, S., Feller, A., Gandorfer, A.,
  Hirzberger, J., et~al. (2010).
\newblock {SUNRISE}: {Instrument}, {Mission}, {Data}, and {First} {Results}.
\newblock \emph{The Astrophysical Journal Letters} 723, L127--L133.
\newblock \doi{10.1088/2041-8205/723/2/L127}
\bibAnnoteFile{solanki_sunrise_2010}

\bibitem[{Solanki et~al.(2013)Solanki, Krivova, and
  Haigh}]{solanki_solar_2013-1}
Solanki, S.~K., Krivova, N.~A., and Haigh, J.~D. (2013).
\newblock Solar {Irradiance} {Variability} and {Climate}.
\newblock \emph{Annual Review of Astronomy and Astrophysics} 51, 311--351.
\newblock \doi{10.1146/annurev-astro-082812-141007}
\bibAnnoteFile{solanki_solar_2013-1}

\bibitem[{Solanki et~al.(2017)Solanki, Riethmüller, Barthol, Danilovic,
  Deutsch, Doerr et~al.}]{solanki_second_2017}
Solanki, S.~K., Riethmüller, T.~L., Barthol, P., Danilovic, S., Deutsch, W.,
  Doerr, H.-P., et~al. (2017).
\newblock The {Second} {Flight} of the {Sunrise} {Balloon}-borne {Solar}
  {Observatory}: {Overview} of {Instrument} {Updates}, the {Flight}, the
  {Data}, and {First} {Results}.
\newblock \emph{The Astrophysical Journal Supplement Series} 229, 2.
\newblock \doi{10.3847/1538-4365/229/1/2}
\bibAnnoteFile{solanki_second_2017}

\bibitem[{Solanki et~al.(2002)Solanki, Schüssler, and
  Fligge}]{solanki_secular_2002}
Solanki, S.~K., Schüssler, M., and Fligge, M. (2002).
\newblock Secular variation of the {Sun}'s magnetic flux.
\newblock \emph{Astronomy and Astrophysics} 383, 706--712.
\newblock \doi{10.1051/0004-6361:20011790}
\bibAnnoteFile{solanki_secular_2002}

\bibitem[{Sotnikova(1978)}]{sotnikova_statistical_1978}
Sotnikova, R.~T. (1978).
\newblock Statistical analysis of the intensity fluctuations of the undisturbed
  chromosphere in the {H}-alpha and {CA} {II} {K} lines.
\newblock \emph{Soviet Astronomy Letters} 4, 246--249.
\newblock ADS Bibcode: 1978SvAL....4..246S
\bibAnnoteFile{sotnikova_statistical_1978}

\bibitem[{Steinegger et~al.(1996{\natexlab{a}})Steinegger, Brandt, and
  Haupt}]{steinegger_sunspot_1996}
Steinegger, M., Brandt, P.~N., and Haupt, H.~F. (1996{\natexlab{a}}).
\newblock Sunspot irradiance deficit, facular excess, and the energy balance of
  solar active regions.
\newblock \emph{Astronomy and Astrophysics} 310, 635--645
\bibAnnoteFile{steinegger_sunspot_1996}

\bibitem[{Steinegger et~al.(1996{\natexlab{b}})Steinegger, Vazquez, Bonet, and
  Brandt}]{steinegger_energy_1996}
Steinegger, M., Vazquez, M., Bonet, J.~A., and Brandt, P.~N.
  (1996{\natexlab{b}}).
\newblock On the {Energy} {Balance} of {Solar} {Active} {Regions}.
\newblock \emph{The Astrophysical Journal} 461, 478.
\newblock \doi{10.1086/177075}
\bibAnnoteFile{steinegger_energy_1996}

\bibitem[{Suo(2020)}]{suo_full-disk_2020}
Suo, L. (2020).
\newblock A full-disk image standardization of the chromosphere observation at
  {Huairou} {Solar} {Observing} {Station}.
\newblock \emph{Advances in Space Research} 65, 1054--1061.
\newblock \doi{10.1016/j.asr.2019.10.035}
\bibAnnoteFile{suo_full-disk_2020}

\bibitem[{Svalgaard and Schatten(2016)}]{svalgaard_reconstruction_2016}
Svalgaard, L. and Schatten, K.~H. (2016).
\newblock Reconstruction of the {Sunspot} {Group} {Number}: {The} {Backbone}
  {Method}.
\newblock \emph{Solar Physics} 291, 2653--2684.
\newblock \doi{10.1007/s11207-015-0815-8}
\bibAnnoteFile{svalgaard_reconstruction_2016}

\bibitem[{Tapping and Morton(2013)}]{tapping_next_2013}
Tapping, K.~F. and Morton, D.~C. (2013).
\newblock The {Next} {Generation} of {Canadian} {Solar} {Flux} {Monitoring}.
\newblock In \emph{Journal of {Physics} {Conference} {Series}}. vol. 440,
  012039.
\newblock \doi{10.1088/1742-6596/440/1/012039}.
\newblock ADS Bibcode: 2013JPhCS.440a2039T
\bibAnnoteFile{tapping_next_2013}

\bibitem[{Teston and Creasey(1997)}]{teston_proba_1997}
Teston, F. and Creasey, R. (1997).
\newblock {PROBA} - {Project} for {Onboard} {Autonomy}.
\newblock In \emph{Data {Systems} in {Aerospace} - {DASIA} 97}, ed. T.-D.
  Guyenne (European Space Agency), vol. 409, 109
\bibAnnoteFile{teston_proba_1997}

\bibitem[{Tähtinen et~al.(2022)Tähtinen, Virtanen, Pevtsov, and
  Mursula}]{tahtinen_reconstructing_2022}
Tähtinen, I., Virtanen, I., Pevtsov, A., and Mursula, K. (2022).
\newblock Reconstructing solar magnetic fields from historical observations.
  {VIII}. {AIA} 1600 Å contrast as a proxy of solar magnetic fields.
\newblock \emph{Astronomy \& Astrophysics} \doi{10.1051/0004-6361/202141164}
\bibAnnoteFile{tahtinen_reconstructing_2022}

\bibitem[{Tlatov et~al.(2015)Tlatov, Dormidontov, Kirpichev, Pashchenko, and
  Shramko}]{tlatov_synoptic_2015}
Tlatov, A.~G., Dormidontov, D.~V., Kirpichev, R.~V., Pashchenko, M.~P., and
  Shramko, A.~D. (2015).
\newblock Synoptic and fast events on the sun according to observations at the
  center and wings of the {Ca} {II} {K} line at the {Kislovodsk} {Mountain}
  station patrol telescope.
\newblock \emph{Geomagnetism and Aeronomy} 55, 961--968.
\newblock \doi{10.1134/S0016793215070245}.
\newblock Number: 7
\bibAnnoteFile{tlatov_synoptic_2015}

\bibitem[{Tlatov et~al.(2009)Tlatov, Pevtsov, and Singh}]{tlatov_new_2009}
Tlatov, A.~G., Pevtsov, A.~A., and Singh, J. (2009).
\newblock A {New} {Method} of {Calibration} of {Photographic} {Plates} from
  {Three} {Historic} {Data} {Sets}.
\newblock \emph{Solar Physics} 255, 239--251.
\newblock \doi{10.1007/s11207-009-9326-9}
\bibAnnoteFile{tlatov_new_2009}

\bibitem[{Tlatov and Tlatova(2019)}]{tlatov_polar_2019}
Tlatov, A.~G. and Tlatova, K.~A. (2019).
\newblock Polar {Activity} of the {Sun} and {Latitudinal} {Activity} {Drifts}
  in {Cycles} 15–24.
\newblock \emph{Geomagnetism and Aeronomy} 59, 6.
\newblock \doi{10.1134/S0016793219080218}
\bibAnnoteFile{tlatov_polar_2019}

\bibitem[{Tripathi et~al.(2017)Tripathi, Ramaprakash, Khan, Ghosh, Chatterjee,
  Banerjee et~al.}]{tripathi_solar_2017}
Tripathi, D., Ramaprakash, A.~N., Khan, A., Ghosh, A., Chatterjee, S.,
  Banerjee, D., et~al. (2017).
\newblock The {Solar} {Ultraviolet} {Imaging} {Telescope} on-board
  {Aditya}-{L1}.
\newblock \emph{Current Science} 113, 616.
\newblock \doi{10.18520/cs/v113/i04/616-619}
\bibAnnoteFile{tripathi_solar_2017}

\bibitem[{Usoskin et~al.(2016)Usoskin, Kovaltsov, Lockwood, Mursula, Owens, and
  Solanki}]{usoskin_new_2016}
Usoskin, I.~G., Kovaltsov, G.~A., Lockwood, M., Mursula, K., Owens, M., and
  Solanki, S.~K. (2016).
\newblock A {New} {Calibrated} {Sunspot} {Group} {Series} {Since} 1749:
  {Statistics} of {Active} {Day} {Fractions}.
\newblock \emph{Solar Physics} 291, 2685--2708.
\newblock \doi{10.1007/s11207-015-0838-1}.
\newblock Number: 9-10
\bibAnnoteFile{usoskin_new_2016}

\bibitem[{Vaquero et~al.(2007)Vaquero, Gallego, Acero, and
  García}]{vaquero_spectroheliographic_2007}
Vaquero, J.~M., Gallego, M.~C., Acero, F.~J., and García, J.~A. (2007).
\newblock Spectroheliographic {Observations} in {Madrid} (1912 -- 1917).
\newblock In \emph{The {Physics} of {Chromospheric} {Plasmas}}, eds.
  P.~Heinzel, I.~Dorotovic, and R.~J. Rutten. vol. 368 of \emph{Astronomical
  {Society} of the {Pacific} {Conference} {Series}}, 17--20
\bibAnnoteFile{vaquero_spectroheliographic_2007}

\bibitem[{Vaquero et~al.(2016)Vaquero, Svalgaard, Carrasco, Clette, Lefèvre,
  Gallego et~al.}]{vaquero_revised_2016}
Vaquero, J.~M., Svalgaard, L., Carrasco, V. M.~S., Clette, F., Lefèvre, L.,
  Gallego, M.~C., et~al. (2016).
\newblock A {Revised} {Collection} of {Sunspot} {Group} {Numbers}.
\newblock \emph{Solar Physics} 291, 3061--3074.
\newblock \doi{10.1007/s11207-016-0982-2}
\bibAnnoteFile{vaquero_revised_2016}

\bibitem[{Vaquero and Vázquez(2009)}]{vaquero_sun_2009}
Vaquero, J.~M. and Vázquez, M. (2009).
\newblock \emph{The {Sun} {Recorded} {Through} {History}: {Scientific} {Data}
  {Extracted} from {Historical} {Documents}}, vol. 361 of \emph{Astrophysics
  and {Space} {Science} {Library}} (New York, NY: Springer New York)
\bibAnnoteFile{vaquero_sun_2009}

\bibitem[{Veronig et~al.(2000)Veronig, Steinegger, Otruba, Hanslmeier,
  Messerotti, Temmer et~al.}]{veronig_automatic_2000}
Veronig, A., Steinegger, M., Otruba, W., Hanslmeier, A., Messerotti, M.,
  Temmer, M., et~al. (2000).
\newblock Automatic {Image} {Processing} in the {Frame} of a {Solar} {Flare}
  {Alerting} {System}.
\newblock In \emph{Hvar {Observatory} {Bulletin}}. vol.~24, 195--205.
\newblock ISSN: 0351-2657
\bibAnnoteFile{veronig_automatic_2000}

\bibitem[{Veronig et~al.(2021)Veronig, Jain, Podladchikova, Pötzi, and
  Clette}]{veronig_hemispheric_2021}
Veronig, A.~M., Jain, S., Podladchikova, T., Pötzi, W., and Clette, F. (2021).
\newblock Hemispheric sunspot numbers 1874–2020.
\newblock \emph{Astronomy \& Astrophysics} 652, A56.
\newblock \doi{10.1051/0004-6361/202141195}
\bibAnnoteFile{veronig_hemispheric_2021}

\bibitem[{Volobuev(2009)}]{2009SoPh..258..319V}
Volobuev, D.~M. (2009).
\newblock The {Shape} of {The} {Sunspot} {Cycle}: {A} {One}-{Parameter} {Fit}.
\newblock \emph{Solar Physics} 258, 319--330.
\newblock \doi{10.1007/s11207-009-9429-3}.
\newblock ADS Bibcode: 2009SoPh..258..319V
\bibAnnoteFile{2009SoPh..258..319V}

\bibitem[{Waldmeier(1968)}]{waldmeier_swiss_1968}
Waldmeier, M. (1968).
\newblock The {Swiss} {Federal} {Observatory}, {Zürich}.
\newblock \emph{Solar Physics} 5, 423--426.
\newblock \doi{10.1007/BF00147155}
\bibAnnoteFile{waldmeier_swiss_1968}

\bibitem[{Walton et~al.(1998)Walton, Chapman, Cookson, Dobias, and
  Preminger}]{walton_processing_1998}
Walton, S.~R., Chapman, G.~A., Cookson, A.~M., Dobias, J.~J., and Preminger,
  D.~G. (1998).
\newblock Processing {Photometric} {Full}-{Disk} {Solar} {Images}.
\newblock \emph{Solar Physics} 179, 31--42.
\newblock \doi{10.1023/A:1005070932205}
\bibAnnoteFile{walton_processing_1998}

\bibitem[{Walton et~al.(2003)Walton, Preminger, and
  Chapman}]{walton_contribution_2003}
Walton, S.~R., Preminger, D.~G., and Chapman, G.~A. (2003).
\newblock The {Contribution} of {Faculae} and {Network} to {Long}-{Term}
  {Changes} in the {Total} {Solar} {Irradiance}.
\newblock \emph{The Astrophysical Journal} 590, 1088--1094.
\newblock \doi{10.1086/375022}
\bibAnnoteFile{walton_contribution_2003}

\bibitem[{Wang et~al.(2018)Wang, Liu, Zhu, Tao, Kautz, and
  Catanzaro}]{wang_high-resolution_2018}
Wang, T.-C., Liu, M.-Y., Zhu, J.-Y., Tao, A., Kautz, J., and Catanzaro, B.
  (2018).
\newblock High-{Resolution} {Image} {Synthesis} and {Semantic} {Manipulation}
  with {Conditional} {GANs}.
\newblock \emph{arXiv:1711.11585 [cs]} ArXiv: 1711.11585
\bibAnnoteFile{wang_high-resolution_2018}

\bibitem[{Wang and Sheeley(1992)}]{wang_potential_1992}
Wang, Y.-M. and Sheeley, N.~R., Jr. (1992).
\newblock On potential field models of the solar corona.
\newblock \emph{The Astrophysical Journal} 392, 310--319.
\newblock \doi{10.1086/171430}
\bibAnnoteFile{wang_potential_1992}

\bibitem[{Watanabe(2014)}]{watanabe_solar-c_2014}
Watanabe, T. (2014).
\newblock The {Solar}-{C} {Mission}.
\newblock In \emph{Space {Telescopes} and {Instrumentation} 2014: Optical,
  Infrared, and Millimeter Wave} (Proc. of SPIE), vol. 9143, 91431O.
\newblock \doi{10.1117/12.2055366}
\bibAnnoteFile{watanabe_solar-c_2014}

\bibitem[{Wells et~al.(1981)Wells, Greisen, and Harten}]{wells_fits_1981}
Wells, D.~C., Greisen, E.~W., and Harten, R.~H. (1981).
\newblock {FITS} - a {Flexible} {Image} {Transport} {System}.
\newblock \emph{Astronomy and Astrophysics Supplement Series} 44, 363
\bibAnnoteFile{wells_fits_1981}

\bibitem[{White and Livingston(1978)}]{white_solar_1978}
White, O.~R. and Livingston, W. (1978).
\newblock Solar luminosity variation. {II} - {Behavior} of calcium {H} and {K}
  at solar minimum and the onset of cycle 21.
\newblock \emph{The Astrophysical Journal} 226, 679--686.
\newblock \doi{10.1086/156650}
\bibAnnoteFile{white_solar_1978}

\bibitem[{White and Livingston(1981)}]{white_solar_1981}
White, O.~R. and Livingston, W.~C. (1981).
\newblock Solar luminosity variation. {III} - {Calcium} {K} variation from
  solar minimum to maximum in cycle 21.
\newblock \emph{The Astrophysical Journal} 249, 798--816.
\newblock \doi{10.1086/159338}
\bibAnnoteFile{white_solar_1981}

\bibitem[{White et~al.(1998)White, Livingston, Keil, and
  Henry}]{white_variability_1998}
White, O.~R., Livingston, W.~C., Keil, S.~L., and Henry, T.~W. (1998).
\newblock Variability of the {Solar} {Call} {K} {Line} over the 22 {Year}
  {Hale} {Cycle}.
\newblock In \emph{{ASP} conference series} (Astronomical Society of the
  Pacific), vol. 140, 293
\bibAnnoteFile{white_variability_1998}

\bibitem[{Wöhl(2005)}]{wohl_old_2005}
Wöhl, H. (2005).
\newblock The {Old} {Archives} of {Solar} {Images} of the {Former}
  {Frauenhofer} {Institut} (now: {Kiepenheuer}-{Institut} für {Sonnenphysik},
  {KIS}).
\newblock \emph{Hvar Observatory Bulletin} 29, 319--328
\bibAnnoteFile{wohl_old_2005}

\bibitem[{Wiegelmann et~al.(2014)Wiegelmann, Thalmann, and
  Solanki}]{wiegelmann_magnetic_2014}
Wiegelmann, T., Thalmann, J.~K., and Solanki, S.~K. (2014).
\newblock The magnetic field in the solar atmosphere.
\newblock \emph{Astronomy and Astrophysics Review} 22, 78.
\newblock \doi{10.1007/s00159-014-0078-7}
\bibAnnoteFile{wiegelmann_magnetic_2014}

\bibitem[{Willamo et~al.(2017)Willamo, Usoskin, and
  Kovaltsov}]{willamo_updated_2017}
Willamo, T., Usoskin, I.~G., and Kovaltsov, G.~A. (2017).
\newblock Updated sunspot group number reconstruction for 1749–1996 using the
  active day fraction method.
\newblock \emph{Astronomy \& Astrophysics} 601, A109.
\newblock \doi{10.1051/0004-6361/201629839}
\bibAnnoteFile{willamo_updated_2017}

\bibitem[{Wolf(1850)}]{wolf_mittheilungen_1850-1}
Wolf, R. (1850).
\newblock Mittheilungen über die {Sonnenflecken} {I}.
\newblock \emph{Astronomische Mitteilungen der Eidgenössischen Sternwarte
  Zurich} 1, 3--13
\bibAnnoteFile{wolf_mittheilungen_1850-1}

\bibitem[{Woods et~al.(2000)Woods, Tobiska, Rottman, and
  Worden}]{woods_improved_2000}
Woods, T.~N., Tobiska, W.~K., Rottman, G.~J., and Worden, J.~R. (2000).
\newblock Improved solar {Lyman} $\alpha$ irradiance modeling from 1947 through
  1999 based on {UARS} observations.
\newblock \emph{Journal of Geophysical Research} 105, 27195--27216.
\newblock \doi{10.1029/2000JA000051}
\bibAnnoteFile{woods_improved_2000}

\bibitem[{Worden et~al.(1998{\natexlab{a}})Worden, White, and
  Woods}]{worden_evolution_1998}
Worden, J.~R., White, O.~R., and Woods, T.~N. (1998{\natexlab{a}}).
\newblock Evolution of {Chromospheric} {Structures} {Derived} from {Ca} {II}
  {K} {Spectroheliograms}: {Implications} for {Solar} {Ultraviolet}
  {Irradiance} {Variability}.
\newblock \emph{The Astrophysical Journal} 496, 998.
\newblock \doi{10.1086/305392}
\bibAnnoteFile{worden_evolution_1998}

\bibitem[{Worden et~al.(1998{\natexlab{b}})Worden, White, and
  Woods}]{worden_plage_1998}
Worden, J.~R., White, O.~R., and Woods, T.~N. (1998{\natexlab{b}}).
\newblock Plage and {Enhanced} {Network} {Indices} {Derived} from {Ca} {II} {K}
  {Spectroheliograms}.
\newblock \emph{Solar Physics} 177, 255--264.
\newblock \doi{10.1023/A:1004921707249}.
\newblock Number: 1-2
\bibAnnoteFile{worden_plage_1998}

\bibitem[{Xu et~al.(2021)Xu, Lei, and Li}]{xu_reconstruction_2021}
Xu, H., Lei, B., and Li, Z. (2021).
\newblock A {Reconstruction} of {Total} {Solar} {Irradiance} {Based} on
  {Wavelet} {Analysis}.
\newblock \emph{Earth and Space Science} 8, e2021EA001819.
\newblock \doi{10.1029/2021EA001819}
\bibAnnoteFile{xu_reconstruction_2021}

\bibitem[{Yeo et~al.(2014)Yeo, Krivova, Solanki, and
  Glassmeier}]{yeo_reconstruction_2014}
Yeo, K.~L., Krivova, N.~A., Solanki, S.~K., and Glassmeier, K.~H. (2014).
\newblock Reconstruction of total and spectral solar irradiance from 1974 to
  2013 based on {KPVT}, {SoHO}/{MDI}, and {SDO}/{HMI} observations.
\newblock \emph{Astronomy and Astrophysics} 570, A85.
\newblock \doi{10.1051/0004-6361/201423628}
\bibAnnoteFile{yeo_reconstruction_2014}

\bibitem[{Yeo et~al.(2020{\natexlab{a}})Yeo, Solanki, and
  Krivova}]{yeo_how_2020}
Yeo, K.~L., Solanki, S.~K., and Krivova, N.~A. (2020{\natexlab{a}}).
\newblock How faculae and network relate to sunspots, and the implications for
  solar and stellar brightness variations.
\newblock \emph{Astronomy \& Astrophysics} 639, A139.
\newblock \doi{10.1051/0004-6361/202037739}
\bibAnnoteFile{yeo_how_2020}

\bibitem[{Yeo et~al.(2020{\natexlab{b}})Yeo, Solanki, Krivova, Rempel, Anusha,
  Shapiro et~al.}]{yeo_dimmest_2020}
Yeo, K.~L., Solanki, S.~K., Krivova, N.~A., Rempel, M., Anusha, L.~S., Shapiro,
  A.~I., et~al. (2020{\natexlab{b}}).
\newblock The {Dimmest} {State} of the {Sun}.
\newblock \emph{Geophysical Research Letters} 47, e2020GL090243.
\newblock \doi{10.1029/2020GL090243}
\bibAnnoteFile{yeo_dimmest_2020}

\bibitem[{Yeo et~al.(2017)Yeo, Solanki, Norris, Beeck, Unruh, and
  Krivova}]{yeo_solar_2017}
Yeo, K.~L., Solanki, S.~K., Norris, C.~M., Beeck, B., Unruh, Y.~C., and
  Krivova, N.~A. (2017).
\newblock Solar {Irradiance} {Variability} is {Caused} by the {Magnetic}
  {Activity} on the {Solar} {Surface}.
\newblock \emph{Physical Review Letters} 119.
\newblock \doi{10.1103/PhysRevLett.119.091102}
\bibAnnoteFile{yeo_solar_2017}

\bibitem[{Zharkova et~al.(2003)Zharkova, Ipson, Zharkov, Benkhalil, Aboudarham,
  and Bentley}]{zharkova_full-disk_2003}
Zharkova, V.~V., Ipson, S.~S., Zharkov, S.~I., Benkhalil, A., Aboudarham, J.,
  and Bentley, R.~D. (2003).
\newblock A full-disk image standardisation of the synoptic solar observations
  at the {Meudon} {Observatory}.
\newblock \emph{Solar Physics} 214, 89--105.
\newblock \doi{10.1023/A:1024081931946}
\bibAnnoteFile{zharkova_full-disk_2003}

\bibitem[{Zhu et~al.(2020)Zhu, Lin, Wang, and Yang}]{zhu_new_2020}
Zhu, G., Lin, G., Wang, D., and Yang, X. (2020).
\newblock A {New} {Approach} for the {Regression} of the {Center} {Coordinates}
  and {Radius} of the {Solar} {Disk} {Using} a {Deep} {Convolutional} {Neural}
  {Network}.
\newblock \emph{The Astrophysical Journal} 902, 72.
\newblock \doi{10.3847/1538-4357/abb2a0}
\bibAnnoteFile{zhu_new_2020}

\bibitem[{Zirin(1974)}]{zirin_studies_1974}
Zirin, H. (1974).
\newblock Studies of {K} line filtergrams.
\newblock \emph{Solar Physics} 38, 91--108.
\newblock \doi{10.1007/BF00161827}.
\newblock Number: 1
\bibAnnoteFile{zirin_studies_1974}

\bibitem[{Zuccarello et~al.(2011)Zuccarello, Contarino, and
  Romano}]{zuccarello_solar_2011}
Zuccarello, F., Contarino, L., and Romano, P. (2011).
\newblock Solar observations carried out at the {INAF} - {Catania}
  {Astrophysical} {Observatory}.
\newblock \emph{Contributions of the Astronomical Observatory Skalnate Pleso}
  41, 85--91
\bibAnnoteFile{zuccarello_solar_2011}

\end{thebibliography}

%%% Make sure to upload the bib file along with the tex file and PDF
%%% Please see the test.bib file for some examples of references

%\section*{Figure captions}

%%% Please be aware that for original research articles we only permit a combined number of 15 figures and tables, one figure with multiple subfigures will count as only one figure.
%%% Use this if adding the figures directly in the mansucript, if so, please remember to also upload the files when submitting your article
%%% There is no need for adding the file termination, as long as you indicate where the file is saved. In the examples below the files (logo1.eps and logos.eps) are in the Frontiers LaTeX folder
%%% If using *.tif files convert them to .jpg or .png
%%%  NB logo1.eps is required in the path in order to correctly compile front page header %%%

\end{document}